\documentclass{jfm}
\usepackage{mathrsfs,array,multirow}
\usepackage[normalem]{ulem}
\begin{document}

\newtheorem{lemma}{Lemma}
\newtheorem{corollary}{Corollary}

\shorttitle{Gravity-Capillary Lumps} 
\shortauthor{Masnadi and Duncan} 

\title{The Generation of Gravity-Capillary Solitary Waves by a Pressure Source Moving at a Trans-critical Speed}

\author
 {
 Naeem Masnadi\aff{1}
  \and 
James H. Duncan\aff{1}
 \corresp{\email{duncan@umd.edu}},
  }

\affiliation
{
\aff{1}
Department of Mechanical Engineering, University of Maryland, USA 20742
}

\maketitle

\begin{abstract}

The unsteady response of a water free surface to a localized pressure source moving at constant speed $U$ in the range $0.95c_\mathrm{min} \lesssim U \leq 1.02 c_\mathrm{min}$, where $c_\mathrm{min}$ is the minimum phase speed of linear gravity-capillary waves in deep water, is investigated through experiments and numerical simulations. This unsteady response state, which consists of a V-shaped pattern behind the source and features periodic shedding of pairs of depressions from the tips of the V, was first observed qualitatively by Diorio \emph{et al.} (\emph{Phys.\ Rev.\ Let.}, \textbf{103}, 214502, 2009) and called state III.  In the present investigation, cinematic shadowgraph and refraction-based techniques are utilized to measure the temporal evolution of the free surface deformation pattern downstream of the source  as it moves along a towing tank, while numerical simulations of the model equation described by Cho \emph{et al.}  (\emph{J. Fluid Mech.}, \textbf{672}, 288-306, 2011) are used to extend the experimental results over longer times than are possible in the experiments. From the experiments, it is found that  the speed-amplitude characteristics and the shape of the depressions are nearly the same as  those of the freely propagating gravity-capillary lumps of inviscid potential theory. The decay rate of the depressions is measured from  their height-time characteristics, which are well  fitted by an exponential decay law with an order one decay constant.   It is found that the shedding period of the depression pairs decreases with increasing source strength and speed.  As the source speed approaches $c_\mathrm{min}$, this period tends to about 1~s for all source magnitudes.   At the low-speed boundary of state III, a new response with  unsteady  asymmetric shedding of depressions is found.  This response is also predicted by the model equation.

\end{abstract}

\section{Introduction}

Solitary waves are localized disturbances that propagate with permanent form as a result of a balance between the opposing effects of nonlinearity and dispersion. Several types of solitary water waves are known to exist due to the  effects of surface tension and finite depth.  These waves bifurcate from the 
extrema of the linear dispersion curve, $c_p(k)$, where $c_p$ is the wave phase speed, $k=2\pi/\lambda$ is the wave number and $\lambda$  the wavelength.   The most familiar of these waves are the two-dimensional gravity solitary waves in the long-wave-length limit ($kH\rightarrow 0$, where $H$ is the undisturbed water depth), which corresponds to a global maximum ($c_p(0) = \sqrt{gH}$, where $g$ is the acceleration of gravity) of the linear pure gravity wave dispersion curve.  These solitary waves are governed by the celebrated Korteweg-de Vries equation (KdV) in the small amplitude limit. This theory was developed to explain the first scientific observations of these solitary waves by John Scott Russell in 1834 \cite[]{KdV,Russell}.

When surface tension effects are included, there are either one or two extrema of the dispersions curve, depending on the value of the Bond number $Bo=\tau/(\rho g H^{2})$, where  $\tau$ is the surface tension  and $\rho$ is the density of water. In the long wave limit, the phase speed $c_{0}=\sqrt{gH}$ is a local maximum (global minimum) of the linear dispersion relation when Bond number is less (greater) than 1/3 and KdV-type solitary waves of elevation (depression) exist.   Experimental observations of the KdV solitary gravity-capillary waves of both depression and elevation were reported in \cite{Falcon2002}.

It is well known that the dispersion curve also features a minimum at a finite wavenumber in water of infinite depth or finite depth when $Bo<1/3$. In deep water, $Bo = 0.0$, this minimum is $c_\mathrm{min}=(4\tau g/\rho)^{\frac{1}{4}}$ and happens at the wavenumber $k_\mathrm{min}=(\rho g/\tau)^{\frac{1}{2}}$. These values correspond to  $c_\mathrm{min}=23.13$ cm/s and $k_\mathrm{min}=3.67$ cm$^{-1}$ ($\lambda_\mathrm{min} =1.71$~cm) in clean water with $\tau$ taken as 73.0~mN/m.   Another class of two-dimensional solitary waves  that bifurcate at this minimum phase speed was discovered more recently by \cite{LH1989}. 
In the small-amplitude limit, these solitary waves are governed by the nonlinear Schr\"{o}dinger equation and behave as modulated wave-packets with the wave envelope moving with the wave crests at a speed slightly below $c_\mathrm{min}$ \cite[]{Akylas1993,LH1993}. These solitary waves were observed in experiments by \cite{Zhang1995} and \cite{LHZhang1997}. For a review of two-dimensional gravity-capillary solitary waves see \cite{DiasKharif1999}.

Three-dimensional solitary waves (usually referred to as ``lumps'') can also bifurcate at the extrema of the linear phase speed. A necessary condition for a lump to remain localized in all spatial directions is that no other linear wave can co-propagate with the speed of the lump \cite[]{Milewski2005}. This condition can only be achieved at a minimum of the linear phase speed, which can occur at zero or a finite wavenumber depending on the Bond number.

As mentioned above, when $Bo>1/3$, the phase speed has a global minimum at zero wavenumber. Small amplitude three-dimensional solitary waves in this condition are governed by the Kadomtsev-Petviashvili I (KP-I) equation, which is a natural extension of the KdV equation to  three dimensions, or by the Benny-Luke equation with surface tension. Both of these equations admit depression lump solutions \cite[]{Berger2000}. However, the condition for the Bond number restricts the water depth to at most a few millimeters and the viscous effects at the bottom boundary, which are not included in the above theories,  become important.

\begin{figure}
\begin{center}
\includegraphics[width=3.5in]{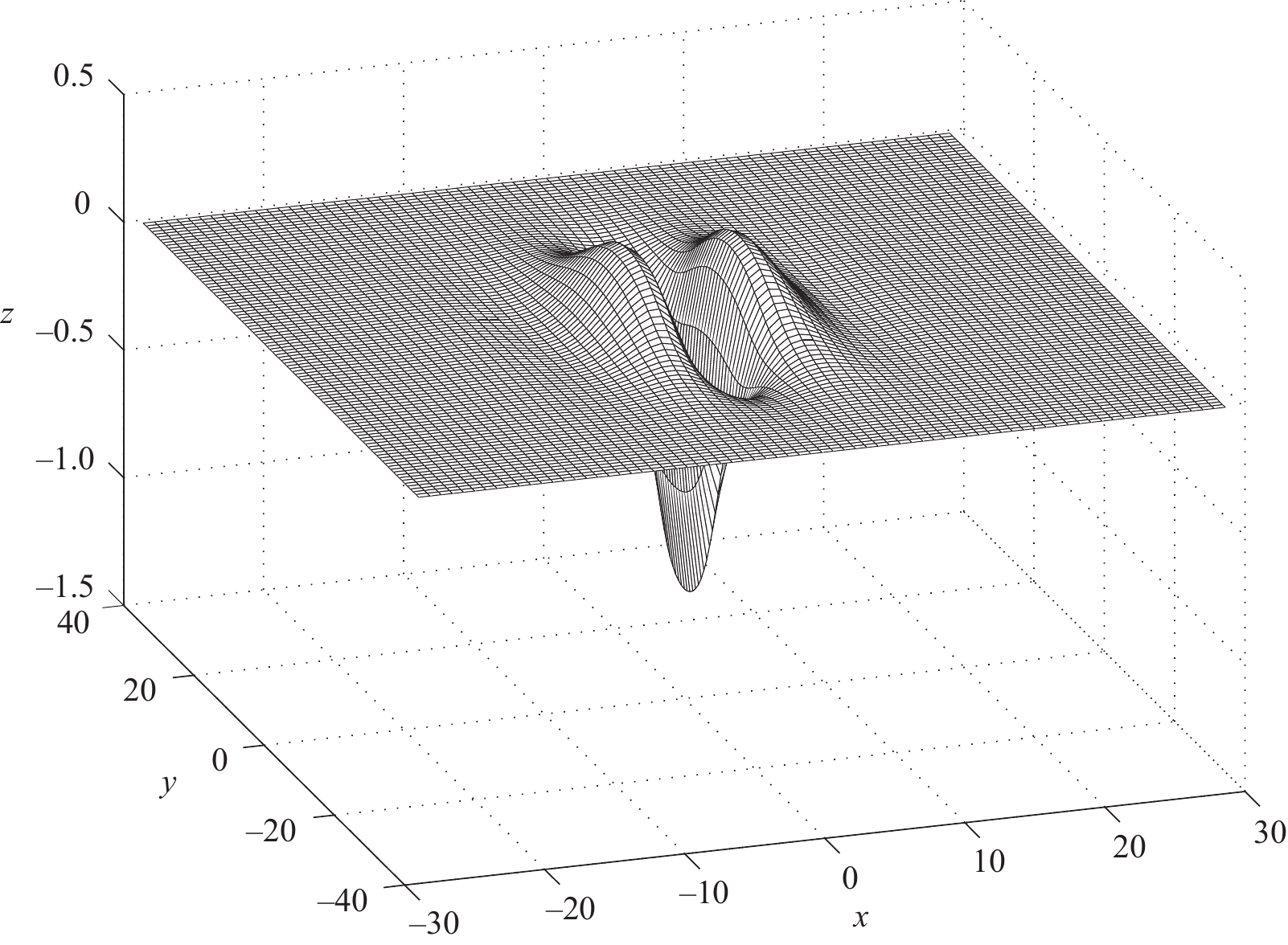}
\caption{A freely propagating three-dimensional lump moving at speed $0.917 c_\mathrm{min}$, from figure 1 of \cite{Parau2005}.  The calculations  use a fully-nonlinear potential flow model.}
\label{fig:Parau_Lump}
\end{center}
\end{figure}

When $Bo<1/3$,  three-dimensional solitary waves of elevation and depression bifurcate at the above-mentioned $c_\mathrm{min}$ that occurs at finite wavenumber. Profiles of free lumps from fully nonlinear potential flow equations were calculated by \cite{Parau2005}. These lumps are fully-localized in all directions and are more extended in the transverse direction (figure \ref{fig:Parau_Lump}). Similar to their 2-D counterparts, in the small-amplitude limit, these waves behave as modulated wavepackets with envelope and crests moving at the same speed \cite[]{Kim2005}. These wave-packet solitary waves can be viewed as a continuation of the long solitary waves in the surface tension dominated regime as $Bo=1/3$ is crossed \cite[]{Milewski2005}. The dynamics of these lumps, including their stability and interactions, have been explored using model equations \cite[]{Akylas2008, Akers2009, Akers2010, Milewski2012}.

Moving a localized free-surface pressure source or bottom topographic feature at speeds near one of the critical speeds, $c_{0}$ or $c_\mathrm{min}$, is a common method of generating solitary waves. These extrema speeds are associated with a resonant condition: The linear solution to a disturbance moving at these speeds becomes unbounded and naturally, the nonlinear effects become important.
In two dimensions, the nonlinear response of a water surface to a steady pressure distribution moving at speeds close to $c_{0}$ features periodic shedding of solitary waves upstream of the disturbance \cite[]{Akylas1984}. The energy input to the system due to the pressure source moving at trans-critical speeds cannot be radiated away since the dispersive effect is relatively weak. The surface response will grow and, since for 2-D gravity waves the higher amplitude waves move faster, the local wave moves ahead of the source and a solitary wave is generated. This process is then repeated periodically \cite[]{Wu1987}.
 
In the regime of strong surface tension ($Bo>1/3$), this problem was considered in three dimensions by \cite{Berger2000}  using a forced generalized Benny-Luke equation. Periodic shedding of 3-D lumps was observed downstream of the source. These lumps moved at an angle with respect to the direction of the source motion. The Benny-Luke equation is inviscid and since viscous effects become important in shallow water flows with $Bo>1/3$, the results may not be physically realizable.

\begin{figure}
\begin{center}
\begin{tabular}{cc}
  (\textit{a})&(\textit{c})\\
  \includegraphics[width=2.5in]{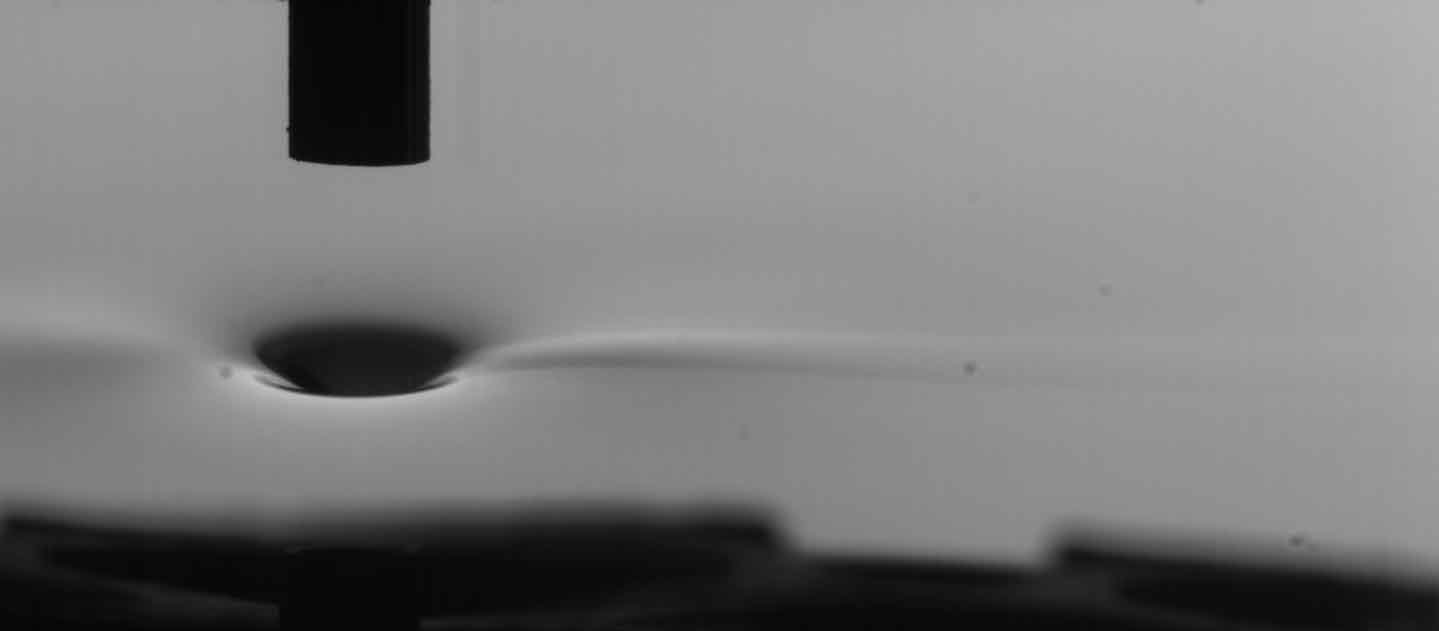}&
  \includegraphics[width=2.5in]{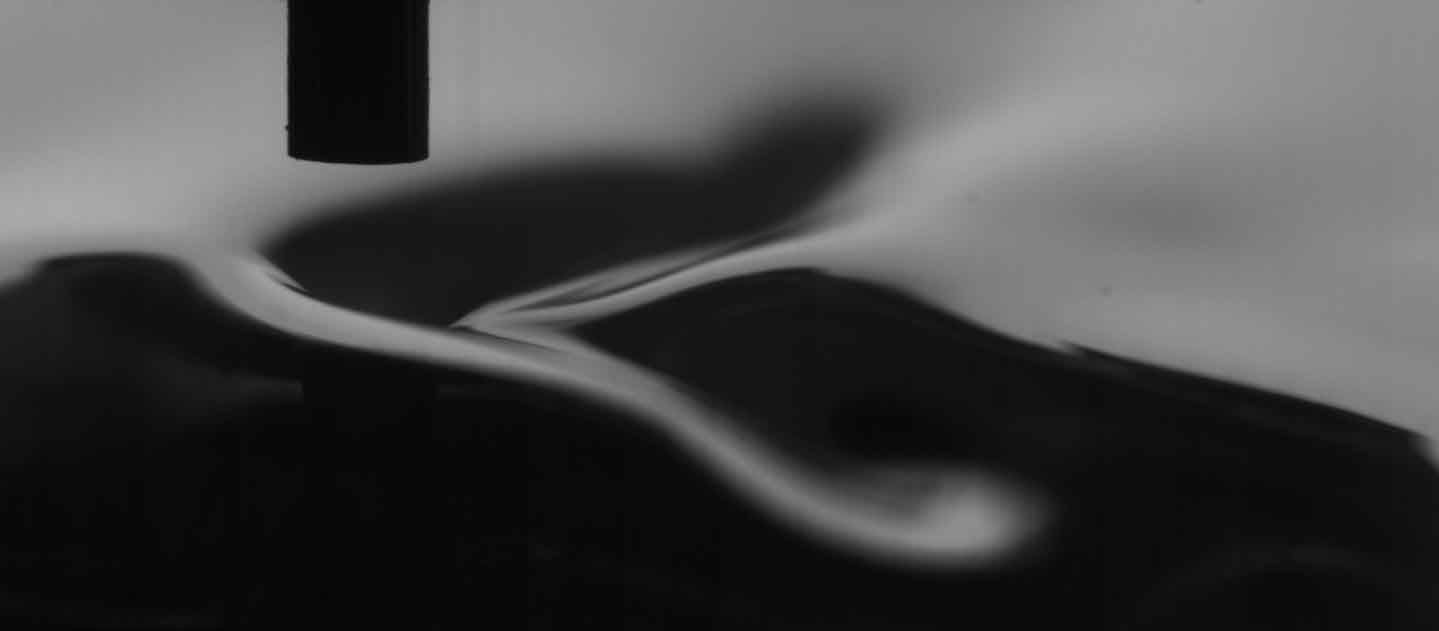}\\
  (\textit{b})&(\textit{d})\\
  \includegraphics[width=2.5in]{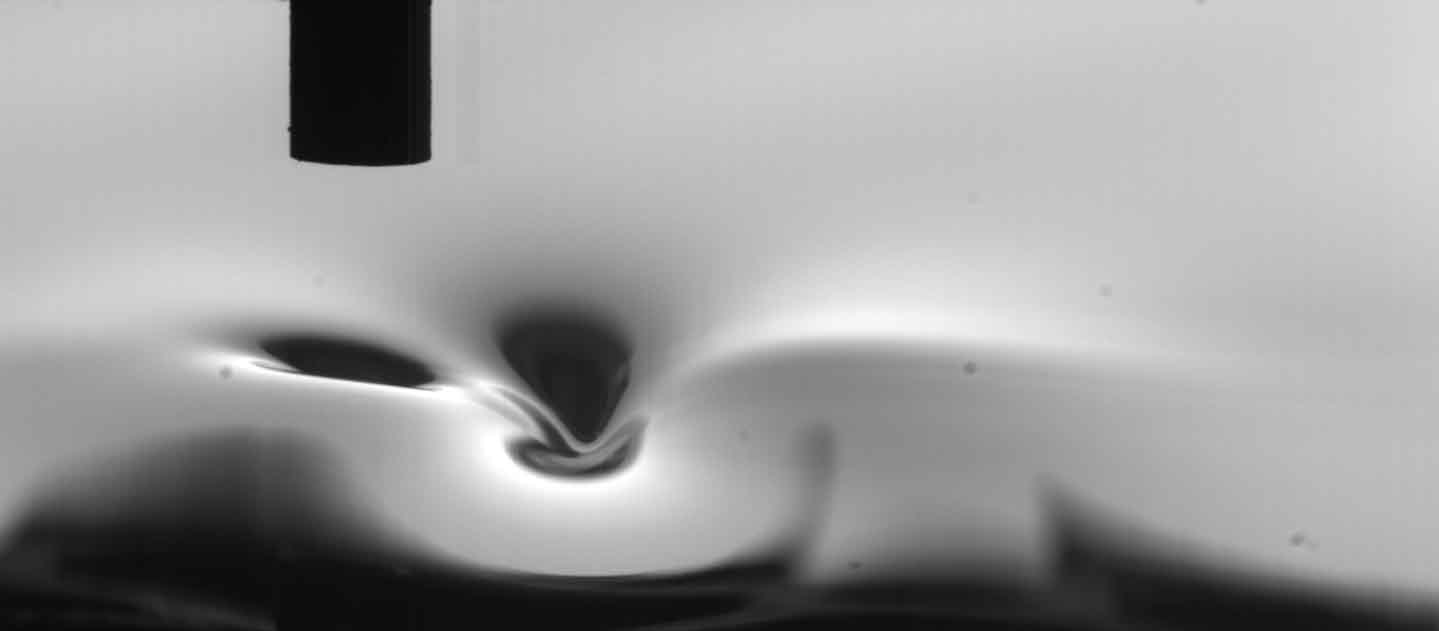}&
  \includegraphics[width=2.5in]{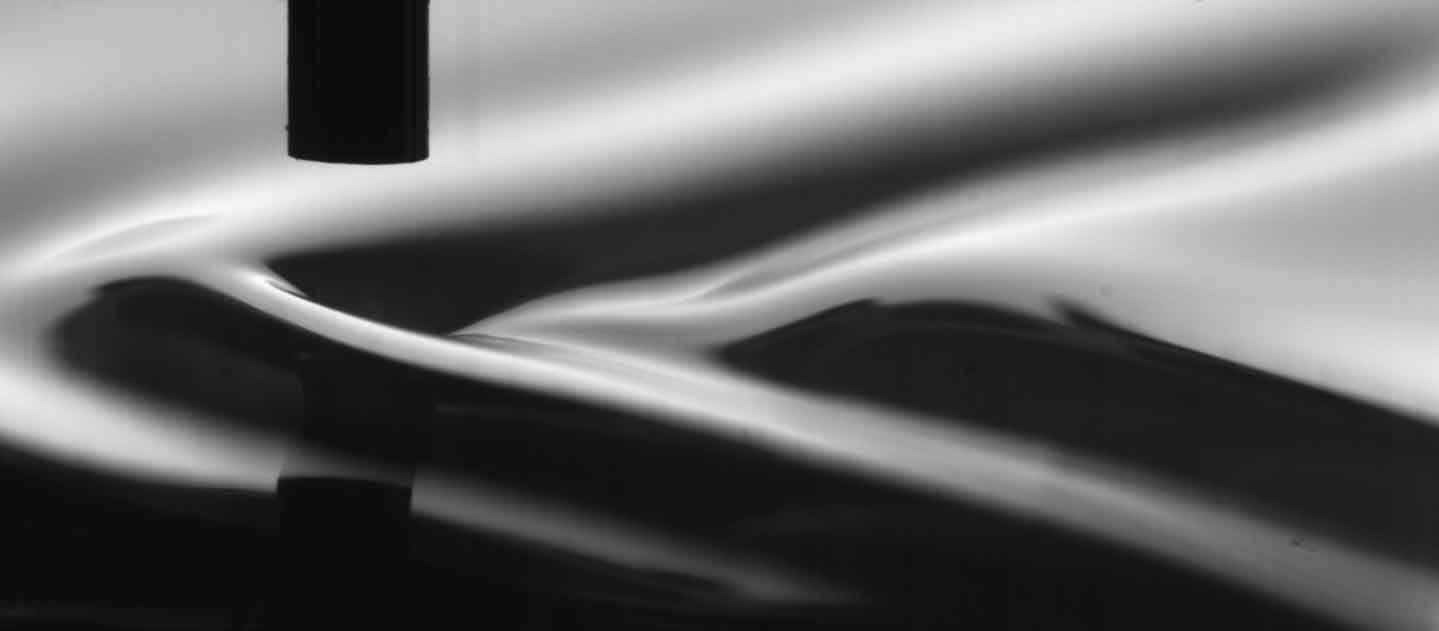}\\
\end{tabular}
\end{center}
\caption{The wave pattern behind a pressure source moving at a constant speed close to $c_\mathrm{min}$, from \cite{DiorioPRL}, figure 1. The images are taken by a high-speed camera positioned to the side of the air-jet tube and above the water surface. The tube is moving from right to left. (\textit{a}) source speed $U =0.905c_\mathrm{min}$, state I: a small depression under the tube. (\textit{b}) $U=0.970c_\mathrm{min}$, state II: wave pattern moves behind the pressure source and is elongated in cross-stream direction. (\textit{c}) $U=0.981c_\mathrm{min}$, state III: unsteady V-shape pattern with lump-like disturbances shedding from the tips of the V. (\textit{d}) $U=1.03c_\mathrm{min}$, linear pattern.}
\label{fig:Diorio}
\end{figure}

In deep water, the surface response to a pressure distribution moving at speeds close to $c_\mathrm{min}$ was investigated experimentally by \cite{DiorioPRL}. In this study, the wave pattern behind a small pressure distribution created by a vertical air jet  that was set to move horizontally at constant speed below the minimum phase speed was observed by a variety of photographic measurement techniques. They identified three response states for the wave pattern as the towing speed of the pressure source approached $c_\mathrm{min}$ (figure \ref{fig:Diorio}). At low speeds, the response is similar to that found when the air jet is stationary (state I) and is essentially a steady circular depression under the pressure source. As the speed is increased, an abrupt transition occurs and a steady wave-like state in the form of a solitary depression is formed behind the pressure source (state II). The amplitude of the water surface depression is increased significantly compared to that in state I. The wave pattern is steady and is elongated in the cross-stream direction. This pattern resembles the freely propagating lumps calculated by \cite{Parau2005} (see figure \ref{fig:Parau_Lump}). At even higher translation speeds but still less than $c_\mathrm{min}$, a third response state (state III) appears and the wave pattern becomes unsteady and features a V-shaped pattern with periodic shedding of lump-like disturbances from the tips of the V. As the speed is increased above $c_\mathrm{min}$, the pattern becomes similar to the steady, V-shaped, linear pattern with surface tension-dominated waves ahead of the source and gravity dominated waves behind it. \cite{DiorioJFM} reported detailed measurements of the state II response and found a one-to-one relation between the amplitude and the phase speed of the waves (a known feature of solitary waves) and compared it to the bifurcation curves from theory. The techniques employed by \cite{DiorioJFM} were suited only to steady wave patterns and therefore, they were not able to make measurements of the state III response or to accurately determine the transition between state II and state III. \cite{ChoJFM} used a model equation with quadratic nonlinearity and a viscous dissipation term to capture the steady and unsteady behavior of the wave pattern generated by a pressure source moving at a speed near $c_\mathrm{min}$. Their results compared favorably with the experimental results of \cite{DiorioJFM}. It was found that viscous damping plays an important role in the formation of both the steady response pattern in state II and the unsteady pattern in state III.

In the present paper, detailed measurements of the unsteady wave pattern behind a pressure source moving with speeds close to $c_\mathrm{min}$ and with a range of pressure magnitudes are presented. The water depth is chosen so that the Bond number is  much smaller than 1/3 and the problem is considered to be in deep water. These measurements were used for two purposes: (i) to explore the state III response including the periodic shedding and propagation of unsteady localized depressions and (ii) to investigate the effect of viscous damping on these depressions.

The paper is structured as follows: The experimental details and the measurement techniques are described in \S \ref{sec:ExpDetails} and \S \ref{sec:Measurement_methods}, respectively.  This is followed in \S \ref{sec:numerical_model} by a  brief overview of the model equation described in \cite{ChoJFM} and the numerical procedure used to solve it. The experimental and numerical results are then presented and discussed in  \S \ref{sec:Results}.  Finally, the conclusions of this study are given in \S \ref{sec:Conclusions}.

\section{Experimental set-up}\label{sec:ExpDetails}

\begin{figure}
\begin{center}
\includegraphics[width=5in]{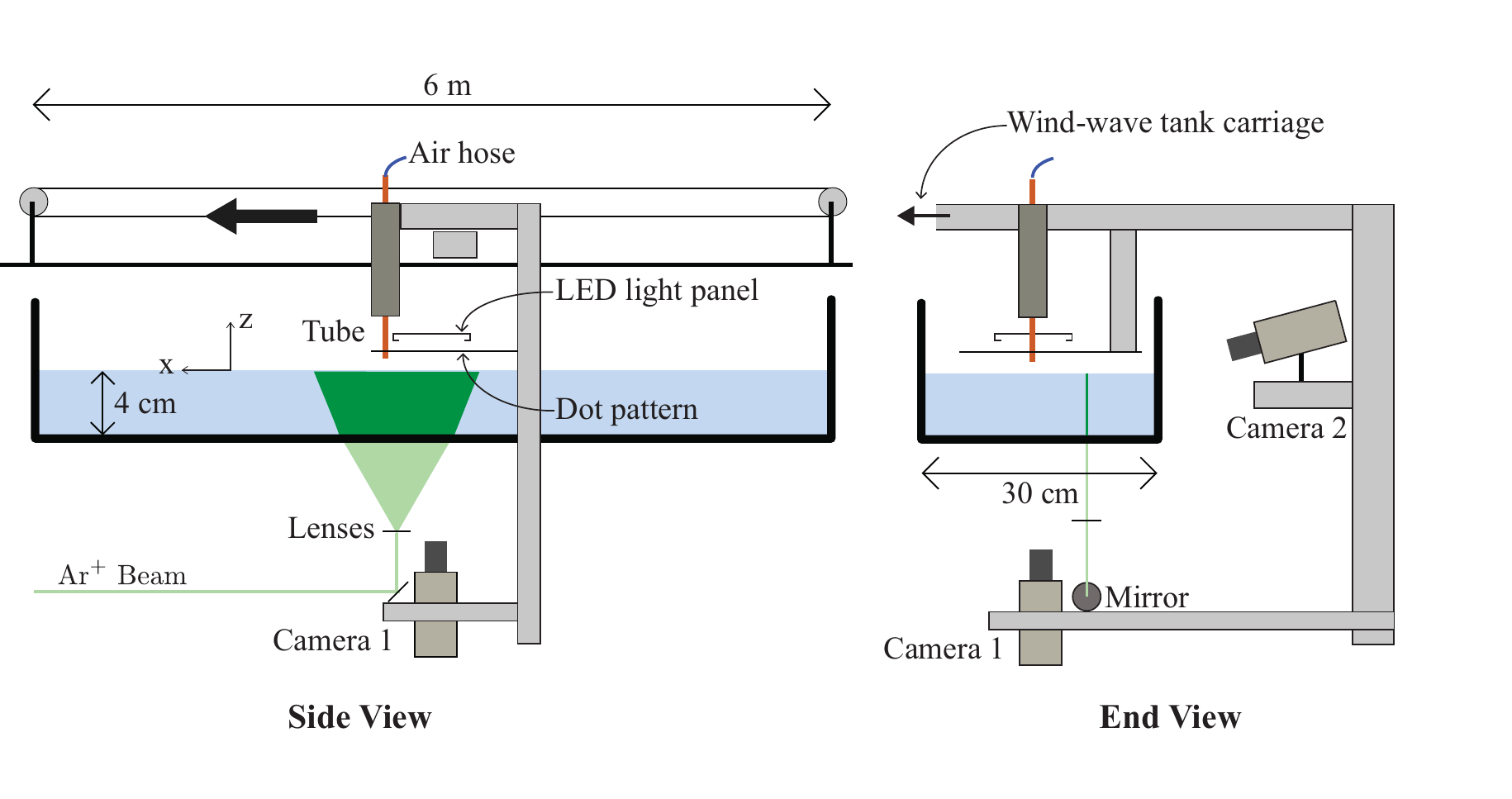}
\end{center}
\vspace{-0.05in}
\caption{ Schematic of the experimental set-up. The tank has clear walls and is positioned about 2.0~m above the laboratory floor.  The water surface is excited by a vertically oriented jet of air exiting from a small-diameter tube that is mounted on the moving carriage.  Two high-speed cameras that are also mounted on the carriage are used to measure the wave pattern around the air jet impingement point.  Camera 1 views, through the tank bottom and the wavy water free surface, a dot pattern held just above the water surface.  Camera 2 views, from the side of the tank,  the intersection of a vertically oriented laser light sheet and the water surface.  }
\label{fig:Lump_tank}
\end{figure}

The experiments were performed in an open-surface towing tank that is 6~m long, 30~cm wide\footnote{That the tank is sufficiently wide to eliminate the effects of the tank sidewall on the results is addressed in \S \ref{sec:Results}.} and 7~cm deep (see figure \ref{fig:Lump_tank}). The side walls and bottom of the tank are made of 0.64-cm-thick clear polycarbonate sheets which are supported by an external aluminium frame. The aluminium frame  is attached to the steel frame of a large wind-wave tank (14.8~m long, 1.2~m wide and 2.4~m high) such that the towing tank is at an elevation of about 2~m above the laboratory floor.  An instrument carriage travels along the length of the wind-wave tank and so along the length of the towing tank. The carriage is supported by four hydrostatic oil bearings that ride on a pair of  precision rails positioned on the top of each sidewall of the wind-wave tank. This bearing system provides a very low vibration level.  The carriage rails where leveled by comparing the carriage height with the calm water surface in the towing tank and after adjustment the carriage height varied by at most  0.4~mm along the length of the tank.  The carriage is driven  by steel cables that are in turn driven by a servo motor.  The carriage motion is controlled by a computer-based feedback system that employs a carriage position sensor that runs the length of the tank. In all the experiments presented in this paper, the carriage is set to accelerate to a constant speed within the first 25.4~cm of the translation and decelerate to zero speed in the last 25.4~cm of its motion. This restriction creates an average acceleration and deceleration of about 10~cm/s$^2$ for the range of towing speeds used. Evaluation of the carriage position sensor data show that the carriage position at any instant in time is repeatable within $\pm 0.5$~mm from run to run and the average speed of the carriage during its constant speed motion has a relative error of less than 0.01~percent.  In the following, the carriage speed $U$ is indicated  by the dimensionless parameter, $\alpha = U/c_\mathrm{min}$.

A pressure disturbance is made on the water surface by blowing air through a 2.5-mm-ID tube that is attached to the instrument carriage.  The tube is oriented vertically with the bottom end of the tube positioned $7.0\pm0.1$~mm above the water surface. The tube is embedded in silicon and sandwiched between two aluminum plates to increase its stiffness.  The pressure source for the jet is a 37-l  compressed air tank that is mounted on the carriage and is connected to the air-jet tube via a system consisting of a pressure regulator, a needle valve, a flow meter and flexible hoses.  The compressed air tank is pressurized to about 2.7~bar before each experimental run.  When the carriage is stationary, the air flow makes a small axisymmetric depression on the water surface directly under the tip of the air-jet tube. The Reynolds numbers based on the average air flow velocity in the tube and the tube ID were in the range from 800 to 1200 and the flow seemed to be laminar, although no conditioning was done to the flow before entering the tube.

Since surface tension effects are of utmost importance in this experiment, the following water preparation and surface tension measurement procedures were followed to minimize and monitor   surfactant levels. First, at the beginning of the day of each series of experiments, a separate water treatment tank with an approximate volume of 1~m$^{3}$ was filled with filtered tap water. Then, chlorine was added to the water to reach a concentration of about 10~ppm. The water was then circulated through a diatomaceous earth filter for about eight hours before the start of the experiments on that day. The water was then dechlorinated by adding an  appropriate amount of hydrogen peroxide.  This latter step is necessary because high chlorine levels degrade the fluorescein dye used in the Laser Induced Fluorescence (LIF) surface profile measurements described below.  The dye is then added to the water at a concentration of about 0.7~ppm.  This cleaned dye solution was then used to fill the towing tank to a depth of approximately $H=4$~cm (corresponding to a Bond number of $Bo=4.65\times10^{-3}$). The towing tank includes a water surface skimming system and for ten minutes before each experimental run, water was pumped from the treatment tank through the towing tank and to the drain via the skimmer. Periodically during the experiments on each day, samples of the towing tank water were extracted and surface tension isotherms of the samples were measured using a KSV NIMA Langmuir trough.  In these measurements, the surface tension in the trough is measured with a Wilhelmy plate as the local water surface  is compressed by moving Teflon barriers  toward the measurement site at a constant rate.  These barriers barely touch the water surface.  The resulting surface compression increases the number of surfactant molecules per unit water surface area while desorption of the surfactant to the bulk fluid decreases the surface concentration at a slower rate.  It was found that the surface tension before compression was maintained at $73.0\pm0.5$~dynes/cm (the value for clean water) throughout the experiments  and in all cases the surface tension after compression of the water surface area by 75 percent over a 60-second period resulted in a drop in surface tension of less than 0.5~dynes/cm. As the barriers continued to move creating even higher compressions, the surface tension eventually experienced a sudden drop at compressions ranging from 80 to 95 percent.  The water temperature was also measured and maintained at $25\pm1$~C.

\section{Measurement details}
\label{sec:Measurement_methods}

Measurements of the free-surface deformation patterns were carried out using several photography-based techniques. The objective of the first measurement is to obtain the relationship between the airflow rate in the brass tube and the depth of the depression made on the water surface while the carriage was stationary. For these measurements, a still image camera (Nikon D800E, 36 mega-pixel sensor) was placed on a tripod and was oriented horizontally so that it viewed the air-jet tube and water surface from the side. An LED light panel  (made by Phlox) was used for illumination and was placed on the opposite side of the tank from the camera.   The camera was positioned vertically so that the depression under the air-jet tube was viewed from just below the mean water surface. With this configuration, the surface depression appears dark in the images and the image resolution in the focal plane of the depression was 5.9 $\mu$m/pixel.   We will refer to this measurement technique as the shadowgraph method.

\begin{figure}
\begin{center}
\begin{tabular}{ccc}
 \multirow{2} {*}{\includegraphics[width=2.75in]{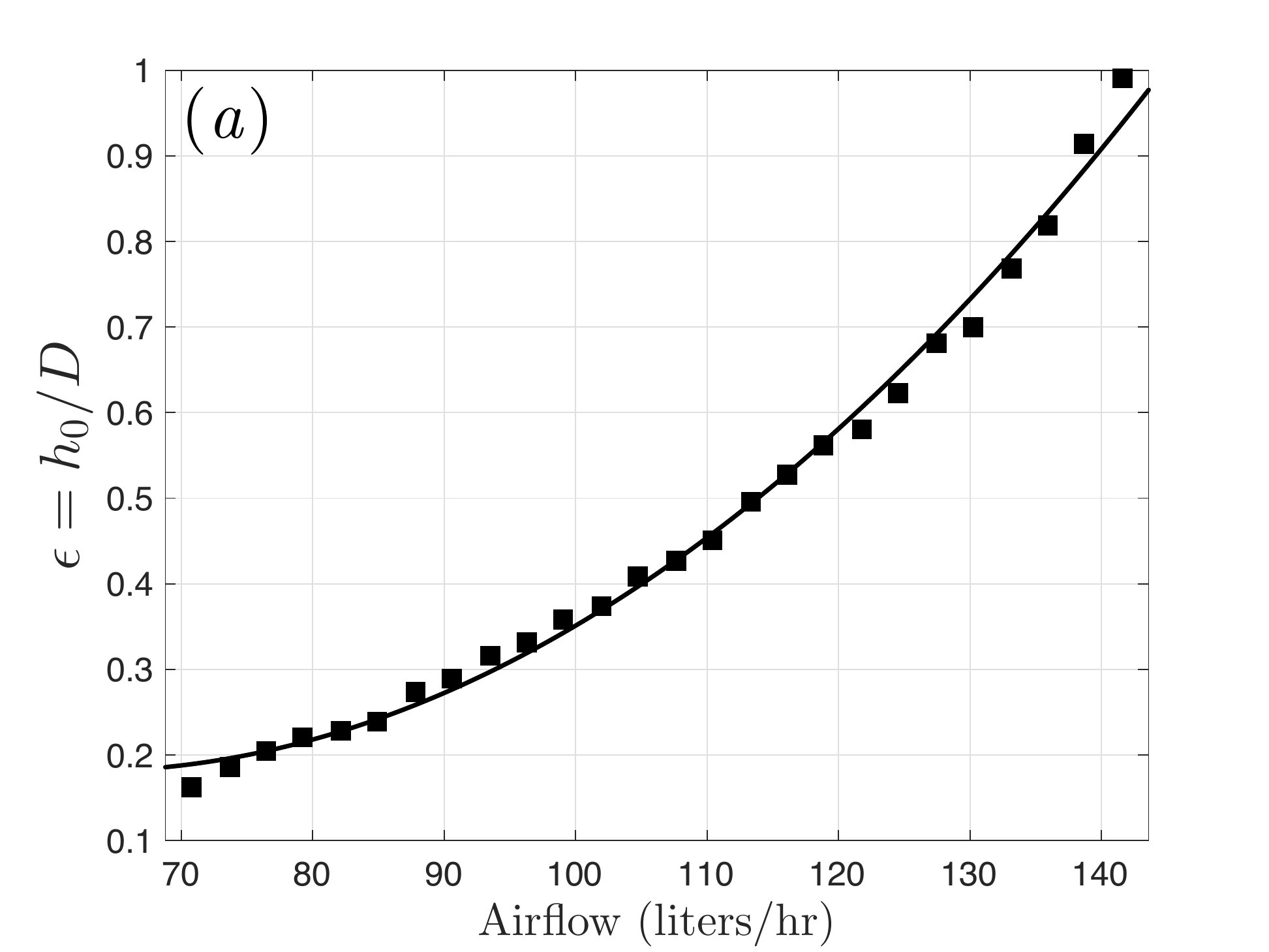}}&&\\
& \includegraphics[width=1in]{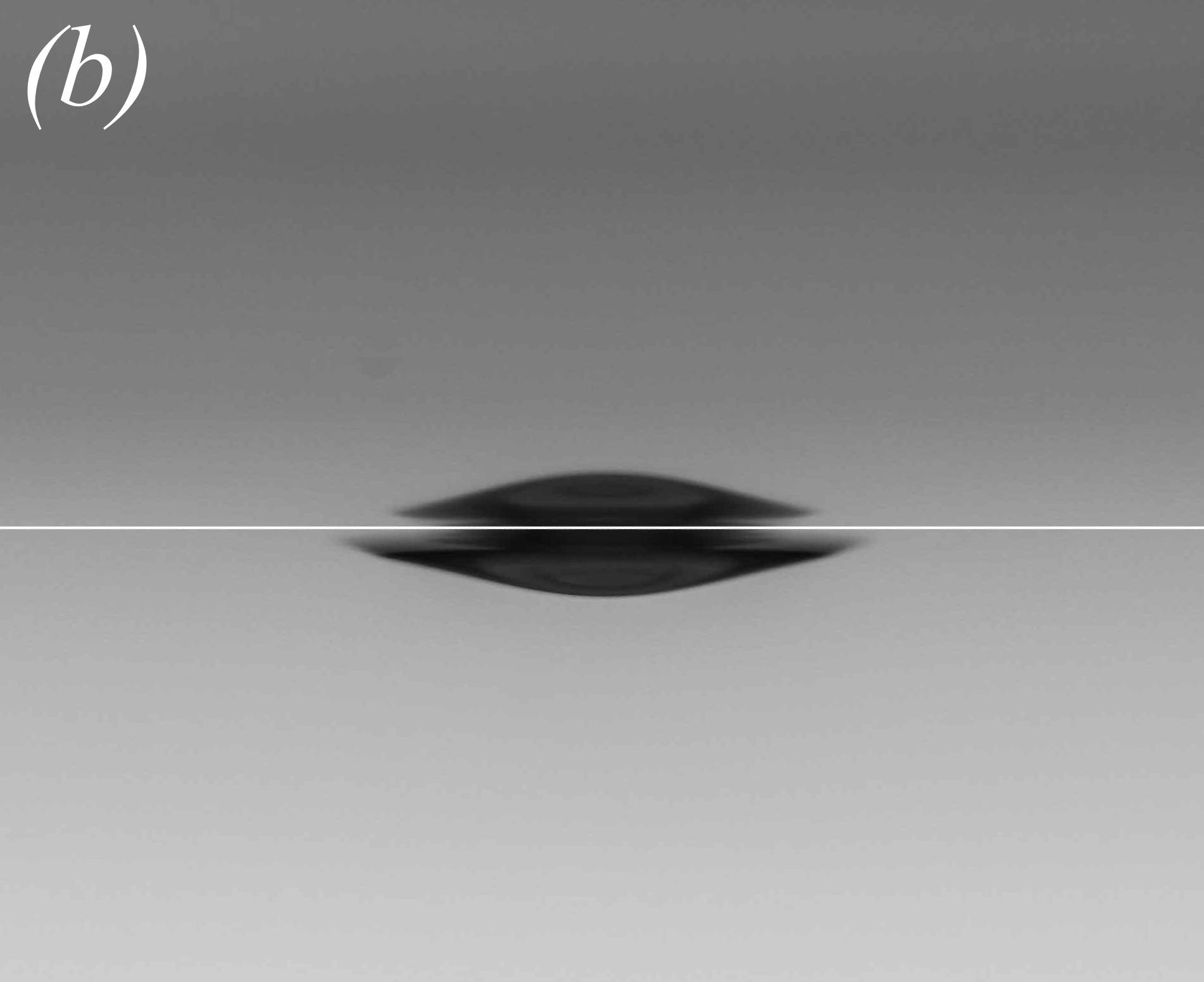}&\includegraphics[width=1in]{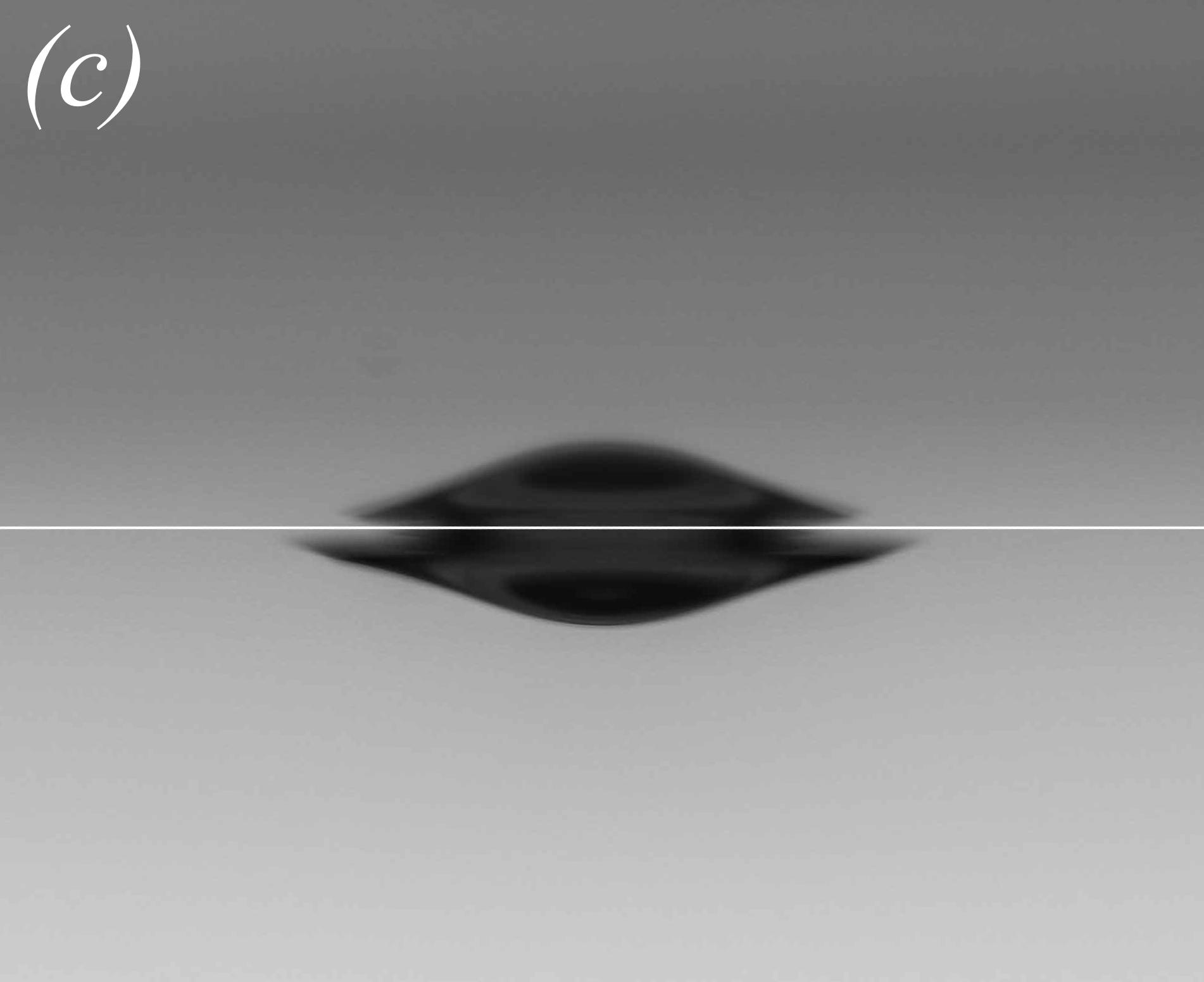}\\
&\includegraphics[width=1in]{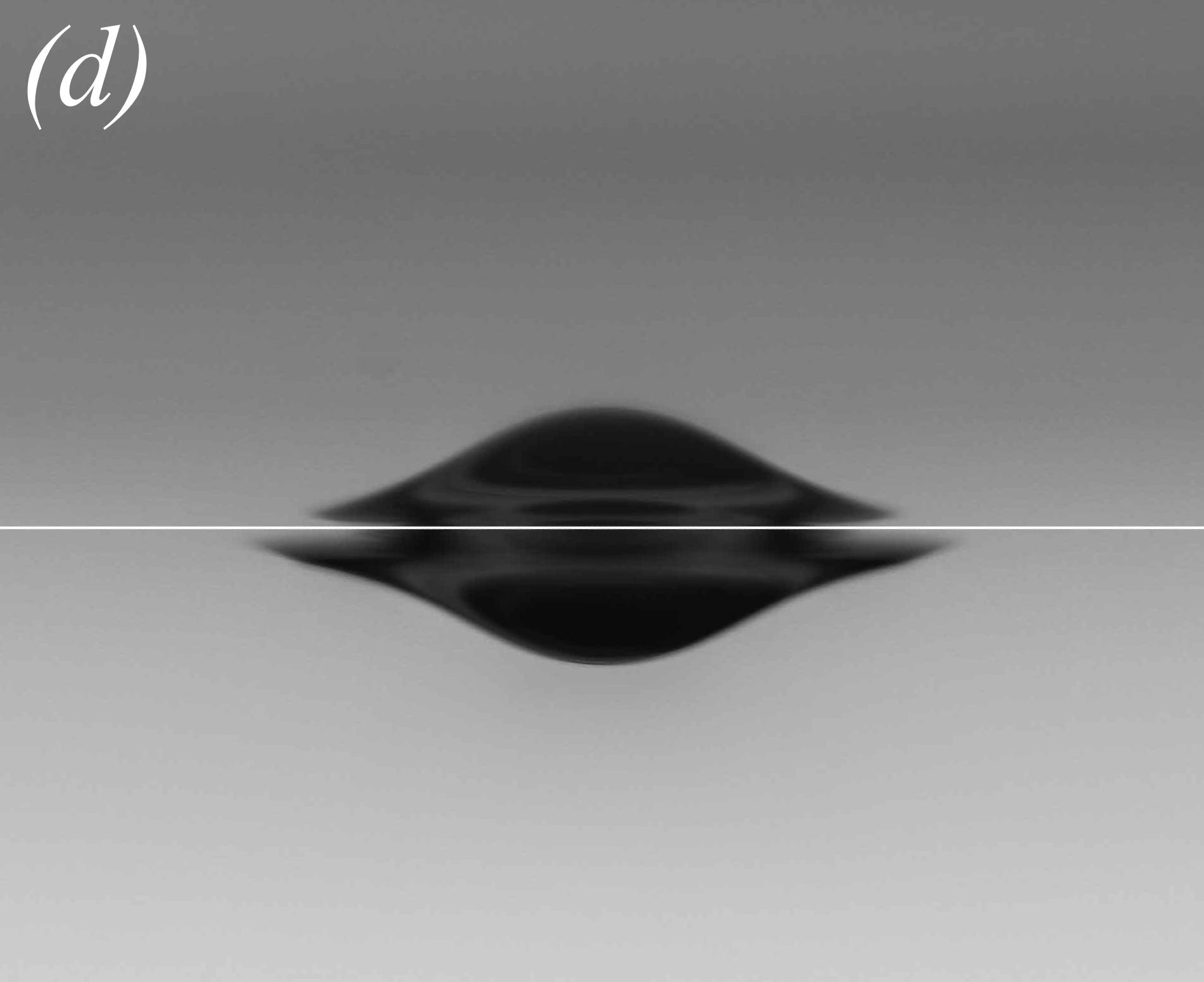}&\includegraphics[width=1in]{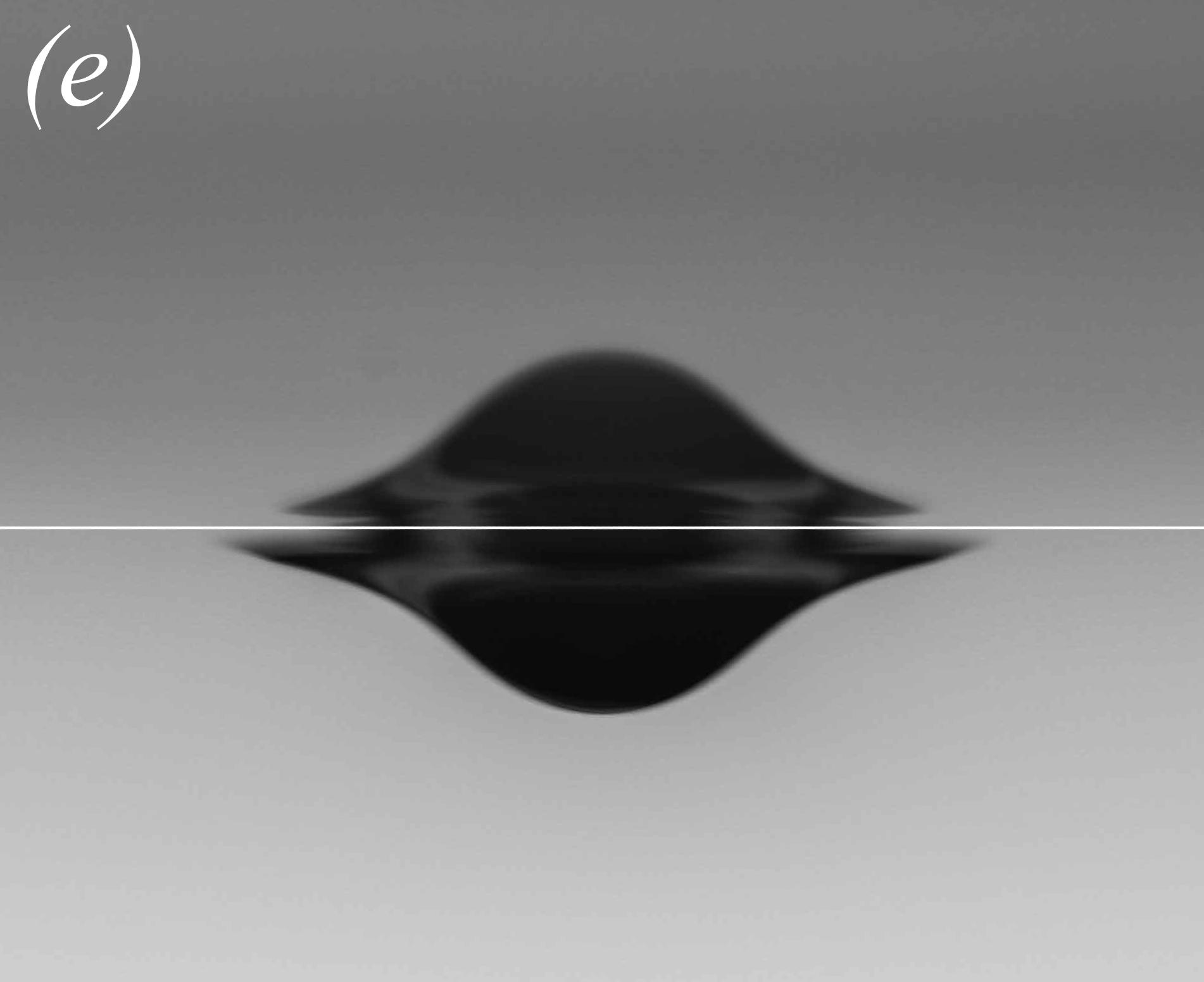}
\end{tabular}
\end{center}
\vspace{0.25in}
\caption{ Characteristics of the dimple when the air-jet tube is stationary. (\textit{a}) Plot of the $\epsilon = h_0/D$ versus air flow rate.  The black squares are experimental data and the solid line is the fit of a second order polynomial to the data.  Images (\textit{b}) through (\textit{e}) are shadowgraphs of the dimple created with the air-jet tube held stationary for air flow rates 85.0, 99.1, 113.3 and 127.4~l/hr, respectively.  The white lines are the locations of the undisturbed water surface.  The part of the image above the white line are the reflections of the dimple in the water surface.  The images are 11.0~mm wide in the object focal plane.}
\label{fig:Dimples}
\end{figure} 

Following the notation of \cite{DiorioJFM} we define the forcing parameter $\epsilon=h_{0}/D$ where $h_{0}$ is the depression depth when the air-jet tube is stationary and $D$ is the internal diameter of the tube. With the above-described measurements, $\epsilon$ was measured to an accuracy of about $\pm 0.01$. A calibration was performed to relate the air-jet flow rate to $\epsilon$ for a range of airflow rates  corresponding to $0.15\leq\epsilon\leq1$. A plot of $\epsilon =h_0/D$ versus air flow rate  and shadowgraph images of the dimple at four air flow rates are given in figure~\ref{fig:Dimples}. Since the air jet flow rate must remain constant during any experimental run, $h_0$ was also measured over a period of  400 seconds after opening the air valve. This time interval is much longer than the duration of individual runs.   It was found that $h_0$  varied by less than five percent.   As mentioned earlier, the carriage tracks are not perfectly level; however, the vertical distance between the tube and the water surface varies by less than four percent.

The shadowgraph method was also used in qualitative observations of the surface response for a wide range of  experimental parameters. In these observations, the still image camera used in the $h_0$ measurements was replaced by a high-speed digital movie camera (Vision Research, Phantom V641, $2560\times1600$ pixel images), which was mounted to the carriage so that it moved along the tank with the air-jet tube. The LED panel was placed at a fixed position about 3~m from the starting point of the carriage motion and on the opposite side of the tank from the camera to provide backlighting. The camera was triggered to capture images as it passed by the LED panel and those images were used to determine the response state (see \S \ref{sec:State_diagram}). 

Quantitative measurements of the unsteady wave pattern behind the air-jet were made via an optical  refraction-based technique (called the refraction method in the following) that  is similar to the method described in \cite{Moisy2009}. In the implementation of this method, a computer-generated image of a pattern of randomly placed black dots is printed on a translucent matte paper (Grafix Matte 0.005 Dura-Lar Film). This pattern is attached to the bottom of a clear plastic plate which is attached to a traverser and held level at an adjustable height above the calm water free surface.   The dot pattern is lit from above by the above-described LED panel and is photographed from under the tank with a vertically oriented high-speed digital movie camera (Phantom V641, Vision Research).  The camera is  mounted 1~m below the tank (camera 1 in figure \ref{fig:Lump_tank}) and a long focal length lens (200~mm) is used to image a small surface area.  As shown in the figure, the dot pattern, the LED panel and the camera are mounted on  the instrument carriage and so move along the tank with the air-jet tube.  The dot pattern was imaged at a resolution of about 26 pixels per millimeter. An optical filter is placed in front of the camera lens to block most of the light from the LIF method described below. The camera is triggered by a photo diode that is placed on one side of the tank.  A low-power laser beam is pointed at the diode from the other side of the tank. When the carriage passes this location, a knife edge mounted on the carriage blocks the laser beam  and the diode sends a signal to trigger the camera. This method was found to be repeatable to within  two frames in the camera image sequences.  For a carriage speed of 23~cm/s and the  camera frame rate of 300~fps used in the experiments, the carriage travels 0.77 mm between frames.    

\begin{figure}
\setlength\fboxsep{0pt}
\begin{center}
\begin{tabular}{cc}
  (\textit{a})&(\textit{b})\\
  \fbox{\includegraphics[width=2.5in]{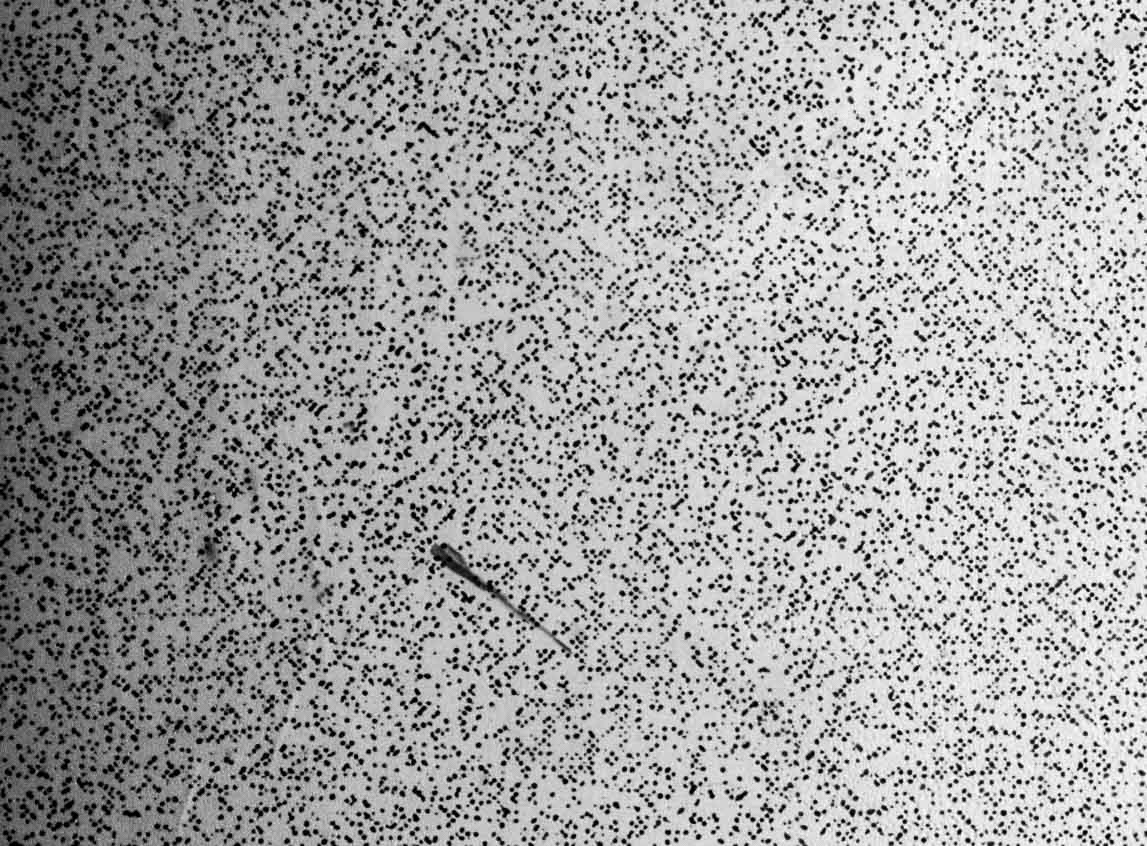}}&
  \fbox{\includegraphics[width=2.5in]{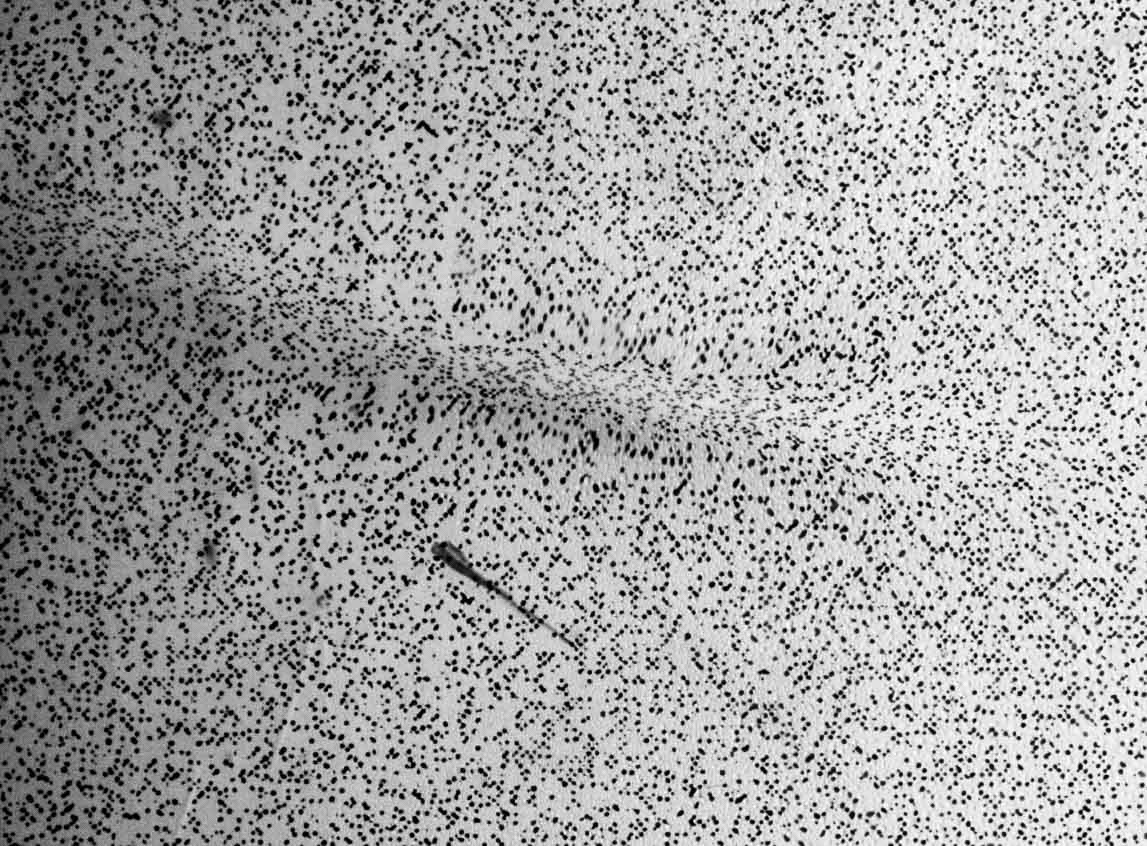}}\\
\end{tabular}
\end{center}
\caption{Sample cropped images of the dot pattern for $\alpha =U/c_\mathrm{min}= 0.994$ and $\epsilon = h_0/D=0.24$. These images are approximately 5~cm wide in the physical plane. The actual field of view is about 10~cm$\times$6~cm. For quantitative measurements of the surface height, the dot pattern is placed very close to the water surface (about 5~mm, as in the images shown here) to make the distortions small. (\textit{a}) A reference image through calm water surface.  (\textit{b}) Corresponding distorted image through disturbed water surface.}
\label{fig:refraction1}
\end{figure}

In the first step of the refraction method measurement procedure, a reference movie is taken when the air jet is off and the water surface is calm. A second movie is then taken through the wavy water surface in a separate run with the air jet on. The measurement technique is based on comparing images that were taken at the same location in the tank in separate runs.  Because the camera is triggered by the fixed laser trip device as described above, a given location in the tank corresponds to a specific frame number in all movies.  If the tank bottom was completely level, flat  and uniform and the carriage remained at exactly the same height above the water surface during the entire run, the images in the reference movie would all be identical and a single reference image taken anywhere in the tank would be sufficient.  The frame-by-frame comparisons between the reference and measurement movies is used to reduce errors caused by whatever slight imperfections may be present.

\begin{figure}
\begin{center}
\begin{tabular}{cc}
  (\textit{a})&(\textit{b})\\
  {\includegraphics[width=2.5in]{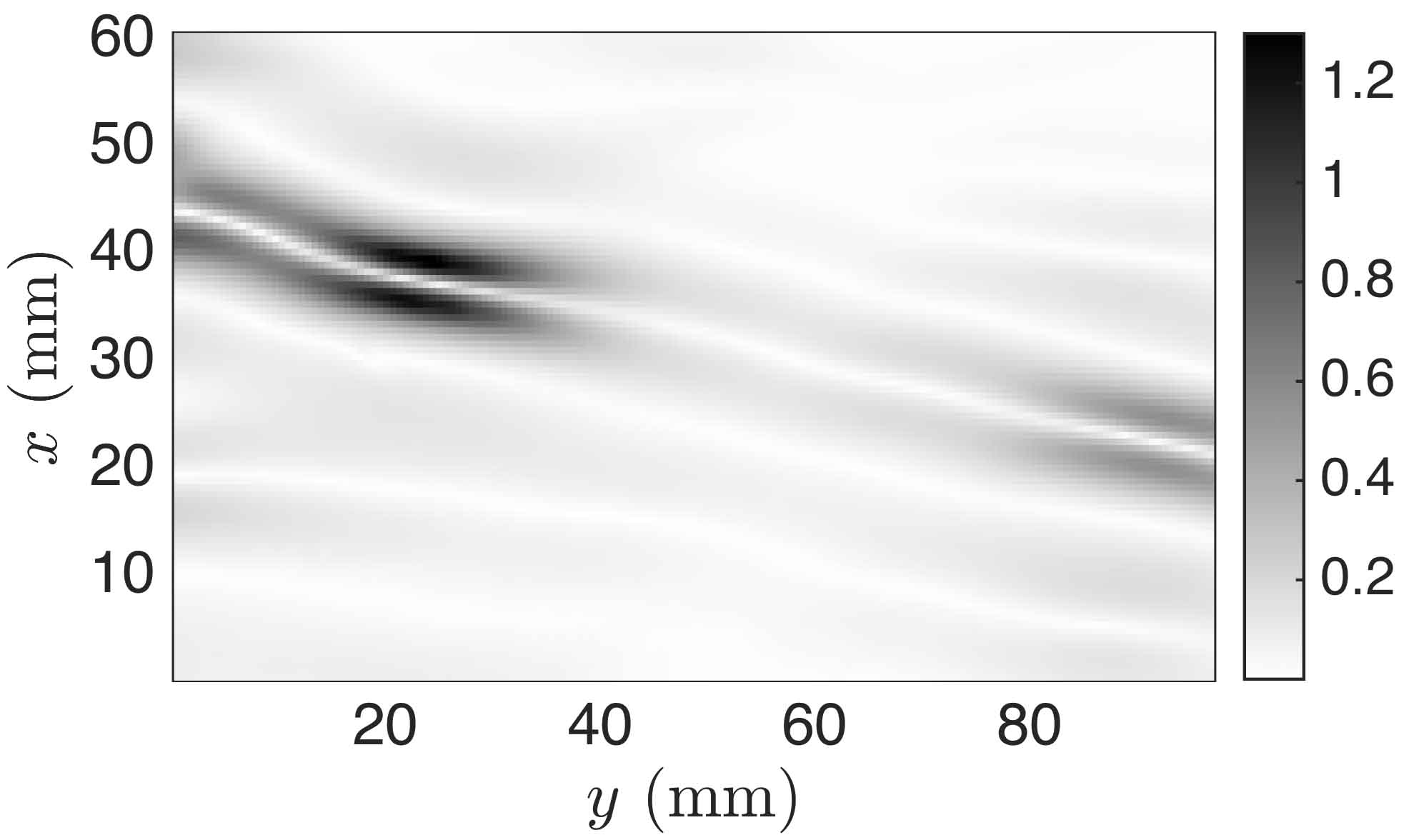}}&
  {\includegraphics[width=2.5in]{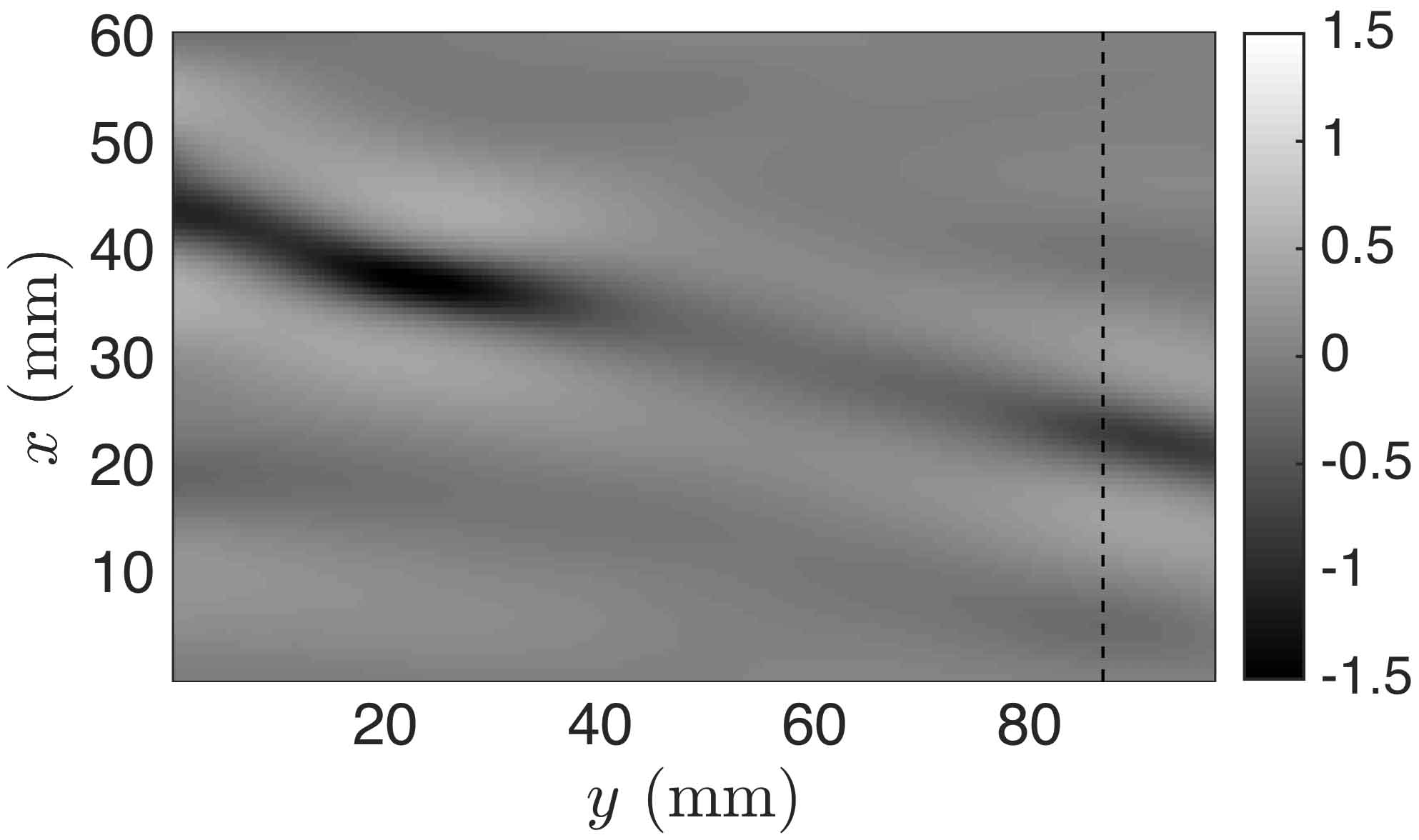}}\\
\end{tabular}
\end{center}
\caption{(\textit{a}) Magnitude of displacement vectors (in mm) for the image pair in figure \ref{fig:refraction1}. (\textit{b}) Free-surface elevation (in mm) for the same image pair. The dashed line is the approximate location of the laser sheet.}
\label{fig:vectors_height_maps}
\end{figure}

A pair of refraction images, one  with the air jet off and one with the jet on,  are shown in figures \ref{fig:refraction1}(\textit{a}) and (\textit{b}), respectively. Comparing these two images, the dots appear to move because of the refraction at the wavy interface in 
image (\textit{b}). This apparent motion is  quantified by using standard  PIV software (DaVis by LaVision) to find the dot displacement vectors. The number and size of the dots in the pattern are designed to follow the recommendations for PIV experiments \citep[see for example][]{AdrianBook}. The magnitude of the displacement vectors depends on the local free surface slope and the local distance between the free surface and the dot pattern. A larger distance gives a higher resolution (displacement) for the measurement technique but can result in ray crossing and spurious vectors. Once the displacement vectors are calculated, approximate surface normal vectors are computed on a uniform grid using  Snell's law and assuming that the local distance between the free surface and the dot pattern is  the distance between the flat water surface and the dot pattern. An inverse gradient algorithm called ``intgrad2'' is then utilized to integrate the surface normal field and obtain  the first estimate of the three-dimensional shape of the surface (see \cite{Moisy2009} for details).   In order to have an absolute measurement of the surface height, a reference height is required as the integration constant. This reference height was measured using a cinematic LIF technique as explained below. Once the first approximation of the surface height map is calculated, the surface normal vectors can be recomputed using the map of calculated local distance  between the water surface and the dot pattern. The procedure to find the surface height map is then repeated. In validation tests, it was  found that three iterations of this procedure are usually enough for convergence.  The magnitude of the dot pattern displacement vectors and the corresponding calculated surface elevation map for the images in figure \ref{fig:refraction1} are shown in figure \ref{fig:vectors_height_maps}.

A cinematic Laser Induced Fluorescence (LIF) method was used to measure the wave height along a line inside the measurement field of the refraction-based method, see figure \ref{fig:Lump_tank}. In order to perform these measurements, Fluorescein dye is added to the water and a thin sheet of light from an argon-ion laser is projected vertically onto the water surface from below.  The light sheet is about 20~cm wide and 1~mm thick and the plane of the  light sheet is oriented in the direction of the carriage motion.    A second camera is mounted on the carriage (camera 2 in figure \ref{fig:Lump_tank}) and views the intersection of the light sheet and the free surface through the tank side wall and from above the surface, looking down with an angle of about 15~degrees from the horizontal.  The image resolution of this camera was 50 pixels per millimeter.
This camera is triggered simultaneously with the refraction method camera via the laser trip device.  

\begin{figure}
\setlength\fboxsep{0pt}
\begin{center}
\begin{tabular}{c}
  (\textit{a})\\
  \fbox{\includegraphics[width=5in]{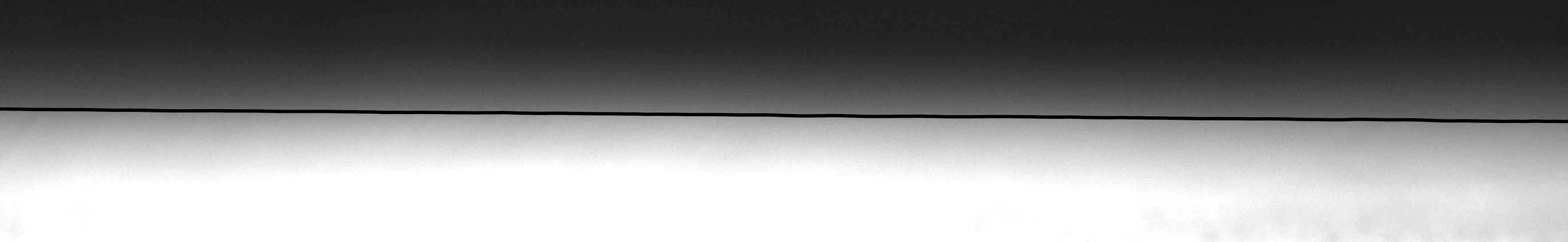}}\\
    (\textit{b})\\
  \fbox{\includegraphics[width=5in]{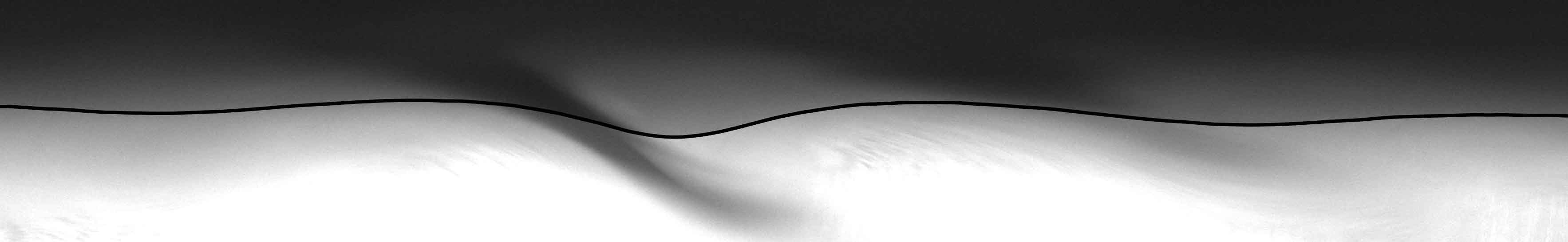}}\\
\end{tabular}
\end{center}
\caption{Sample images from the LIF method. These images are about 5~cm wide. The black lines are the calculated free-surface profiles after image processing. (\textit{a}) Calm water surface. (\textit{b}) Disturbed surface.}
\label{fig:LIF_raw}
\end{figure}

A gradient-based method for edge detection combined with dynamic programming (\cite{Kaas1988,Amini1990,Duncan1999}) is utilized to find the free surface shape along the image of the intersection of the laser sheet with the water surface. This method starts with an initial guess for the profile and tries to minimize a local energy functional to find the optimum profile. The energy functional consists of an internal energy and an external energy. The internal energy term keeps the profile smooth and connected and the external energy moves the profile towards image features (in this case edges). A sequence of about 5500 images are taken in each run. To improve the speed and accuracy of the program, the wave profile from each image is used as the initial guess for the next image in the sequence. The first image in each movie is taken during the carriage acceleration and therefore the free surface is still calm and the program is allowed to search for 20 pixels away from the initial guess. Since the movies are taken at a high frame rate, the change in the shape of the profiles in consecutive images is very small and it was decided to allow the program to search only 1 pixel away from the initial guess after the first image. The structure that holds the water tank blocks the view of the camera at three places during each run. For the first image after each blockage, the search window size is again set to 20 pixels. A sample pair of LIF images, one taken with the air jet off and the other with the jet on, are shown in figures \ref{fig:LIF_raw}(\textit{a}) and (\textit{b}), respectively. These images are for the same instants in time as the dot pattern images in figure \ref{fig:refraction1}. The calculated profiles are also plotted on top of the images. It is estimated that the free surface can be located to within $\pm 1$~pixel ($\pm 0.02$~mm) with this LIF method.  Since the cameras used for the refraction method and the LIF method are triggered simultaneously, there is an LIF image pair, with the air jet on and off, for each refraction image pair.  Thus, for each image pair from camera 1, the surface shape along the laser sheet is known from camera 2 and is used as the integration reference to obtain the free surface elevation map. Because the structure holding camera 1 is so far below the carriage, the camera tends to vibrate during each run creating an oscillating uniform pixel displacement with an amplitude of about 6 pixels primarily in the streamwise direction.  This oscillation creates a fictitious mean slope of the computed surface height maps.  This slope is removed by using the LIF reference line.

\begin{figure}
\begin{center}
\includegraphics[width=3.5in]{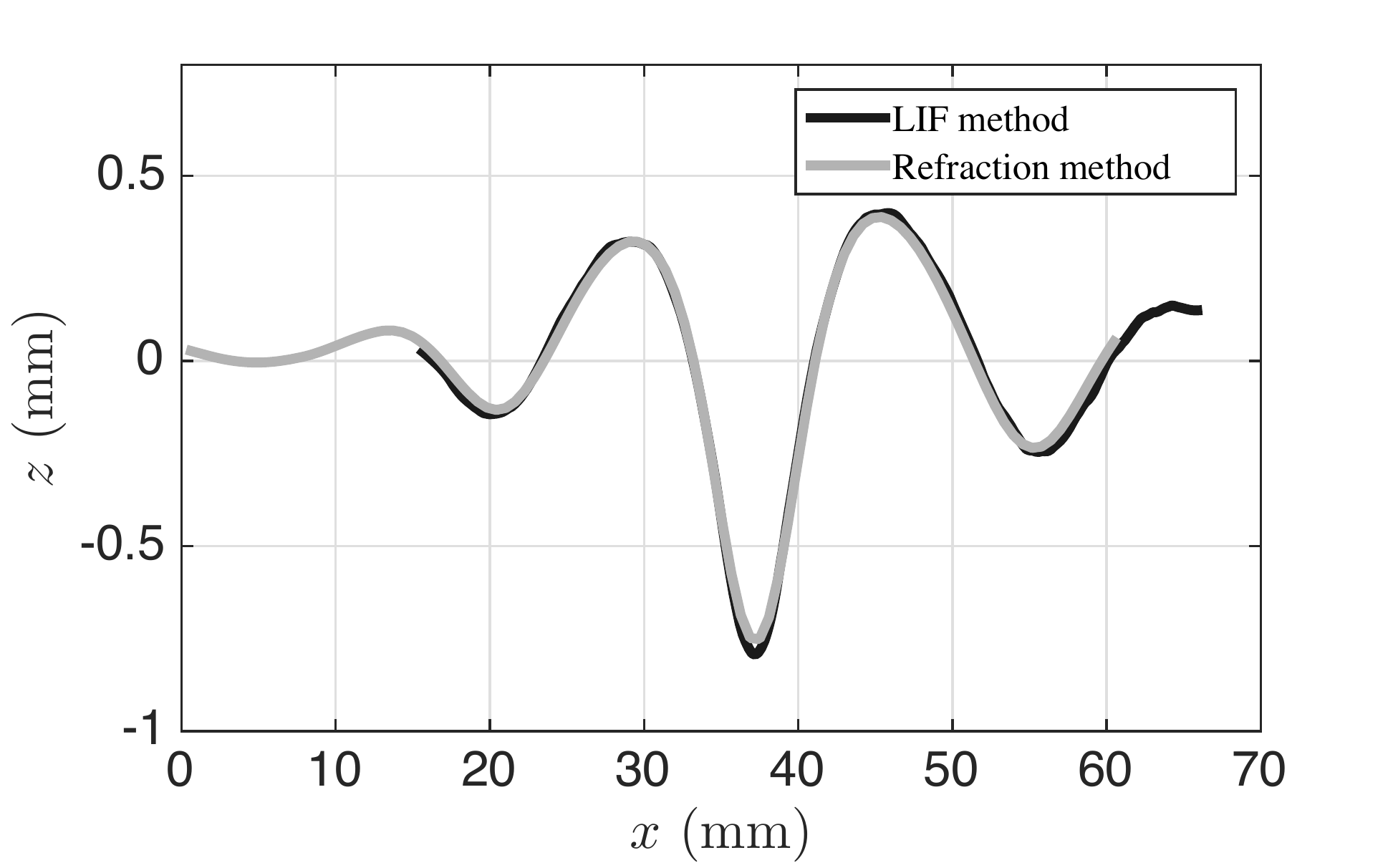}
\caption{Comparison between the results of the LIF method and the refraction method for the images in figures \ref{fig:refraction1} and \ref{fig:LIF_raw}.}
\label{fig:comparison}
\end{center}
\end{figure}

The accuracy of the refraction method can be estimated by comparing the profile along the laser sheet with the LIF profile at the same instant. In figure \ref{fig:comparison}, surface profiles from the sample images of figures \ref{fig:vectors_height_maps} and \ref{fig:LIF_raw} are plotted in the same coordinate system. The maximum difference in surface height between the two methods in a test run (5500 frames) had an RMS value of 0.066~mm or about 6 percent of maximum depth. 

For the quantitative measurements of the state III deformation patterns, the following experimental conditions and measurement settings were used.   In order to obtain accurate measurements with the refraction method, the dot image displacements should be large enough that they can be measured accurately but not so large that the images of single dots become distorted or refracted light rays cross before reaching the dot pattern.    As described above, dot image displacements increase with increasing distance ($h_d$)  between the mean water surface and the dot pattern and with increasing surface slope, which is controlled partially by the forcing parameter ($\epsilon$).  By trial and error, it was determined that accurate measurements were obtained   
with $h_d =5.4$~mm and $\epsilon=0.24$. These conditions were used for all of the full-field quantitative surface deformation pattern measurements.  In these experiments, the towing speed of the carriage was varied between  21.9~cm/s and 23.5~cm/s with a step of 0.1~cm/s. The field of view of the camera 1 was about 60~mm in the streamwise direction and 100~mm in the cross-stream direction and was located on one side of the pressure source, hence looking at half of the wave pattern. For each condition, each ``run" consisted of four movies (5500 frames for each movie at 300 frames per second) including two movies for each camera, one with the air-jet off and one with the air jet on.  
Three ``runs" were carried out and processed for each experimental condition.

The distorted dot pattern images can also be used to obtain a qualitative understanding of the surface shape without image processing.  For these observations, $h_d$ was increased so that the dot image displacements were large and the dot images were distorted into lines at locations with high surface slope. These high-distortion images are used to observe the free surface deformation  in the state II-III boundary and to measure the shedding period of the depressions in state III (see \S \ref{sec:State_diagram} and \S \ref{sec:stateII-III}). For these experiments, $\alpha$ was varied from 0.779 to 1.016 and $\epsilon$ was varied from 0.24 to 0.63.

\section{Numerical model}
\label{sec:numerical_model}

The model equation derived by \cite{Akers2009} and extended by \cite{ChoJFM} was used alongside the experimental results to investigate some features of the unsteady response to a moving pressure source. This equation is devised to capture the main effects of nonlinearity, dispersion and dissipation in the system and the interplay between them. Velocity and length are scaled by $c_\mathrm{min}$ and $1/k_{\mathrm{min}}$, respectively. In a coordinate system moving with the source ($\xi=x+\alpha t$ and $y$), the free-surface elevation $\eta$ is governed by

\begin{equation}
\label{eq:cho}
\eta_{t}-\tilde{\nu}
(\eta_{\xi\xi}+\eta_{yy})+
(\alpha-{\textstyle\frac{1}{2}})\eta_{\xi}-\beta(\eta^{2})_{\xi}-
{\textstyle\frac{1}{4}}\mathscr{H}\left\{\eta_{\xi\xi}+2\eta_{yy}-\eta\right\}
=Ap_{\xi},
\end{equation}
where $\mathscr{H}$ denotes the Hilbert transform, $\tilde{\nu}$ controls the viscous damping and $A$ controls the peak amplitude of the pressure distribution $p$. The coefficient of the nonlinear term is set to
\begin{equation}
\beta=\sqrt{11/2}/8
\end{equation}
to be consistent with the weakly nonlinear theory in the small-amplitude limit. The pressure distribution is assumed to have the Gaussian form 
\begin{equation}
p(\xi,y)=\exp(-2\xi^{2}-2y^{2}).
\end{equation}
The relationship between the lump amplitude and speed for the unforced inviscid equation of \cite{Akers2009} is compared to  fully nonlinear potential flow numerical results by E. Parau and a weakly nonlinear approximation in figure 1 of \cite{ChoJFM}.

Numerical solution of equation \ref{eq:cho} was obtained using the technique described in \cite{ChoThesis}. The computational domain was $-37.7<\xi<37.7$ and $-62.8<y<62.8$ with 512 grid points in each direction and no symmetry was assumed. The time step was set to $\Delta t=0.001$.

\section{Results and Discussion}\label{sec:Results}

\subsection{Qualitative Description of Response and State Diagram}
\label{sec:State_diagram}

\begin{figure}
\setlength\fboxsep{0pt}
\begin{center}
\begin{tabular}{ccc}
  \fbox{\includegraphics[width=1.6in]{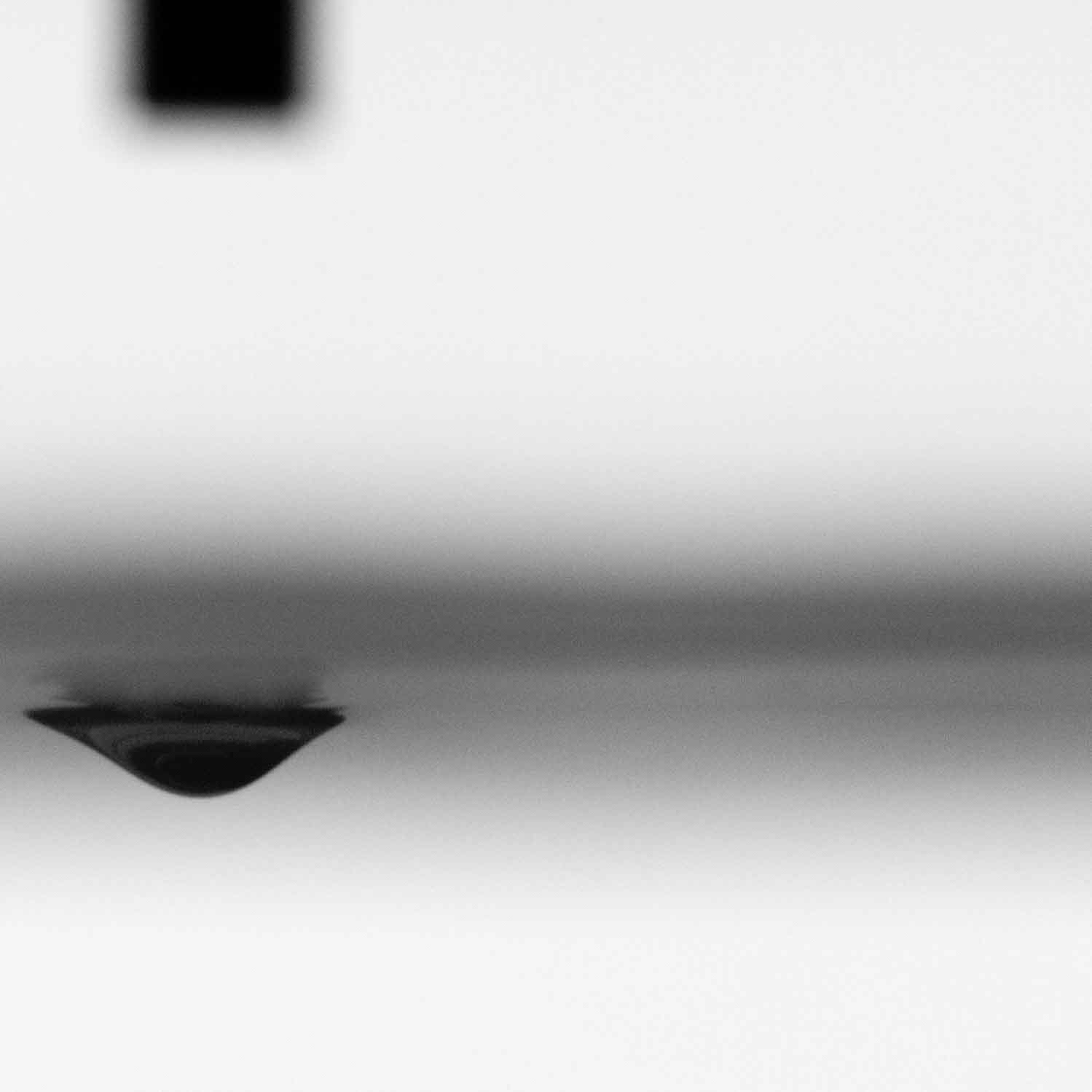}}&
  \fbox{\includegraphics[width=1.6in]{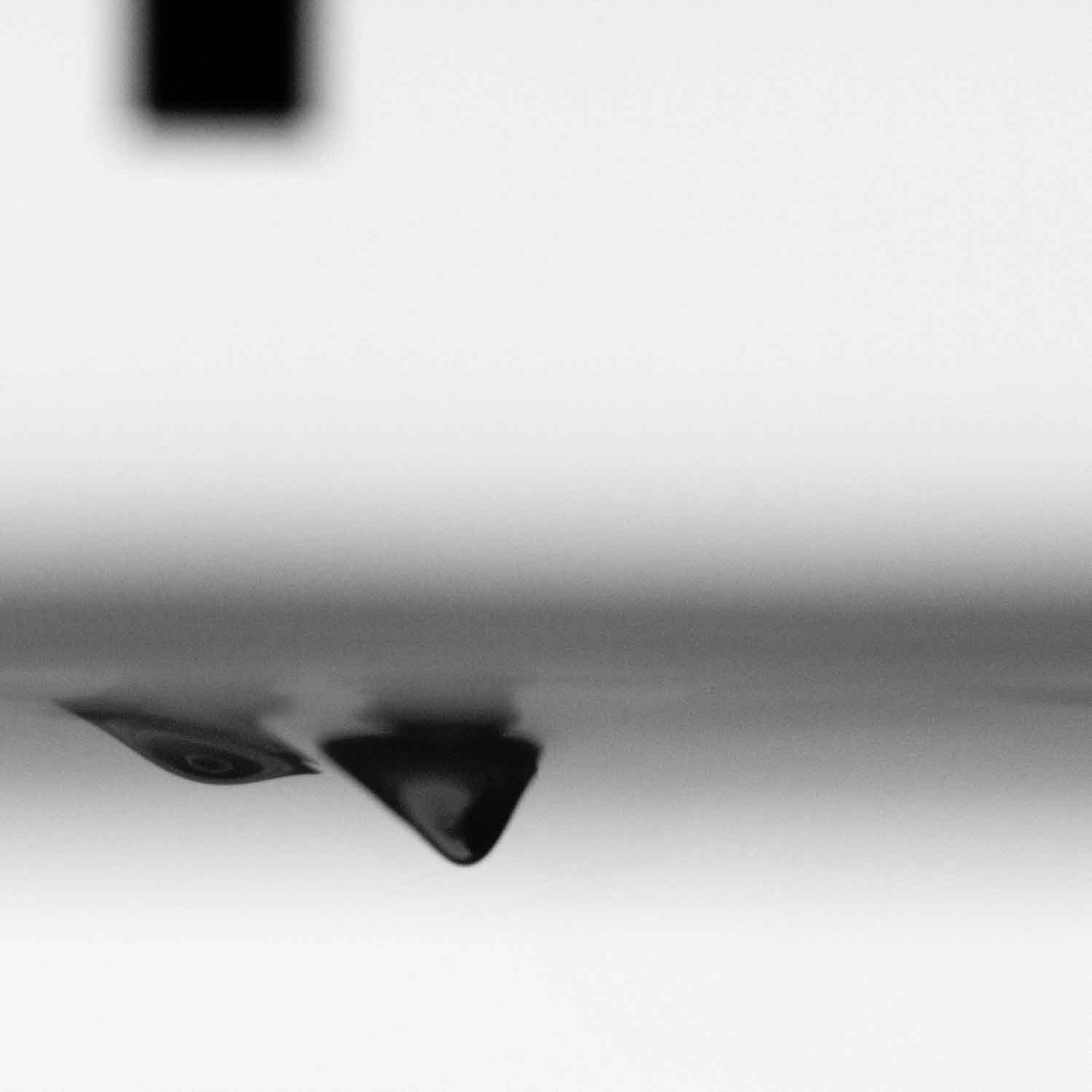}}&
  \fbox{\includegraphics[width=1.6in]{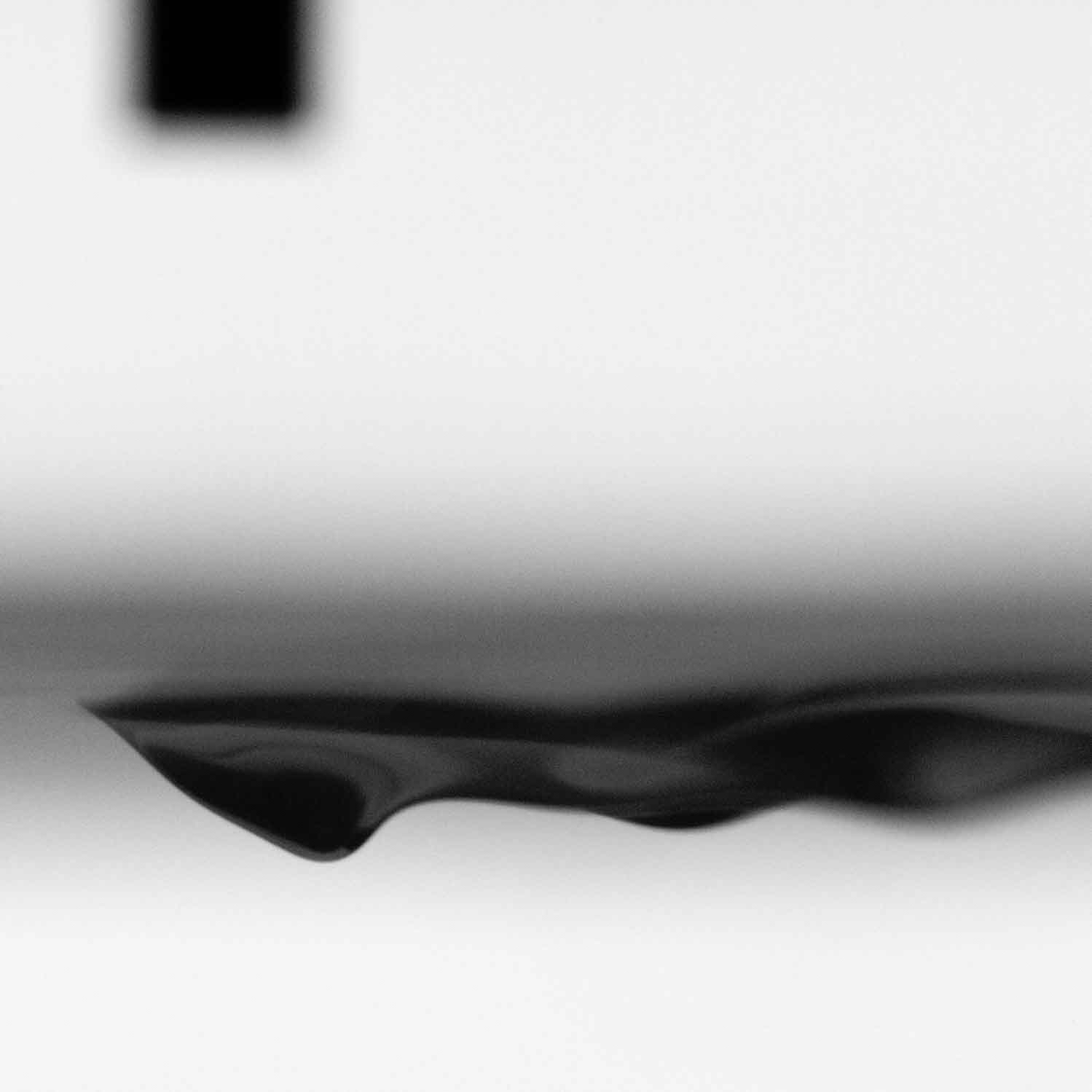}}\\
(\textit{a})&(\textit{b})&(\textit{c})\\
\end{tabular}
\end{center}
\caption{Shadowgraph images from camera 1 for $\epsilon=0.36$ and three values of $\alpha$.  The camera is focused through the water on the depression.  The air-jet tube, which is moving from right to left, is also visible in the upper left of each image, but it is out of focus since it is viewed through air.  (\textit{a}) $\alpha=0.82$. State I response: a small dimple beneath the air-jet tube. (\textit{b}) $\alpha = 0.93$. State II response: a steady lump behind the tube. (\textit{c}) $\alpha = 0.97$.  State III response: An extended unsteady pattern behind the tube. }
\label{fig:shadowgraph}
\end{figure}

\begin{figure}
\begin{center}
\begin{tabular}{c}
(\textit{a})\\
\includegraphics[width=3.5in]{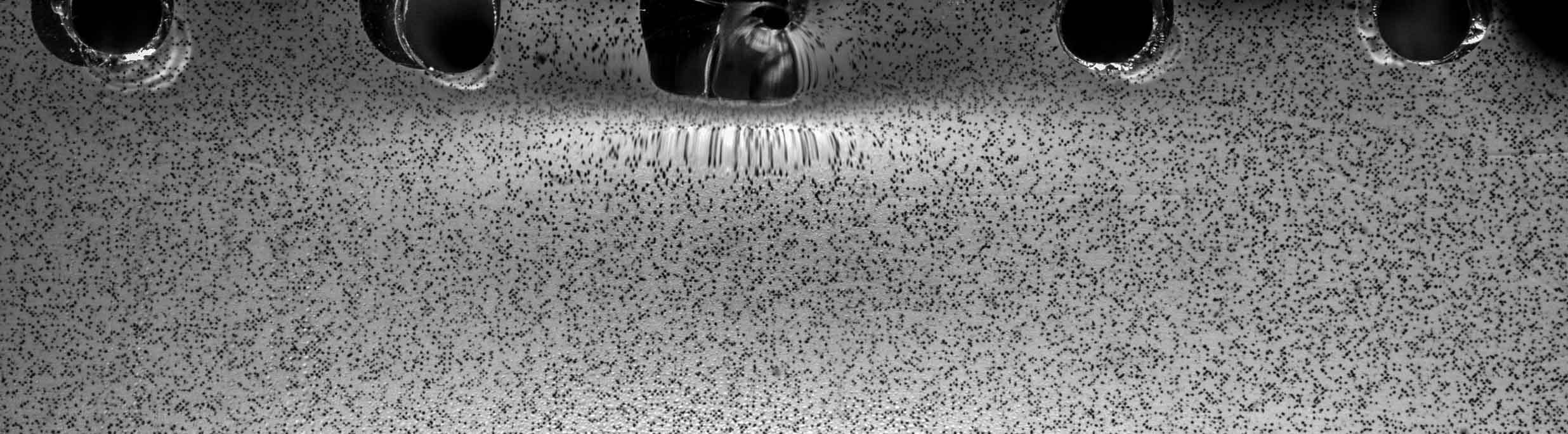}\\
(\textit{b})\\
\includegraphics[width=3.5in]{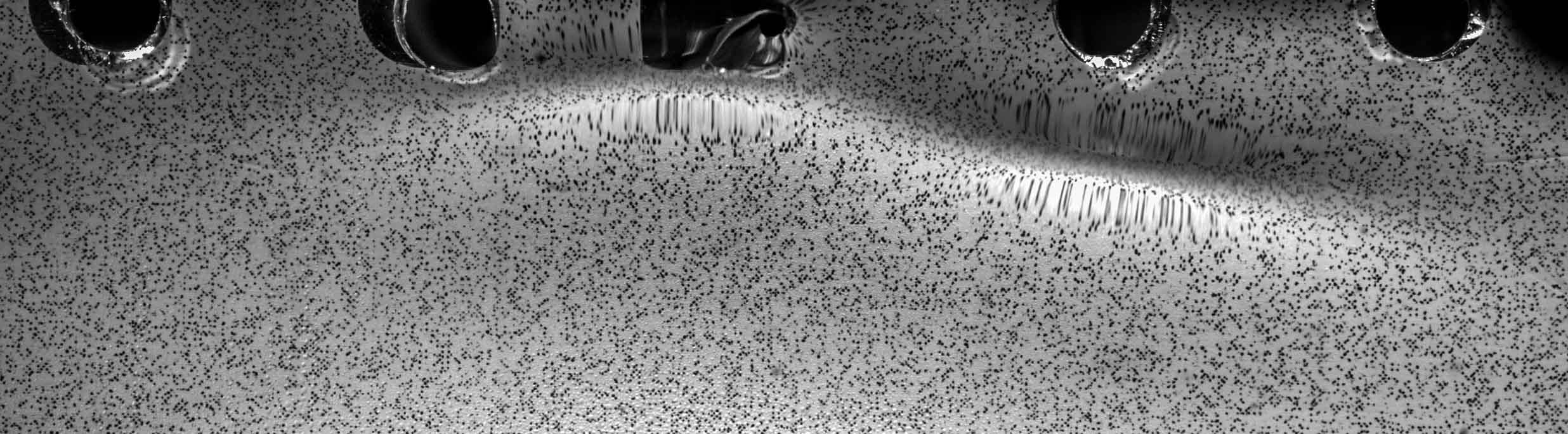}\\
(\textit{c})\\
\includegraphics[width=3.5in]{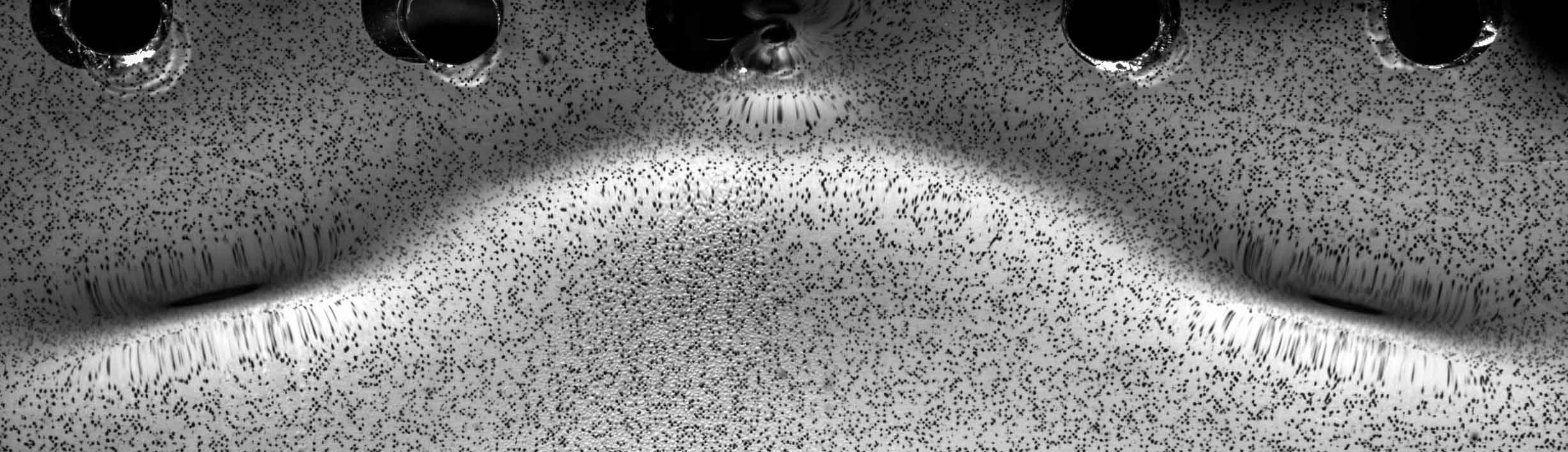}\\
\end{tabular}
\caption{ Refraction images of the deformation pattern behind the air-jet tube for $\epsilon = 0.24$ and $\alpha= 0.930$, 0.964 and 0.986,  corresponding to state II, the state II-III boundary and state III, are shown in  (\textit{a}), (\textit{b}) and (\textit{c}), respectively.   The dark circles along the top of each image are holes where the air-jet tube can penetrate the dot pattern.   In the images shown here, the air-jet tube is located in the center hole and the carriage is moving from the bottom to the top of the page. The distance between the dot pattern and the water surface is relatively large to make the distortions more obvious for qualitative visualization. The width of each image is about 9~cm in the physical plane. The depressions are the dark elongated areas while the adjacent regions where the dots are stretched are regions of high surface slope.}
\label{fig:dotpattern}
\end{center}
\end{figure}

\begin{figure}
\begin{center}
\includegraphics[width=3.5in]{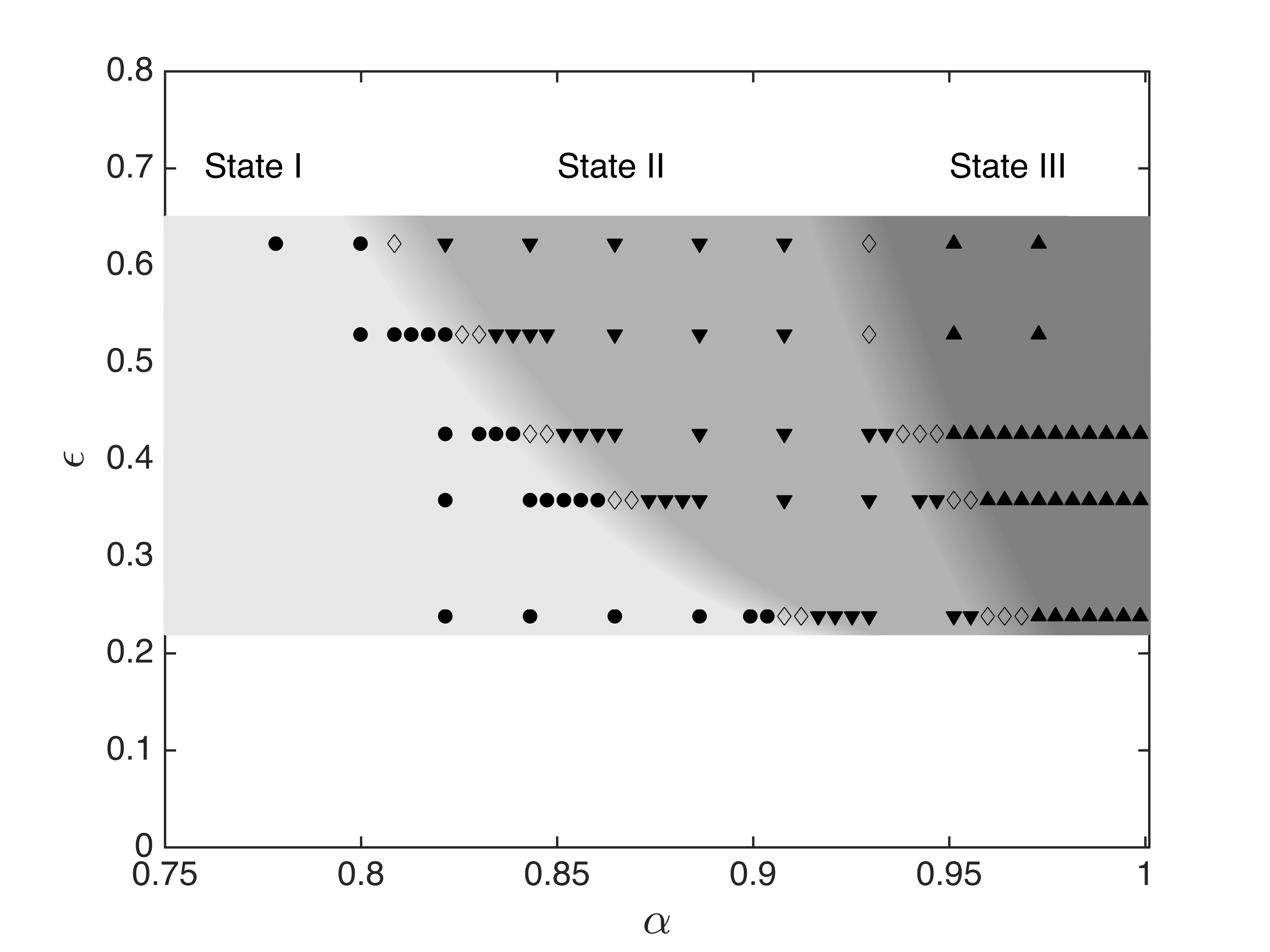}
\caption{State diagram in the $\alpha$--$\epsilon$ plane: $\bullet$ -- state I,  $\blacktriangledown$ -- state II, $\blacktriangle$ -- state III,  and $\diamond$ -- state I-II transition on the left and state II-III transition on the right.  In the state I-II transition region, the response is sometimes in state I and sometimes in state II.  In the state II-III transition region, the deformation pattern is unsteady with asymmetric and irregular shedding of localized depressions. Estimates of the locations of the state boundary regions are indicated by a gradient in the image grey level.}
\label{fig:state_diagram}
\end{center}
\end{figure}

Qualitative observations of the shape of the free-surface deformation pattern were made for a wide range of $\alpha$ and $\epsilon$.  These observations were carried out by using  both the underwater shadowgraph movies and the refraction method movies.  The shadowgraph movies are well suited  for observing state I,  state II and the state I-II boundary  (see figure \ref{fig:shadowgraph}), while the refraction movies are well suited for observing state III and the state II-III boundary (see figure~\ref{fig:dotpattern}).   The final result of this work is the state map in the $\alpha$-$\epsilon$ plane shown in figure~\ref{fig:state_diagram}.  This state map was determined by choosing five values of $\epsilon$ and for each $\epsilon$, starting with an $\alpha$ such that a state I response occurred, increasing $\alpha$ in steps of 0.02 and recording the state as it changed for I to II to III.  The boundary regions between states were then explored in smaller steps of $\alpha$ and frequently with repeated experimental runs.  The state I, II and III responses as seen in shadowgraph images are shown in figure \ref{fig:shadowgraph}(\textit{a}), (\textit{b}) and (\textit{c}), respectively.  The state I response is a steady dimple under the air jet; the state II response consists of a steady dimple under the air jet followed by a steady deeper depression behind the air jet; and the  state III response is an unsteady deformation pattern that is elongated in the downstream direction. Similar observations were reported by \cite{DiorioJFM}.   The state II and state III responses are also shown in the refraction images in  figure~\ref{fig:dotpattern}(\textit{a}) and (\textit{c}), respectively.  In figure~\ref{fig:dotpattern}(\textit{a}), the state II response is difficult to see because it is under and very close behind the air jet at the top center of the image; however, the state III response in figure~\ref{fig:dotpattern}(\textit{c}) is seen much more clearly than in the shadowgraph image in figure~\ref{fig:shadowgraph}(\textit{c}).

Between states I and II and between states II and III, there are boundary regions of finite thickness in $\alpha$ where the response is clearly not consistently one of the three states.  In the state I-II boundary region, the  surface deformation pattern oscillates between states I and II, as discussed by \cite{DiorioJFM}.  In the present work, it is found that the thickness of this boundary region is about 0.02 in $\alpha$.  In the state II-III boundary region,  the surface shape is highly oscillatory (especially at large $\epsilon$ values) and features an extended pattern that is asymmetric with respect to the vertical streamwise plane going through the air jet tube, see figure~\ref{fig:dotpattern}(\textit{b}). In experiments with $\epsilon=0.24$, localized depressions are generated alternately from the left and right sides of the centerline and decay quickly before moving very far from the pressure source. The shedding of depressions in this transition state seems to be irregular and does not have a well-defined period or left/right order. For $\epsilon=0.36$ and 0.43, the surface pattern features large-amplitude localized depressions close to the pressure source. These depressions are highly oscillatory and often seem to disappear as they  radiate small-amplitude waves.  The $\alpha$ values for both the state I-II and state II-III boundaries decrease with increasing~$\epsilon$.

\subsection{The evolution of free-surface shape in state III}
\begin{figure}
\begin{center}
\begin{tabular}{ccc}
  (\textit{a})&(\textit{b})&(\textit{c})\\
  \includegraphics[height=1.15in]{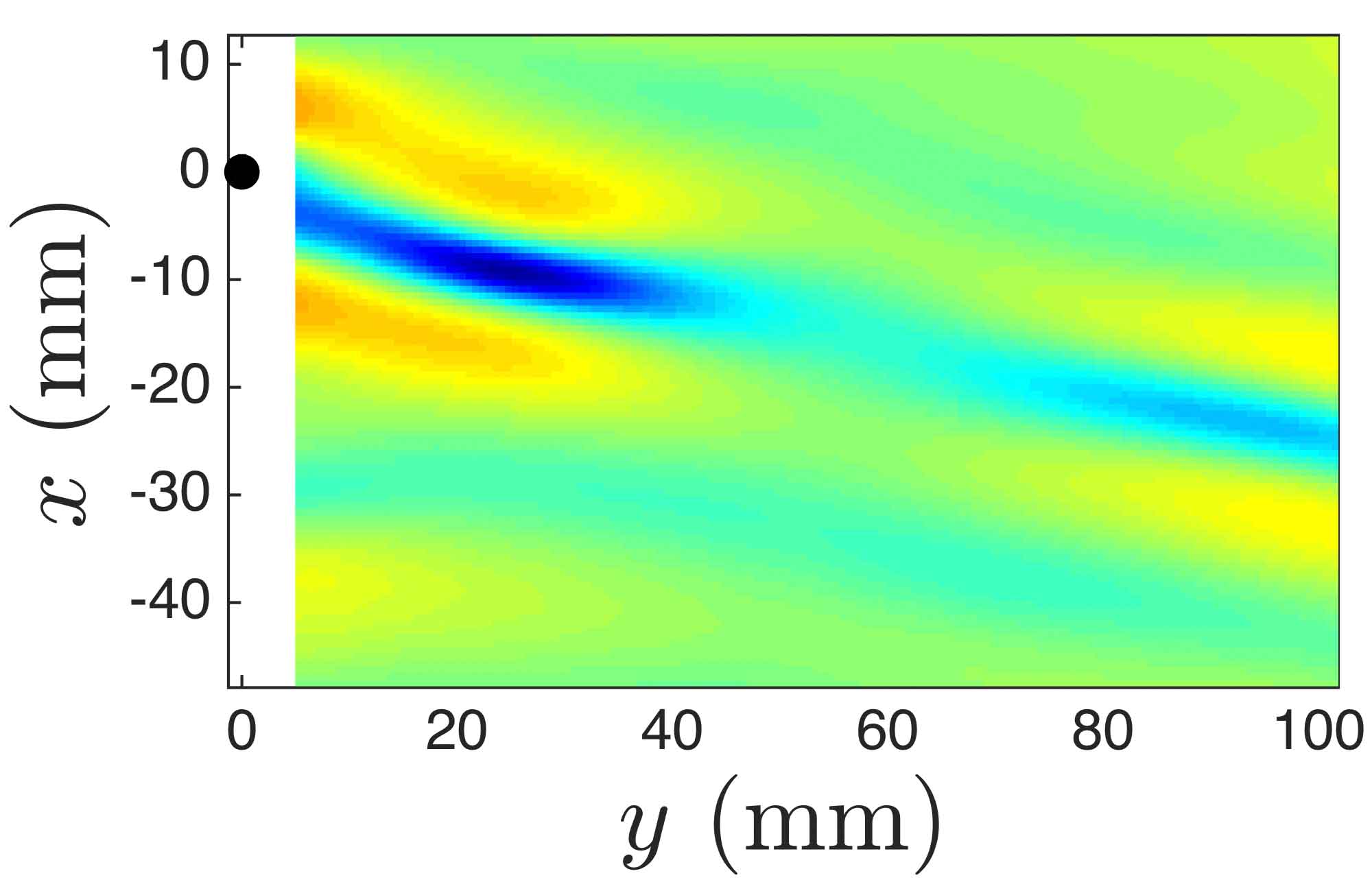}&
  \includegraphics[height=1.15in]{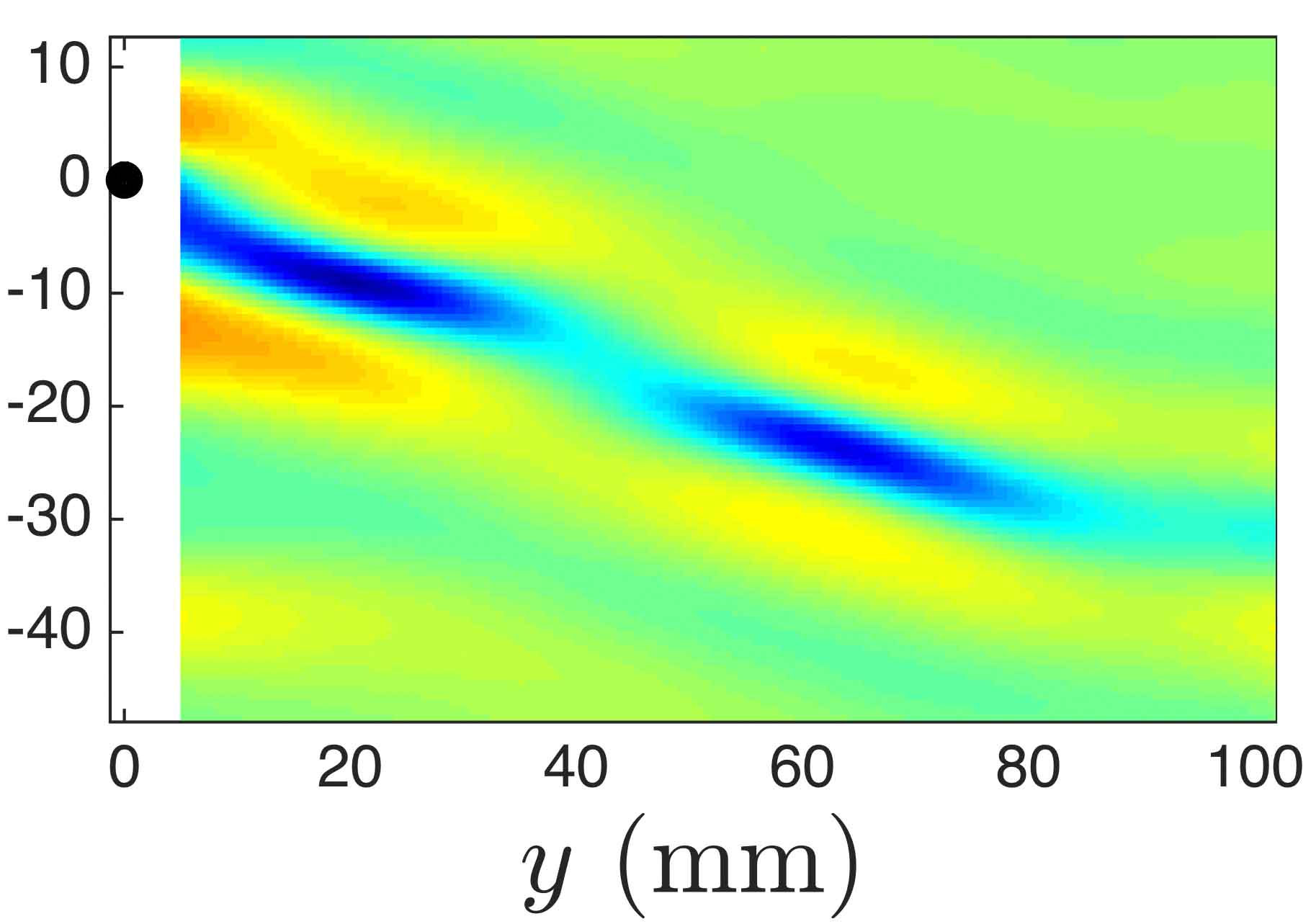}&
  \includegraphics[height=1.15in]{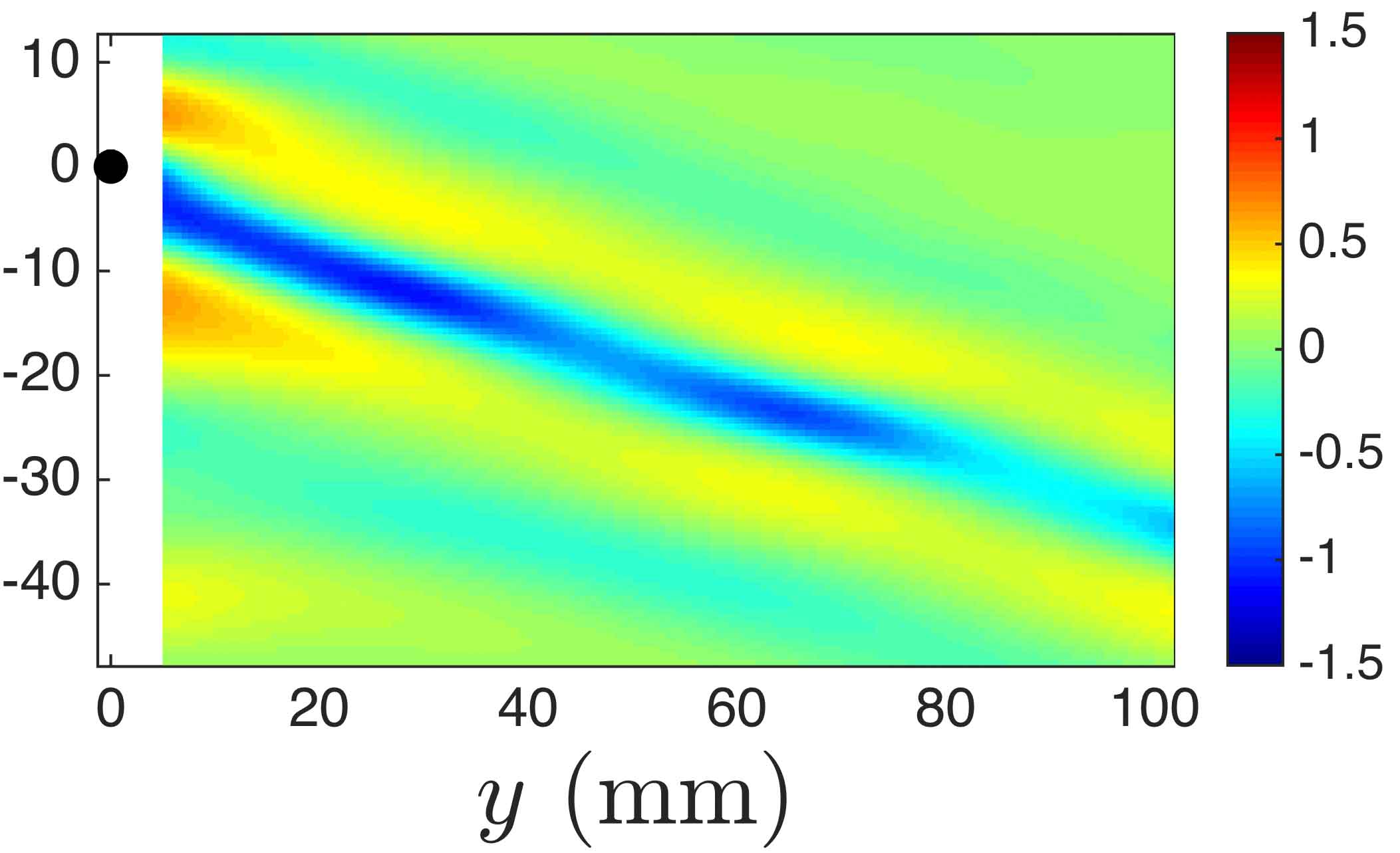}\\
  \includegraphics[height=1.15in]{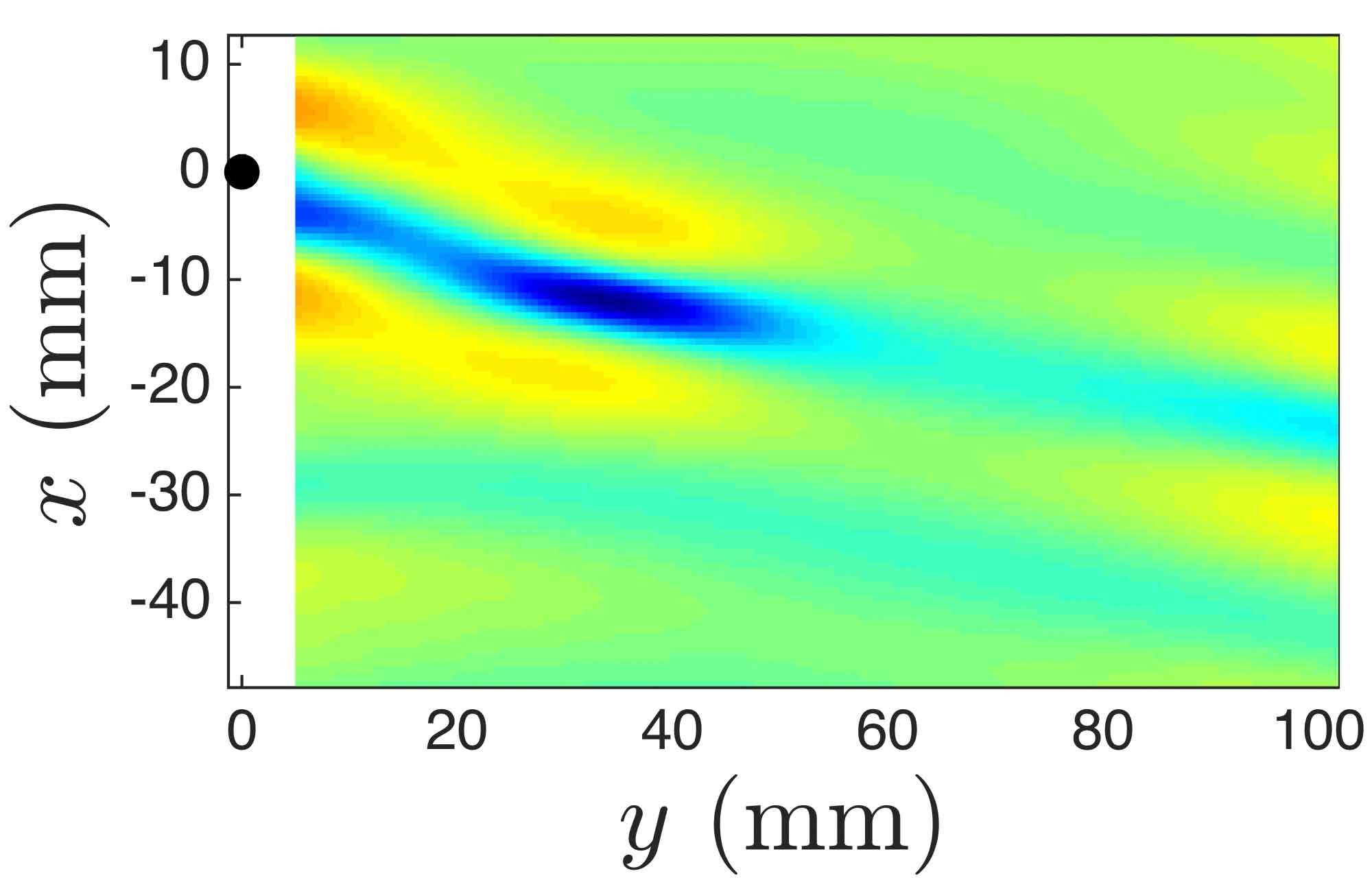}&
  \includegraphics[height=1.15in]{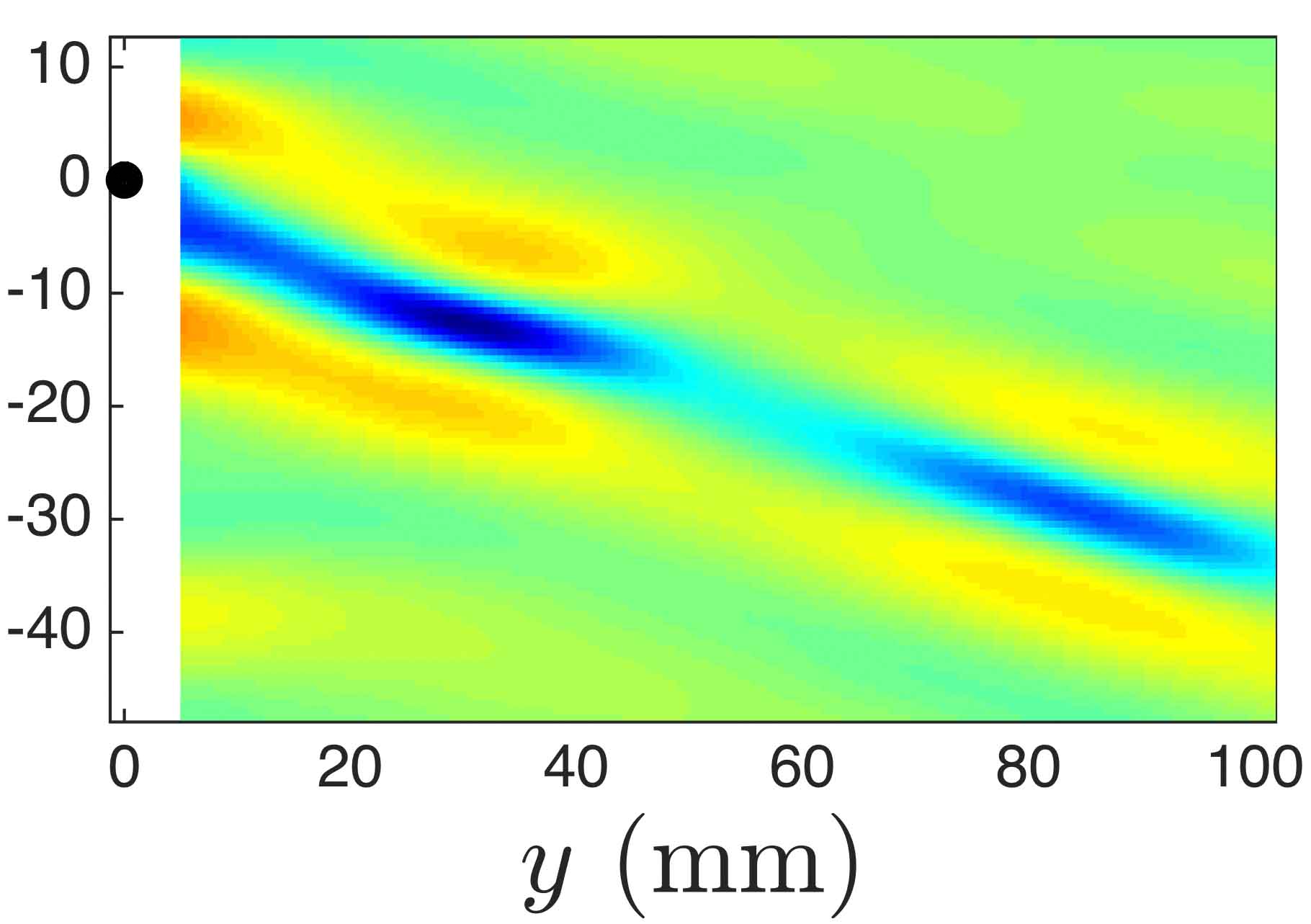}&
  \includegraphics[height=1.15in]{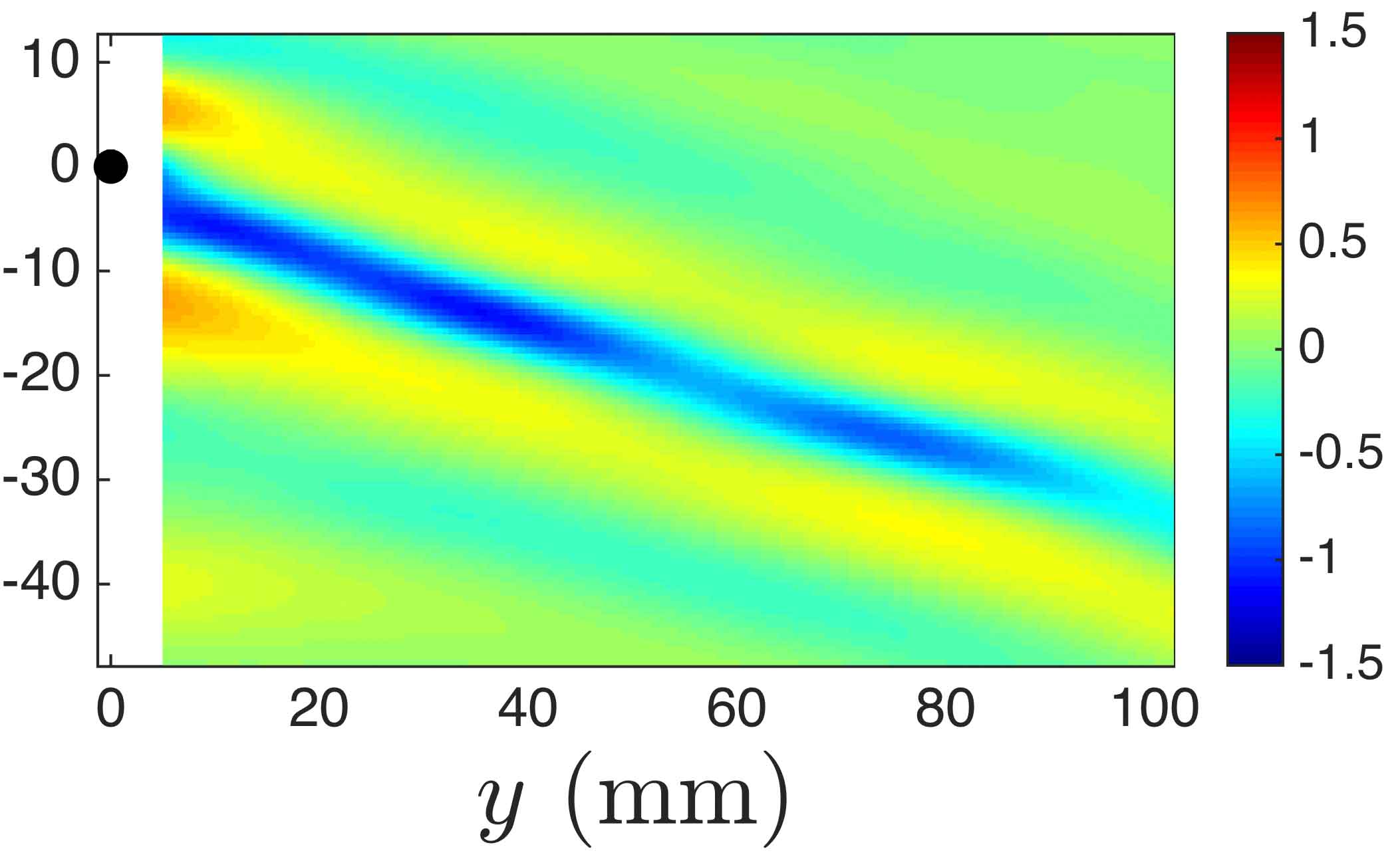}\\
  \includegraphics[height=1.15in]{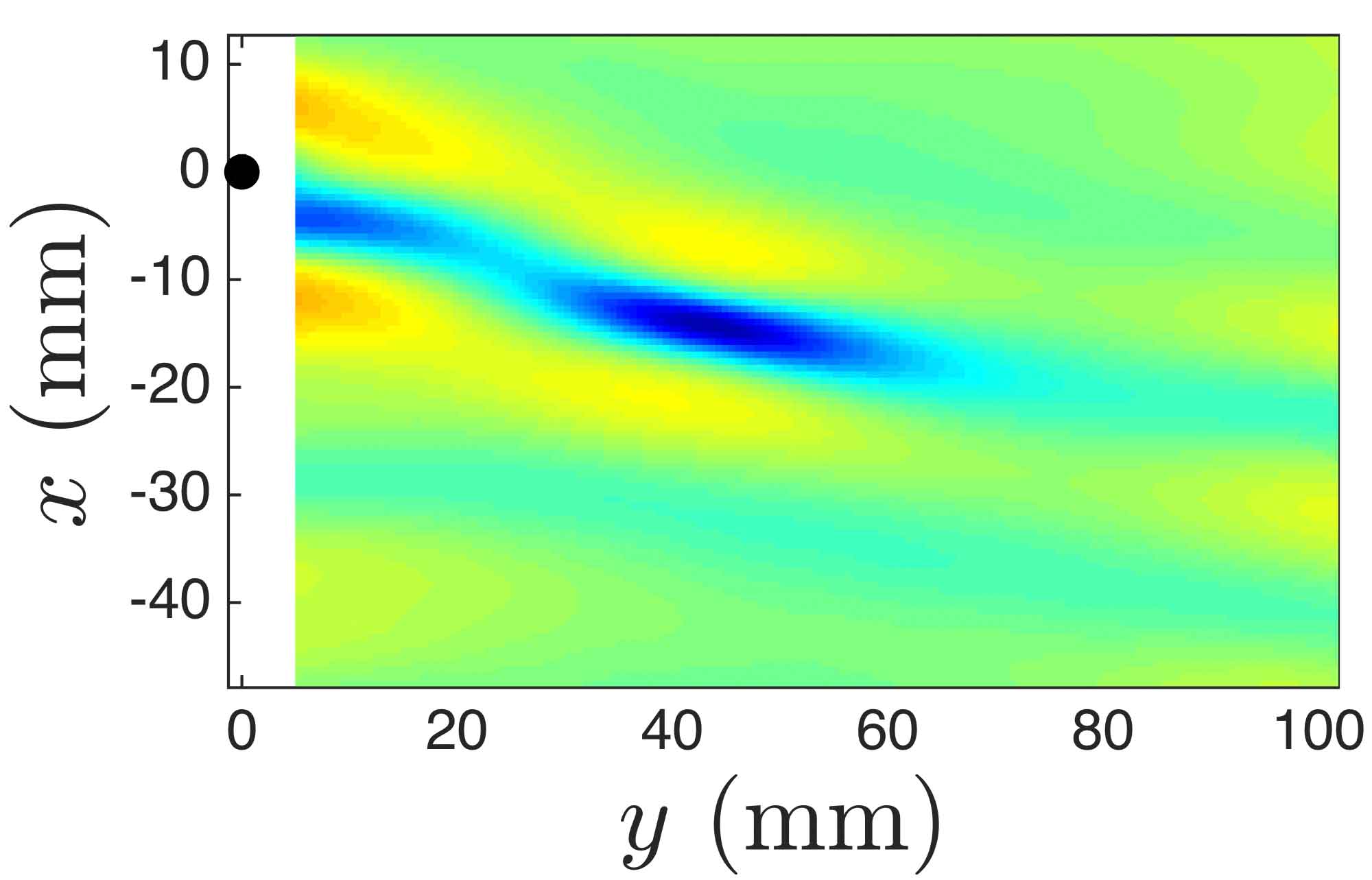}&
  \includegraphics[height=1.15in]{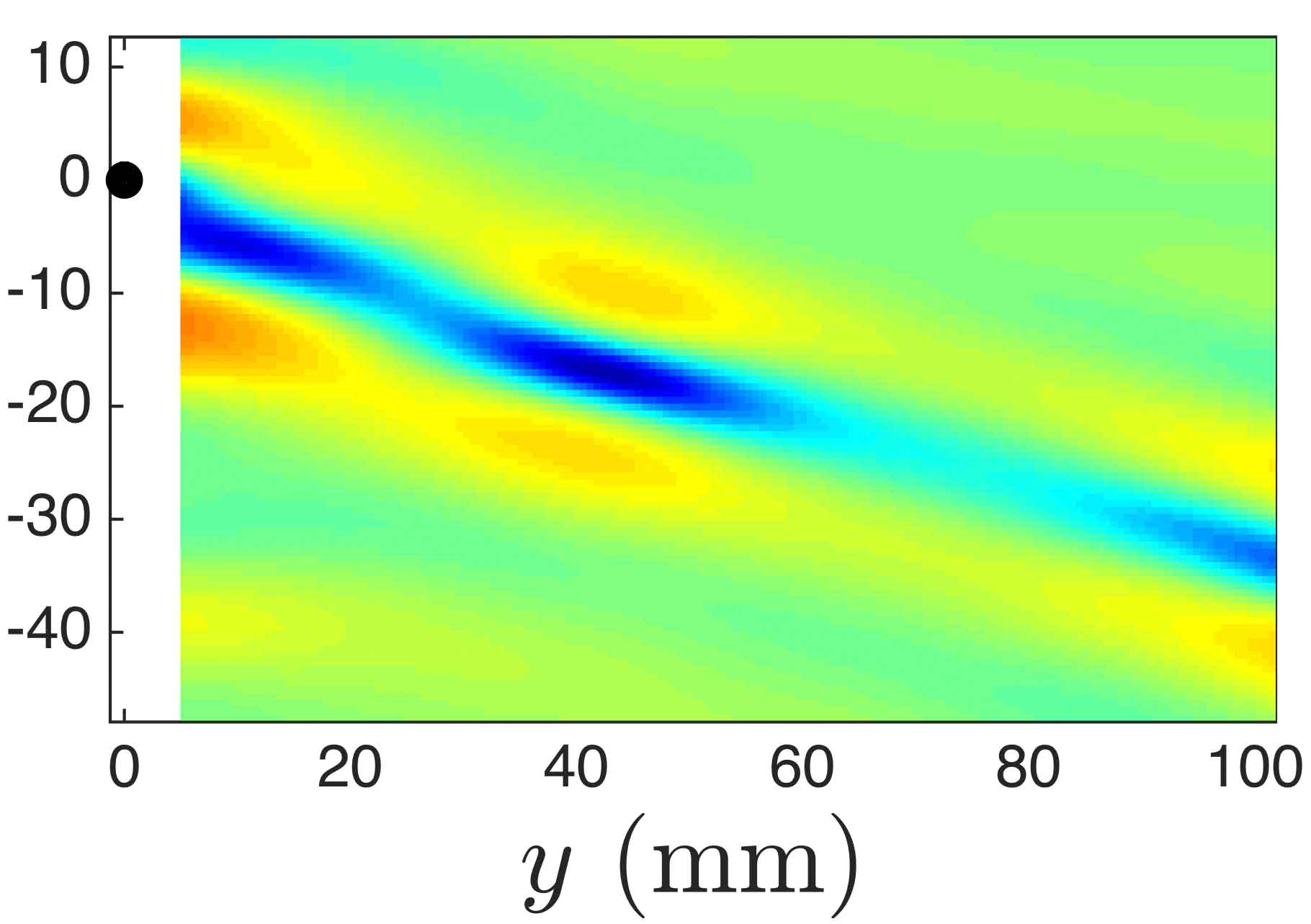}&
  \includegraphics[height=1.15in]{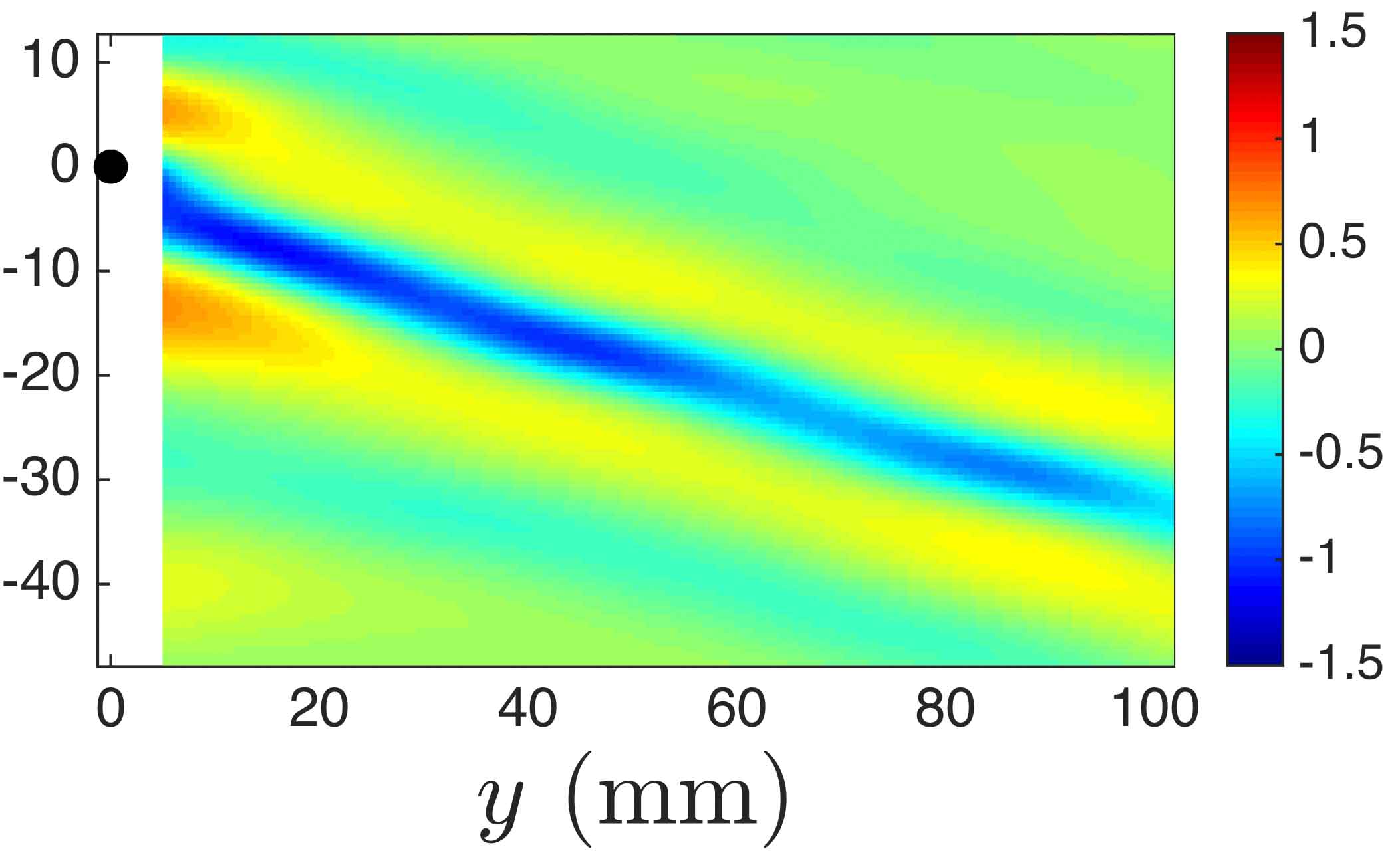}\\
  \includegraphics[height=1.15in]{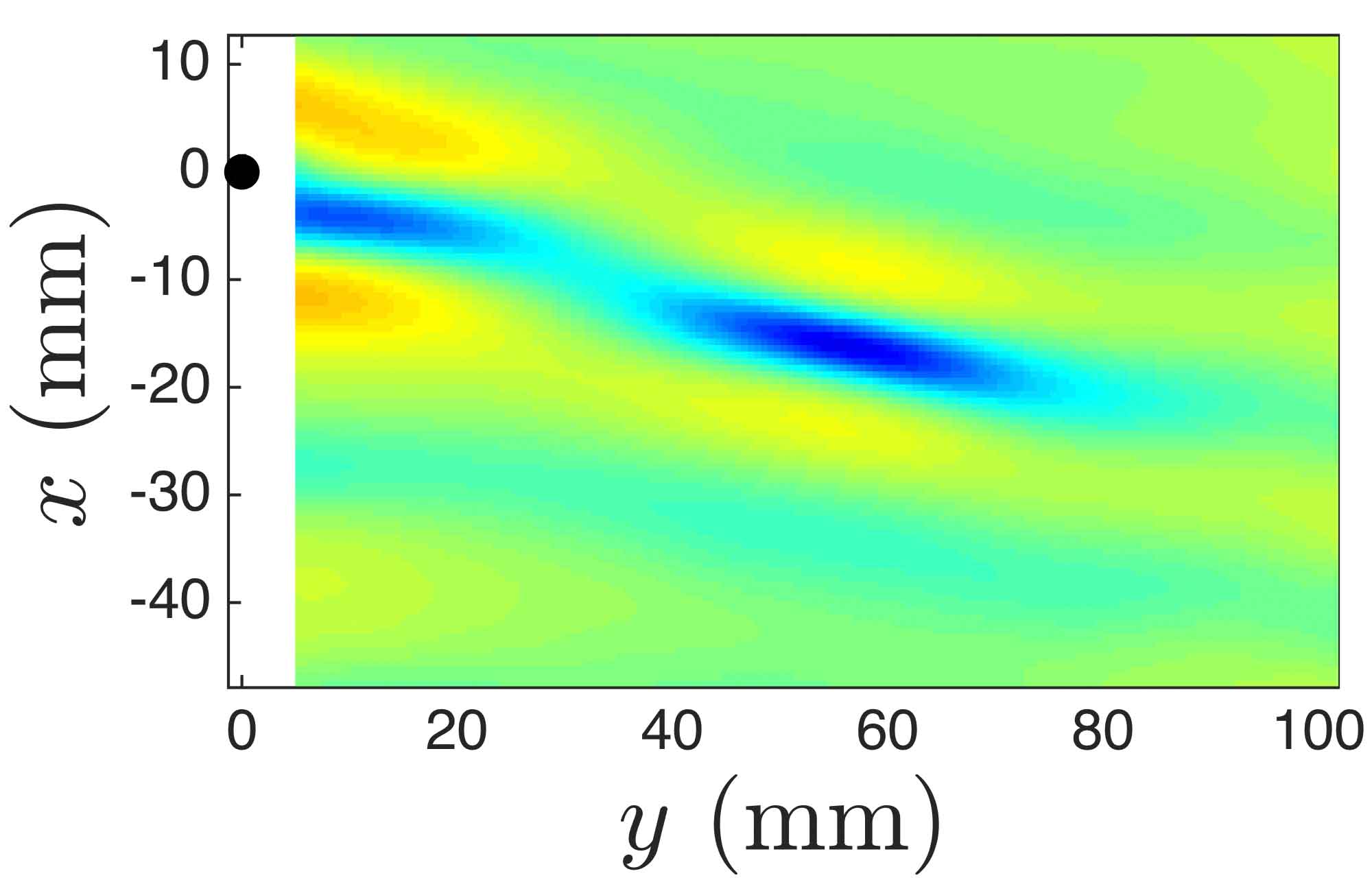}&
  \includegraphics[height=1.15in]{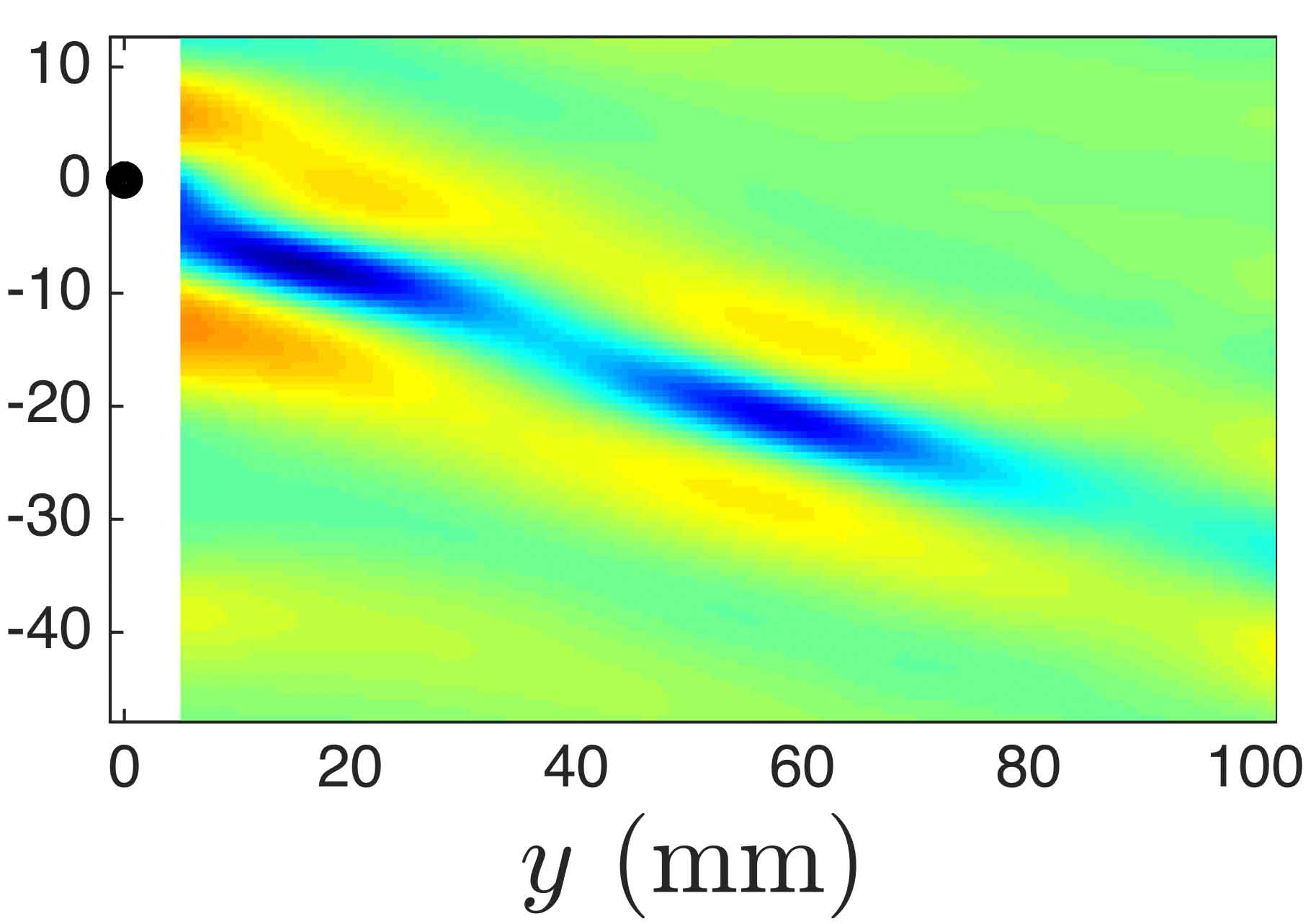}&
  \includegraphics[height=1.15in]{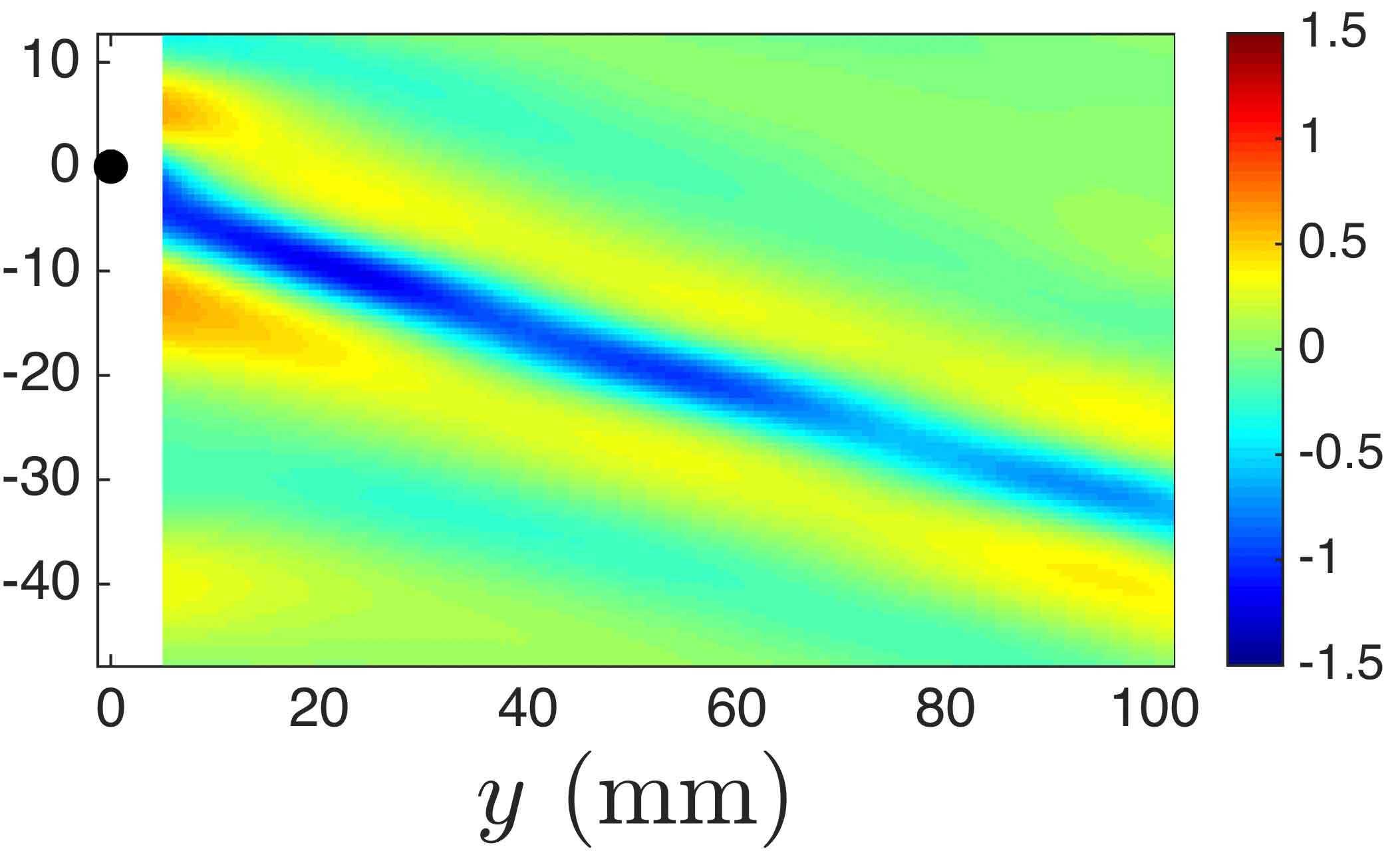}\\
  \includegraphics[height=1.15in]{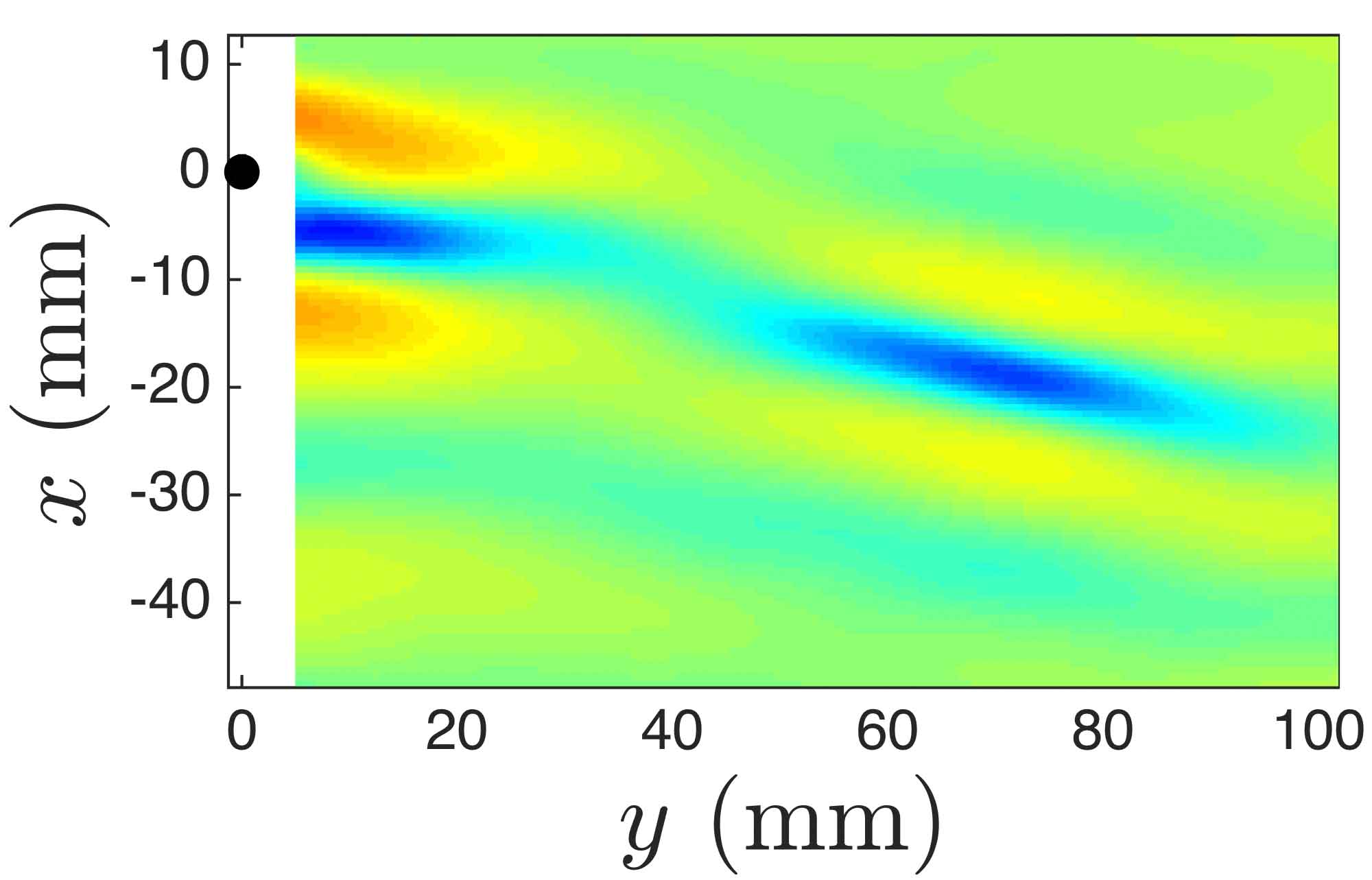}&
  \includegraphics[height=1.15in]{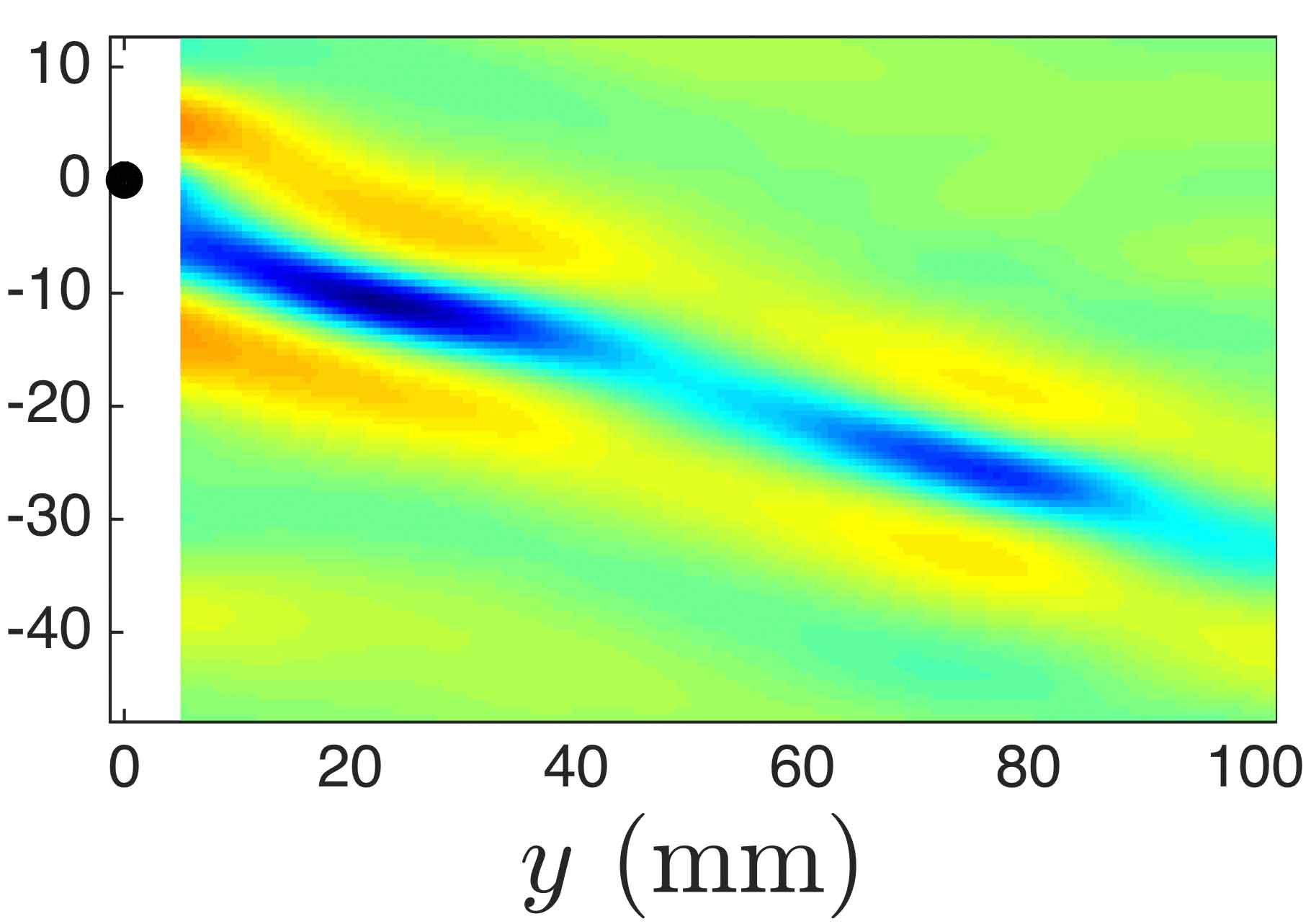}&
  \includegraphics[height=1.15in]{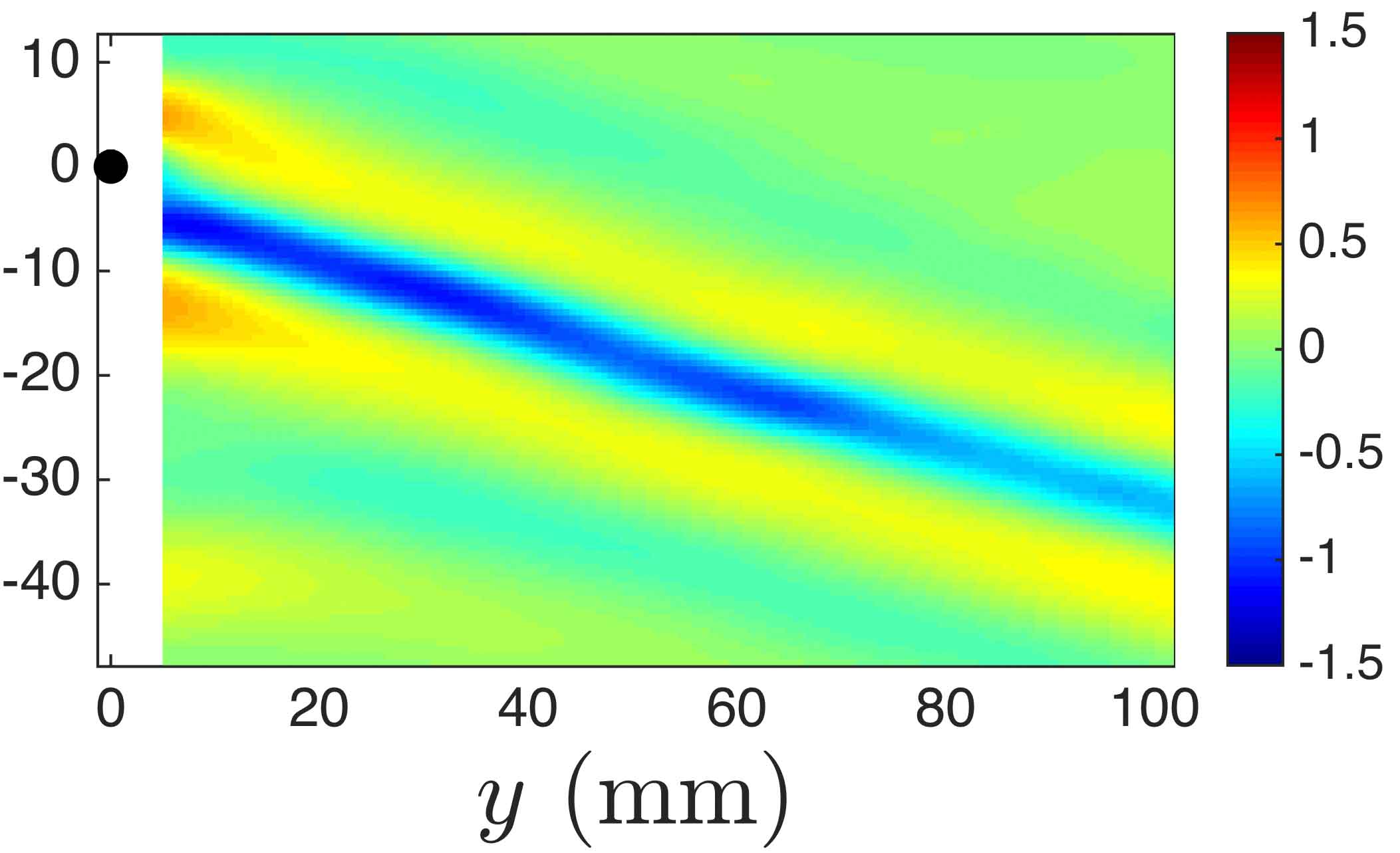}\\
  \includegraphics[height=1.15in]{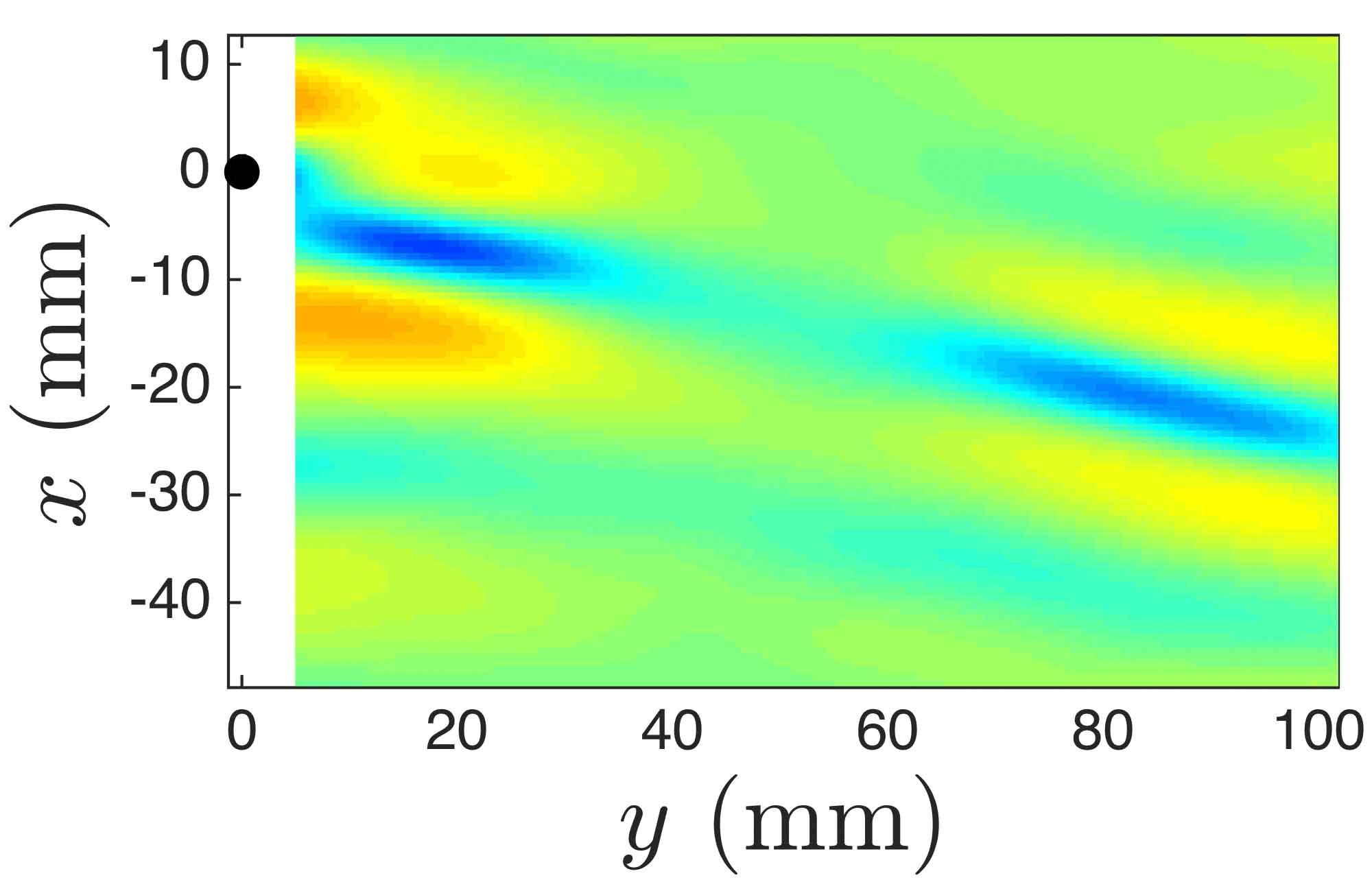}&
  \includegraphics[height=1.15in]{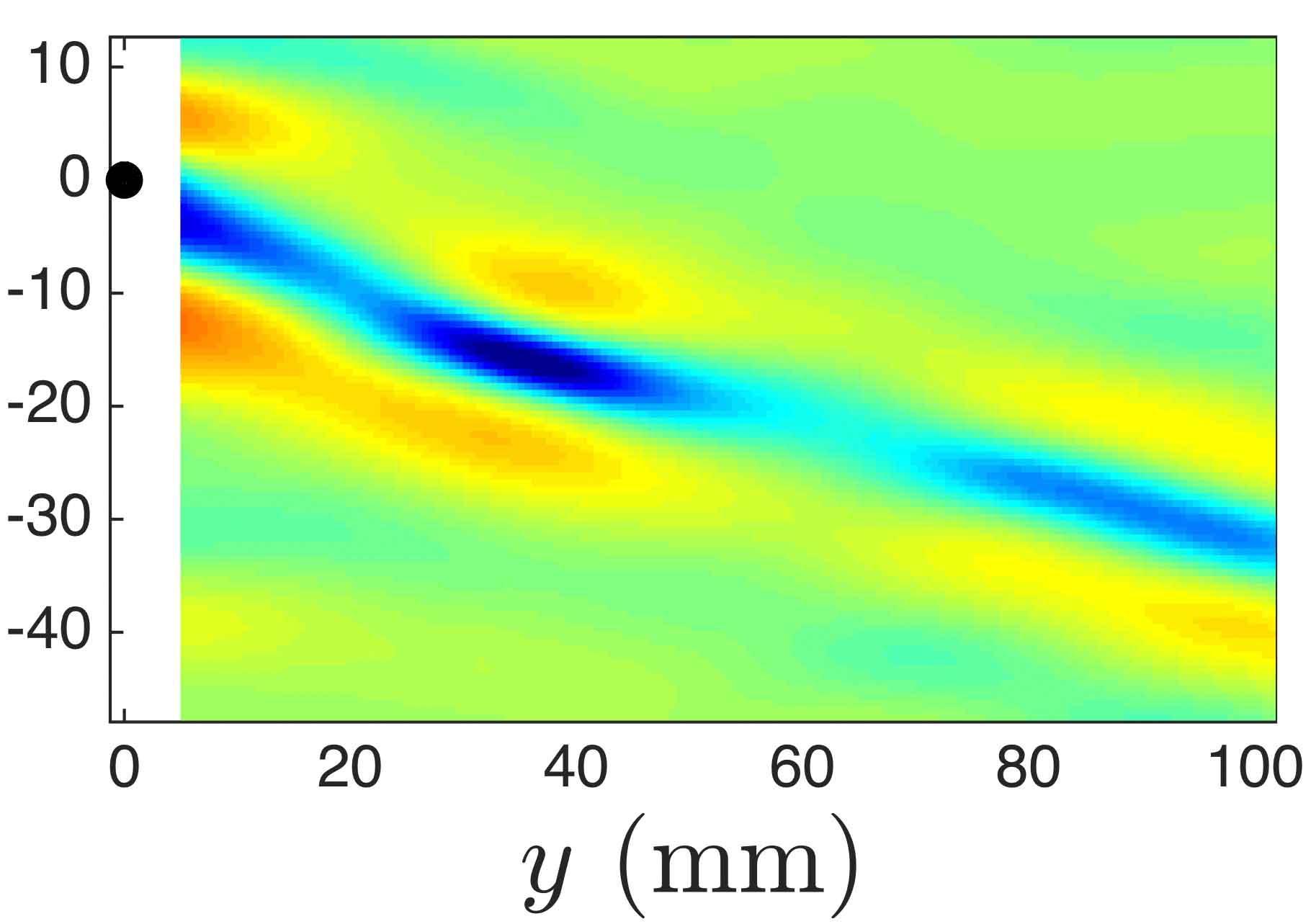}&
  \includegraphics[height=1.15in]{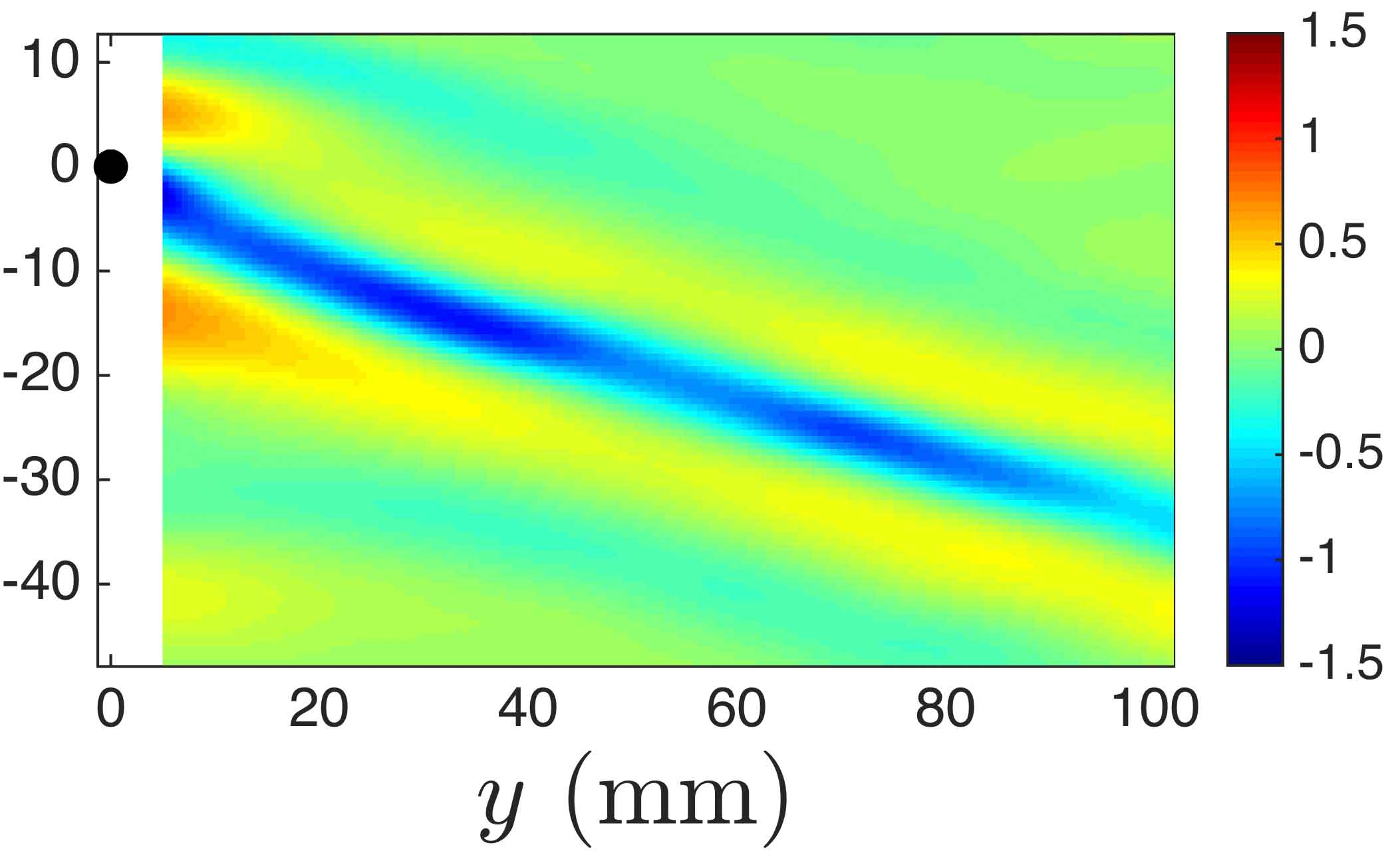}\\
\end{tabular}
\end{center}
\caption{Snapshots of the measurements of the right half of the free-surface elevation pattern (in mm) for $\epsilon=0.24$. The air-jet tube is located at the origin (shown as a black dot) and is moving in the positive $x$-direction. The time separation between figures in each column is 0.3~s. Each column is from one run and all three runs produce a state III response: (\textit{a}) $\alpha=0.986$, (\textit{b}) $\alpha=0.994$, (\textit{c}) $\alpha=1.003$.}
\label{fig:stateIII_snapshots}
\end{figure}

The evolution of the three-dimensional shape of the free surface in state III was measured with the refraction-based method described in \S \ref{sec:Measurement_methods} for $\epsilon = 0.24$ and  $\alpha$ values ranging from 0.946 to 1.016. Sequences of surface elevation patterns for $\alpha=0.986$,~0.994 and 1.003 are shown in columns (\textit{a}), (\textit{b}) and (\textit{c}) of figure \ref{fig:stateIII_snapshots}, respectively, in a reference frame moving with the pressure source. The time separation between the plots in each column is $\Delta t=0.3$ seconds. The pressure source is located at the origin (shown as a black dot) and is moving in the positive $x$-direction in the laboratory reference frame.   In the following, both general observations and quantitative aspects of the deformation pattern are discussed.

In each run, as the carriage reaches its final speed, a V-shaped pattern appears behind the source and a pair of isolated depressions form at the tips of the V.  These depressions, which appear as blue ovals in figure~\ref{fig:stateIII_snapshots}, are oriented with their long axes making an angle of about 15 degrees in the clockwise direction relative to the cross-stream direction.  In the reference frame of the air-jet tube, the depressions move away from the source in approximately  the direction of their long axis; in the laboratory reference frame they propagate in the direction perpendicular to their long axis.   This  angle increases slightly with increasing towing speed.  The maximum depth of each depression increases as it is generated, but as it moves away from the source, the maximum depth decreases and the depressions seem to move at a faster rate. This process is repeated during the entire run as depressions are periodically shed and quickly damp out.

At lower towing speeds, the shedding events are isolated and separate depressions move away from the source and decay (column (\textit{a}) in figure \ref{fig:stateIII_snapshots}), while at higher speeds, the shedding occurs more frequently and new depressions are shed before the previous depressions  have moved far from the source (column (\textit{b}) in figure \ref{fig:stateIII_snapshots}).
Hence, as the towing speed is increased each depression is more likely to interact with its neighbors. The behavior of depressions is quite complicated. For example,  they can suddenly speed up or slow down, change shape or merge with another localized depression. This behavior is probably due to these interactions.

As the towing speed is increased above $c_\mathrm{min}$, the depressions become shallower and the shedding becomes even more frequent. Here, it is difficult to recognize individual isolated depressions, especially at later times during a run (column (\textit{c}) in figure \ref{fig:stateIII_snapshots}). For $\alpha>1.02$, the pattern starts to look similar to the steady V-shape pattern of linear theory. Quantitative measurements of the shedding period are discussed later.	

\begin{figure}
\begin{center}
\includegraphics[width=3in]{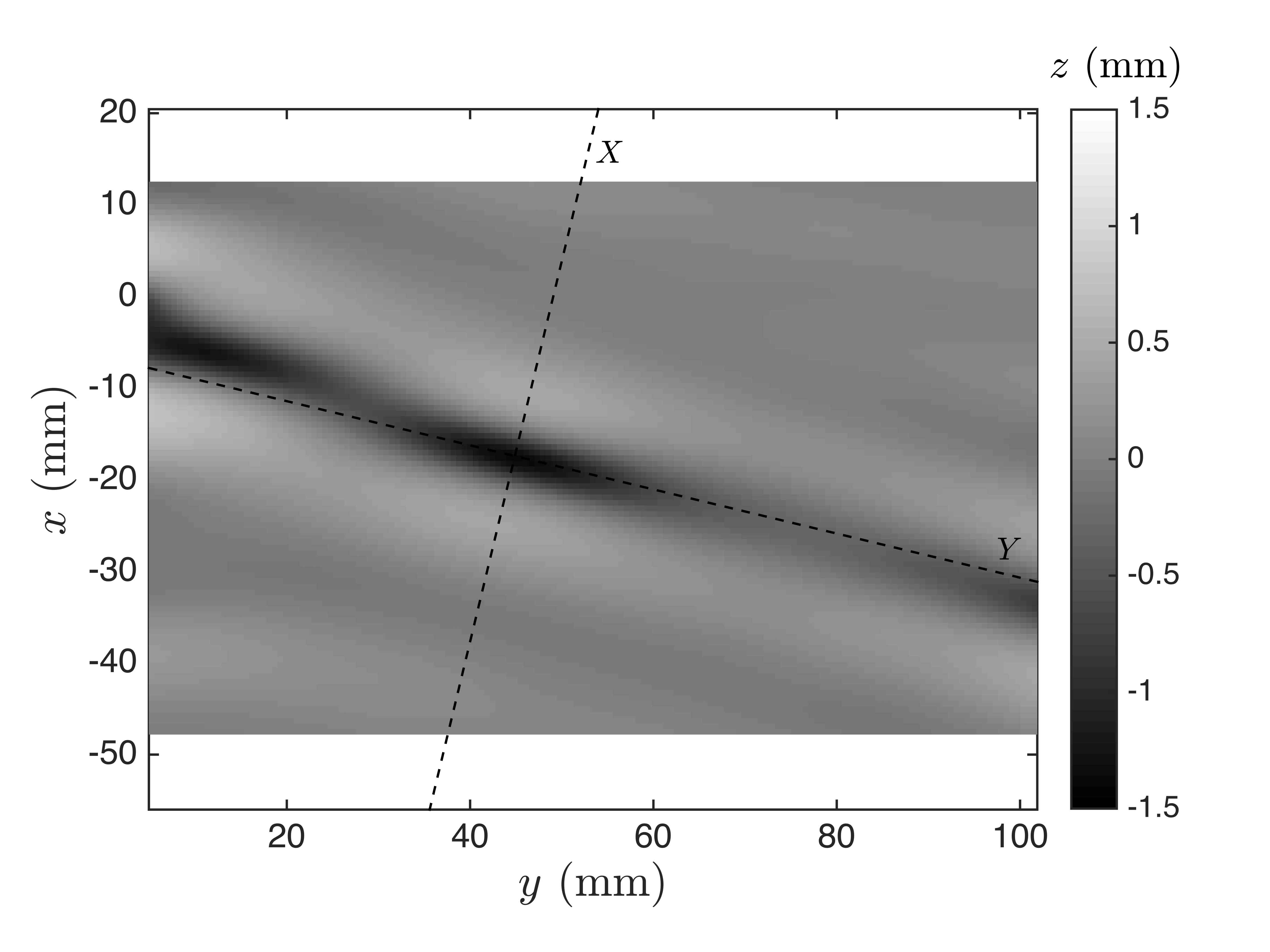}
\caption{The modified coordinate system for a state III localized depression with $\alpha=0.986$ and $\epsilon=0.24$. }
\label{fig:profile_pcolor}
\end{center}
\end{figure}

\begin{figure}
\begin{center}
\begin{tabular}{cc}
  (\textit{a})&(\textit{b})\\
  \includegraphics[trim=0 0 0.1 0.1in,clip=true,width=2.70in]{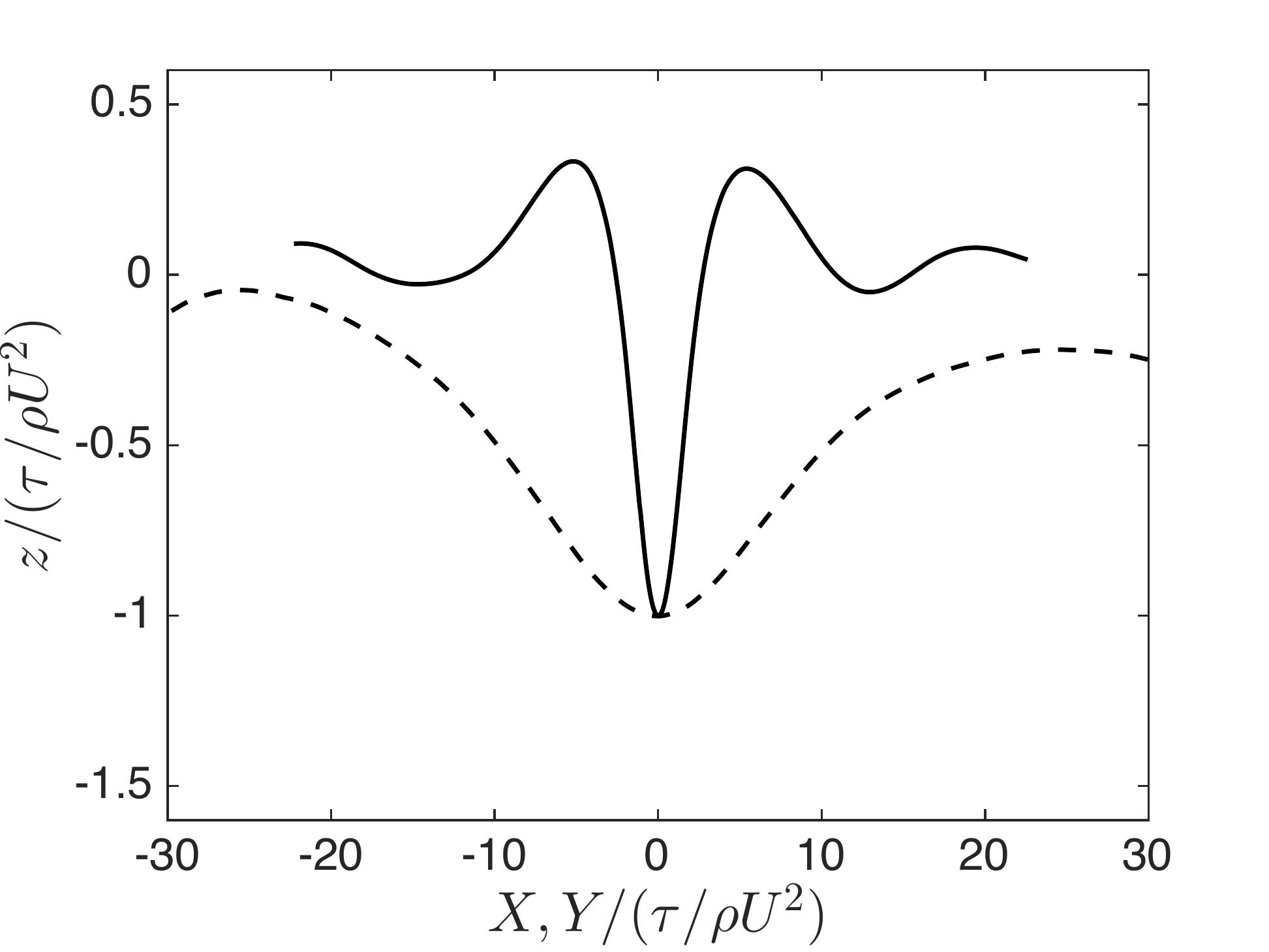}&
  \includegraphics[width=2.4in]{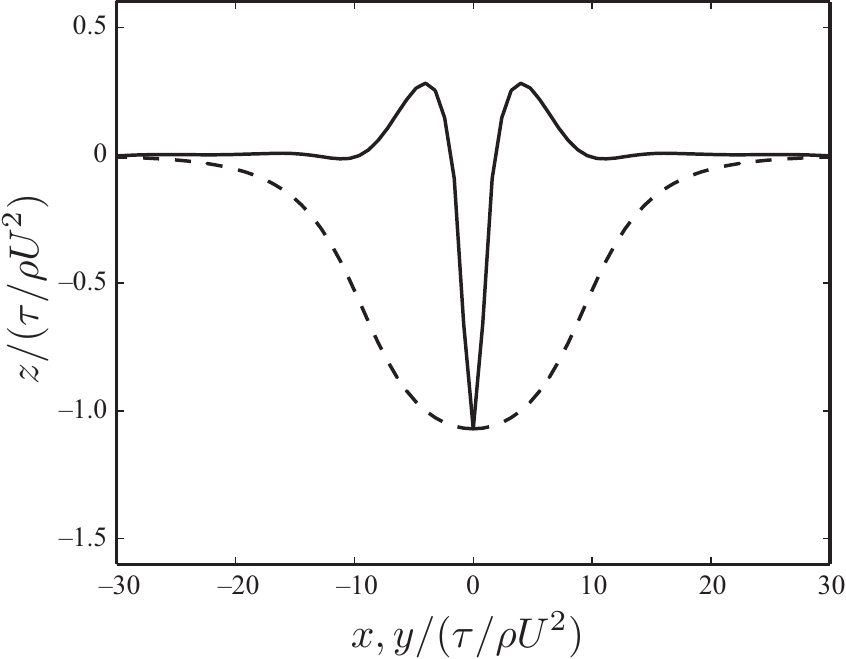}\\
\end{tabular}
\end{center}
\caption{(\textit{a}) Profiles along the main axes of the lump shown in figure \ref{fig:profile_pcolor} in dimensionless form. The solid line is the profile in the streamwise ($X$) direction and the dashed line is the profile in the cross-stream  ($Y$) direction. (\textit{b}) Profiles of a steady freely propagating lump from inviscid potential flow calculations by \cite{Parau2005} for $\alpha=0.919$. }
\label{fig:profile}
\end{figure}

From the surface elevation plots in figure \ref{fig:stateIII_snapshots}, it is observed that the three-dimensional shape of the isolated depressions in state III resembles the shape of the lumps in state II measured by \cite{DiorioJFM} and the freely propagating steady lumps from potential flow calculations by \cite{Parau2005}. To better illustrate the shape of the state III depressions, the two-dimensional profiles along the axes of symmetry of the depressions are calculated. The cross-stream direction (Y-axis) is defined by fitting a line to the location of the maximum depth in each y-position in the vicinity of the maximum depth of the depression. The streamwise direction (X-axis) is perpendicular to the Y-axis and intersects with it at the position of the maximum depth of the depression (see figure \ref{fig:profile_pcolor}). The profiles along the X and Y axes are plotted in figure \ref{fig:profile}-(\textit{a}) and show remarkable similarity to the freely propagating lumps of \cite{Parau2005} shown in figure \ref{fig:profile}-(\textit{b}). Note that the steady lumps in state II and the freely propagating lumps in \cite{Parau2005} are highly localized in all directions and the surface height approaches zero away from the center of the lump.  However, in state III, the depressions are not fully isolated and are probably interacting with one another. This is especially clear in the cross-stream profile (dashed line). 

One of the well-known features of solitary waves is the one-to-one relationship between their speed and amplitude. Resonantly forced lumps of state II move at the same speed as the pressure forcing and a monotonically decreasing relation between their maximum depth and the towing speed was found in the experiments of \cite{DiorioJFM}. A similar relation was found in the inviscid potential flow calculations of freely 
propagating steady lumps by \cite{Parau2007}. One wonders if such a relationship exists for the state III depressions as well.

Unlike the lumps  in state II,  the  speed and depth of a depression in state III are not constant --  each depression undergoes both smooth and erratic changes in these quantities.  The erratic behavior is probably due to interactions with previous and subsequent depressions, see above discussion.  Furthermore, obtaining instantaneous velocity from the position of the maximum depth of a depression often results in noisy data. To overcome this latter issue, a quasi-steady behavior is assumed for the depressions: the velocity for each depression is estimated by locally fitting straight lines to position versus time data as the the depression depth passes through each of  a specified set of equally spaced depths between 0.6~mm and 1.4~mm. To reduce noise in determining the location of the maximum depth of a depression, the surface shape was approximated by fitting a quadratic surface to a small area around the central region of the depression and the location of the maximum depth of the fitted surface was used for further processing. Each depression is tracked in this way until it moves out of the field of measurement or its depth becomes smaller than 0.5 mm.

\begin{figure}
\begin{center}
\includegraphics[width=3.5in]{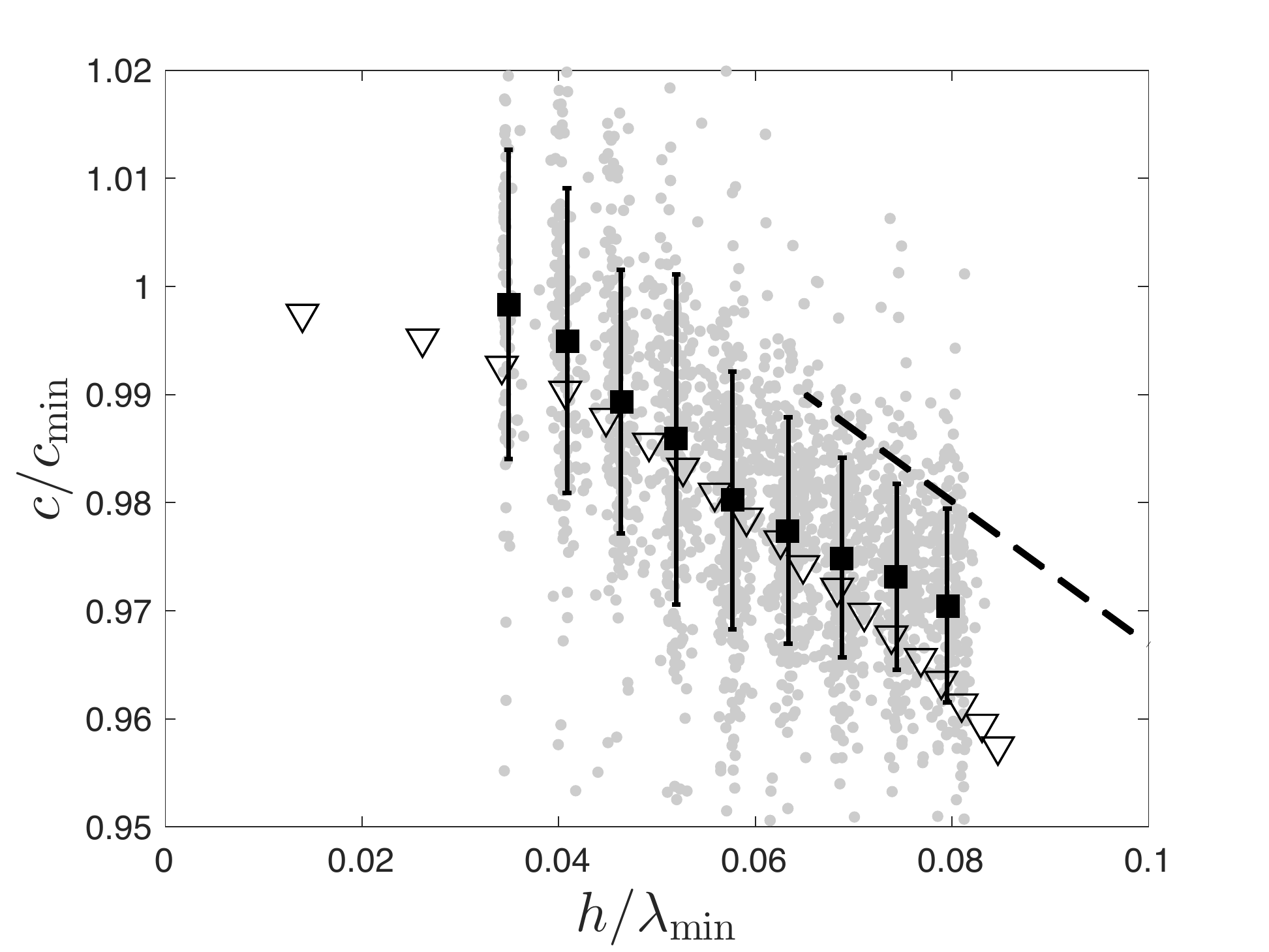}
\caption{Speed of the depressions (normalized by $c_\mathrm{min}$) in the reference frame of the laboratory versus the maximum depth of the depressions (normalized by $\lambda_\mathrm{min}$). The grey dots are the velocity of a given state III depression in each band of depression depths.  $\blacksquare$ Average speed for each depth band. The error bars are the standard deviations. $\triangledown$ Potential flow calculations of \cite{Parau2007} for freely propagating lumps. $--$ Linear fit from the experimental data for forced lumps in state II from \cite{DiorioJFM}.}
\label{fig:v_vs_h}
\end{center}
\end{figure}

This measurement was performed for 373 individual depressions observed in all the runs with $0.977\leq\alpha\leq1.016$. Results are plotted in figure \ref{fig:v_vs_h}. The depression depth, $h$, is normalized by the wavelength of a linear wave at the minimum phase speed ($\lambda_\mathrm{min}$=1.71~cm). In this plot, each grey dot represents the velocity of a lump at one of the prescribed depression  depths. The black squares are the average speeds for each depth and the error bars are the standard deviations. As expected, the grey dots are dispersed over a relatively large area due to the uncertainty in the speed measurement and the erratic behavior of the  depressions. However, in an average sense, the behavior of the unsteady lumps of state III is remarkably similar to that of the inviscid and steady free lumps of \cite{Parau2007} (shown as triangles in the plot) and the forced steady lumps of \cite{DiorioJFM} (shown as the dashed line in the plot).  Given the shape and depth-speed characteristics of the depressions, they appear to be freely propagating gravity-capillary lumps.

\begin{figure}
\begin{center}
\includegraphics[width=3.5in]{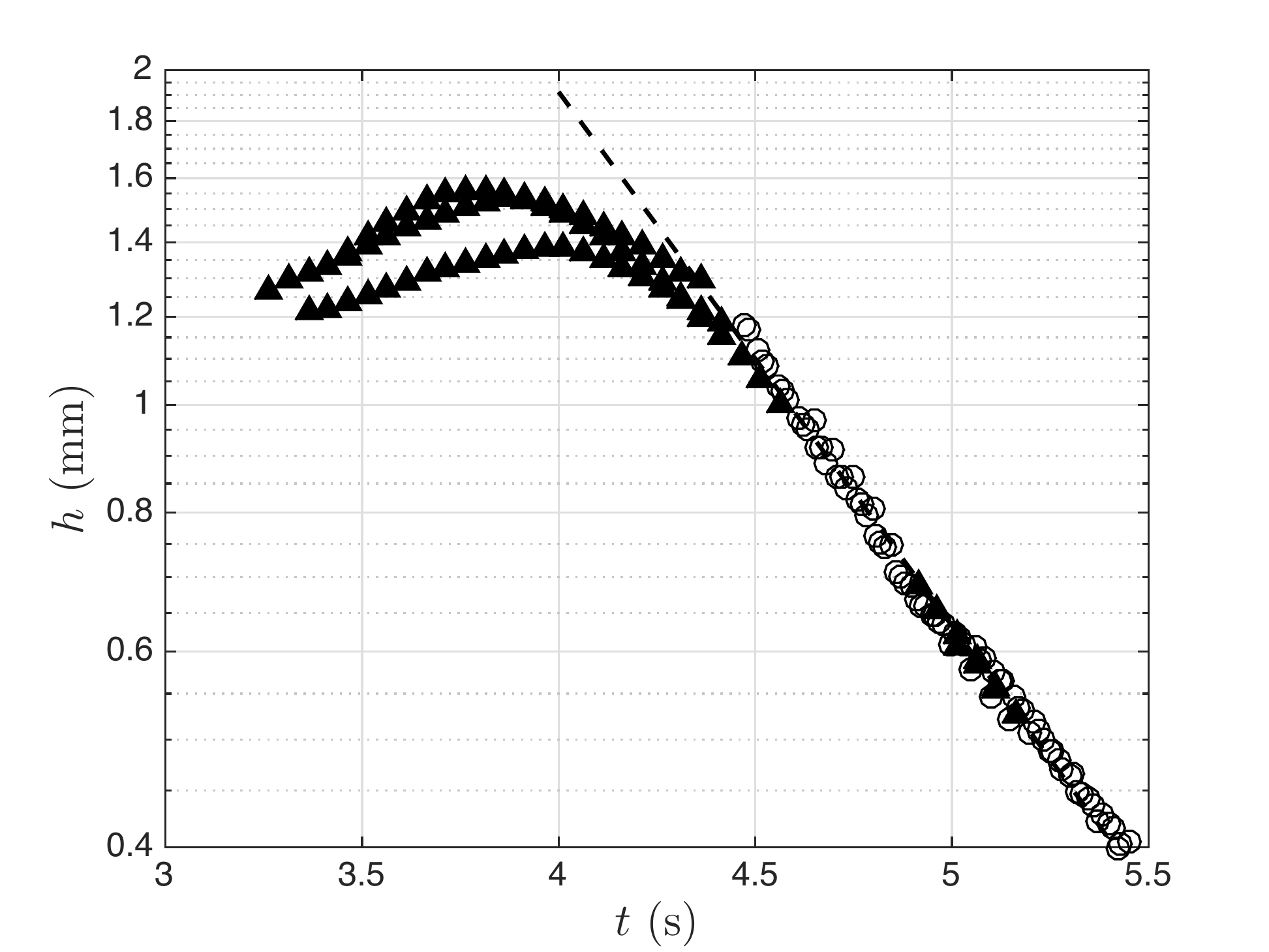}
\caption{Maximum depth of first lumps ($h$) versus time ($t$).  The time $t=0$ is approximately the instant when the carriage starts its motion. Data from separate experimental runs with  $\epsilon=0.24$ and $\alpha=0.986$ are shown. The black triangles (open circles) are from measurements with the camera field of view close to (far from) the pressure source. The dashed line is a least-squares fit of an exponential decay law to the open circles. The exponential decay rate calculated from this fit is $\sigma=1.11$~s$^{-1}$. The accuracy of the depth measurements is $\pm 0.06$~mm.}
\label{fig:decay}
\end{center}
\end{figure}

As mentioned earlier, the lumps in state III first form and then decay quickly as they move away from the pressure source. The decay properties of these lumps were measured by tracking their maximum depth in time.  In order to measure lump behavior that is likely to be  similar to that of freely propagating lumps, conditions where the lump in question was far from any neighbors were chosen. To achieve this, two criteria were met.  First, only the first lump in each run was considered since these lumps propagate into  relatively undisturbed water. Second, an experimental condition  with a long shedding period ($\epsilon=0.24$ and $\alpha=0.986$, see the left column of plots in figure \ref{fig:stateIII_snapshots}) was chosen so the next lump shed was far behind the first lump. 
A plot of  depression depth ($h$ on a logarithmic scale) versus time for several lumps meeting these conditions is shown in figure \ref{fig:decay}. The three sets of black triangles are from three experimental runs in which the camera's field of view was close to the pressure source. For these runs, the depth increases at first as the lump is formed and then starts decreasing once it becomes detached from the local depression region.   It was observed that the first shedding events in separate runs are not exactly repeatable. Therefore, the data points from separate runs are shifted in time (by no more than 0.15 seconds) to form a better match between runs.  The gaps in these data sets for $4.5 \lesssim t \lesssim  5$~s are due to blockage of the camera's line of sight by one of the water tank's structural elements.
In order to track the depth of the lumps for later  times, the camera was moved laterally away from the air-jet tube and the distance between the dot pattern and the water surface was increased to obtain a more accurate measurement of small surface deformations. 
Furthermore, the starting position of the carriage motion was set to a new location so that the tank structure would not block the camera view. The results for the first lump in four separate runs are shown as open circles in figure \ref{fig:decay}, along with the previously discussed data. An exponential decay function in the form of $h(t)=A\exp(-\sigma t)$ was fitted by the least squares method to the data from these four runs and is plotted as a dashed line in the figure. The calculated decay rate is $\sigma=1.11$~s$^{-1}$. Thus, a lump loses about 63 percent of its initial depth after about 0.90~s, a time over which it propagates approximately $0.90c_\mathrm{min} = 20.8$~cm. A similar decay time was predicted from numerical results obtained with the model equation in \cite{ChoJFM}.  This short characteristic decay time may partially account for the scarcity of reported observations of lumps in nature.

\begin{figure}
\begin{center}
\includegraphics[width=3.5in]{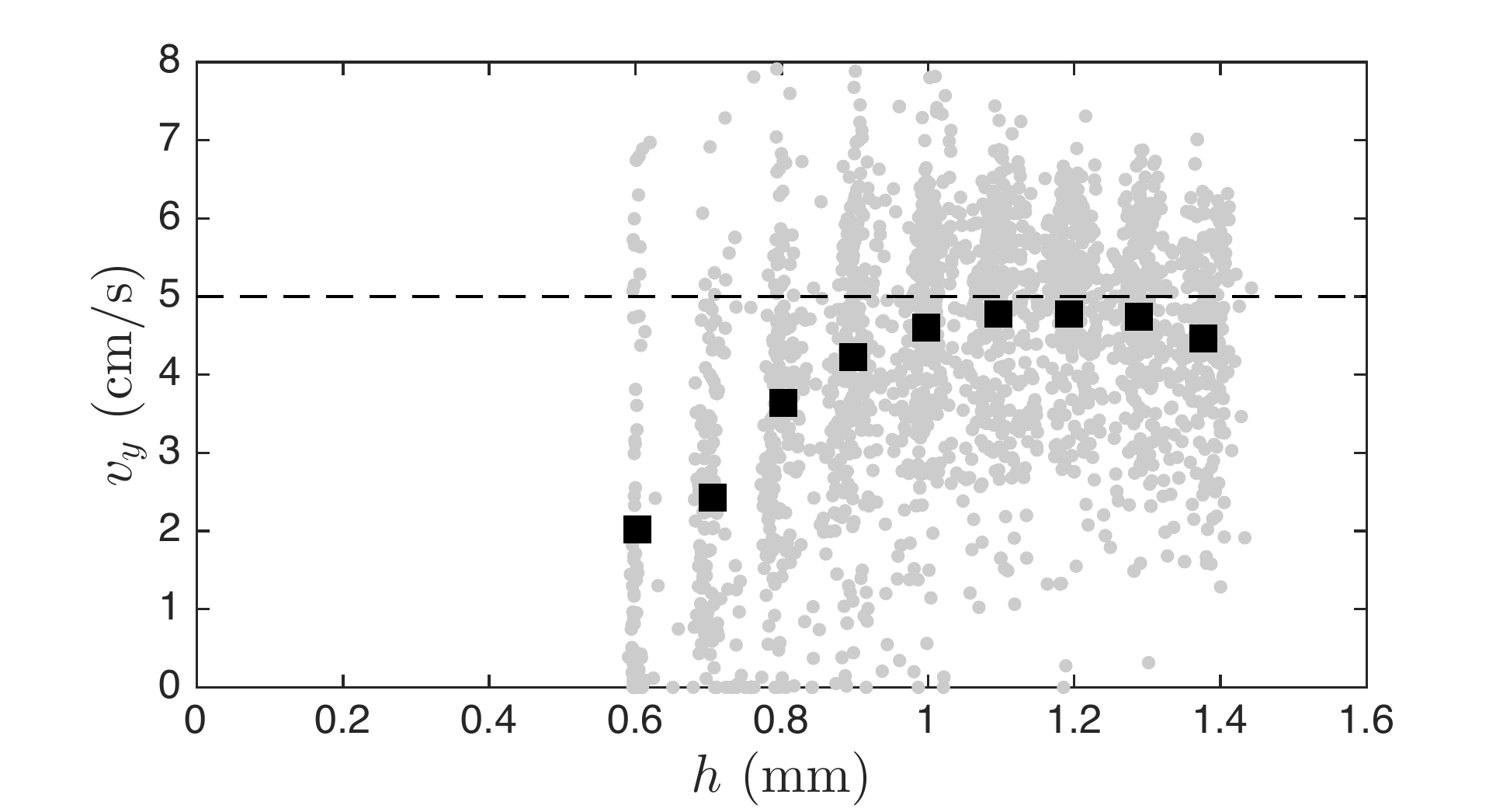}
\caption{Cross-tank component of lump velocity ($v_y$) versus $h$ for all lumps.}
\label{fig:vy_vs_h}
\end{center}
\end{figure}

The measurements of the speed, propagation direction and decay rate of the lumps can now be used to verify that  the tank width is sufficient to avoid the influence of sidewall reflections on the experimental results.  In this verification, the decay of $h$ as a lump travels laterally over a distance of one-half of the tank width is calculated.  The measurements of the cross-tank components ($v_y$) of the lump velocities are shown in Figure~\ref{fig:vy_vs_h}.  As can be seen in the figure, the highest average $v_y$ values are about $5$~cm/s.  Given the tank half-width of 15~cm, the transit time would be about 3~s.  In this time, $h$ would decay by a factor of $e^{-3.0\sigma} = 0.036$.  Thus, the tank wall reflections should have minimal influence on the results.

The effects of $\alpha$ and $\epsilon$ on the shedding period in state III were also investigated by using dot pattern image sequences with large distortions (see figure \ref{fig:dotpattern}\textit{c}).  The instantaneous period ($T$) was taken as the time difference between the passage of two successive lumps across fixed vertical lines near both sides of the images and an average of these $T$ values  over each run was taken as the shedding period $\overline{T}$ for each experimental condition. A reference time, $T^*$, based on linear theory was chosen to nondimensionalize the shedding period data.  In choosing $T^*$, it was noted that in linear theory when a source starts from rest and reaches a steady speed, the unsteadiness of the wave pattern can be partially due to a wave component whose group velocity $c_g$ is equal to the towing speed $U$. In the present scaling idea, only a single time scale is desired,  rather than one for each towing speed,  so this wave component is taken as the one for which $c_g = c_\mathrm{min}$.  For the Bond number in these experiments ($Bo=0.00465$), there are two values of $k$ that satisfy this criterion, $k_1=k_\mathrm{min}$ and  $k_2=0.594~\mbox{cm}^{-1}=0.162k_\mathrm{min}$. At $k_1$, 
$c = c_\mathrm{min}=c_g$ so the period of this wave, which is finite in the laboratory reference frame, is infinite in the reference frame moving with speed $c_\mathrm{min}$. At $k_2$, $T= 0.259~\mbox{s} =3.40T_\mathrm{min}$ in the laboratory reference frame and $0.598$~s in the reference frame moving at speed $c_\mathrm{min}$.  This latter period is  taken as the reference time scale $T^*$.  

\begin{figure}
\begin{center}
\includegraphics[width=3.5in]{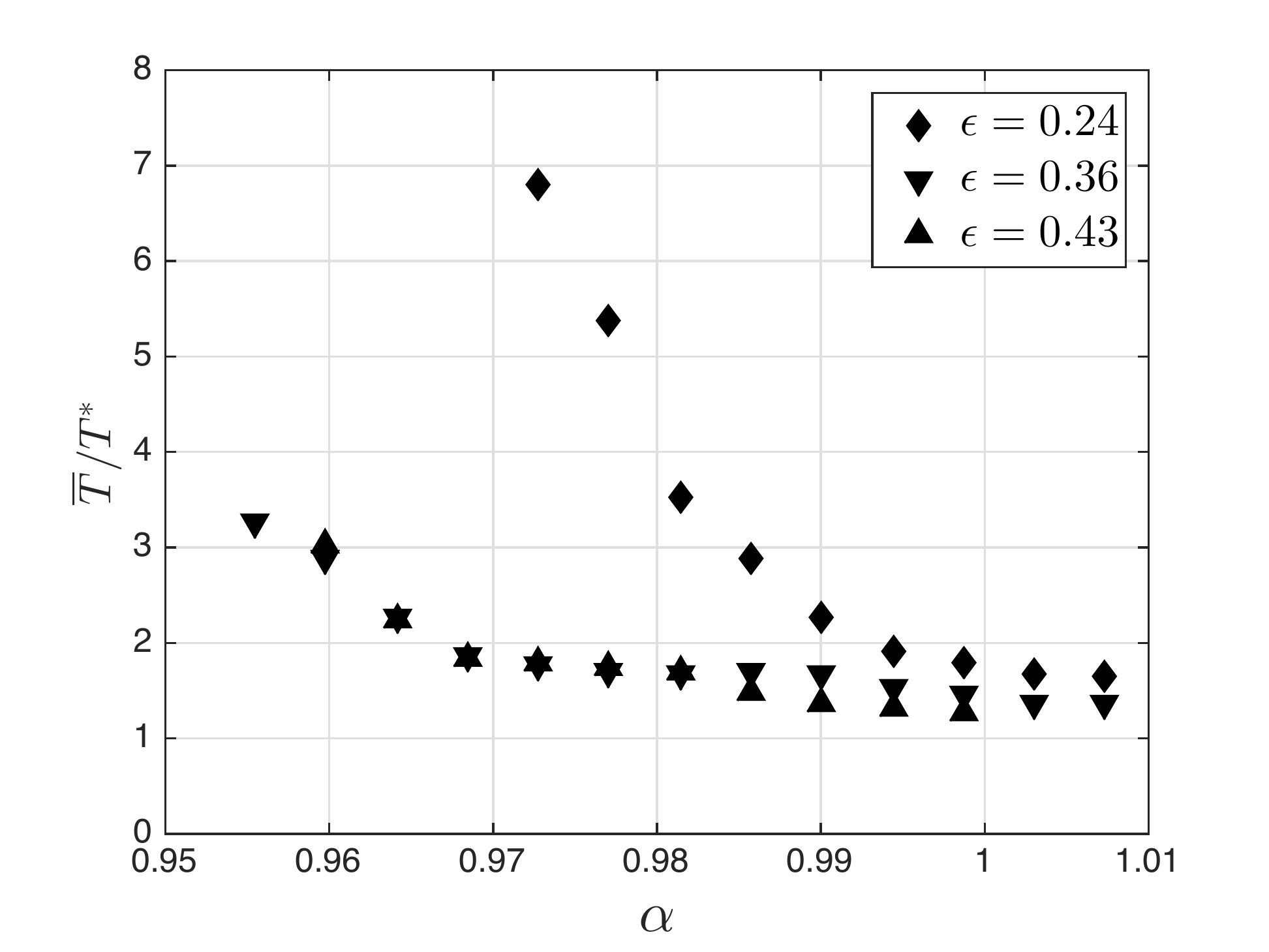}
\caption{The nondimensional shedding period ($\overline{T}/T^*$) of state III lumps versus the towing speed parameter ($\alpha$) for three values of the forcing parameter ($\epsilon$).  $\overline{T}$ is the average lump shedding period and $T^*(= 0.598$~s) is a reference time scale based on linear wave theory as described in the text.   The standard deviation of the  measured periods during a single run  was about $\pm 0.3$~s for low values of $\alpha$ and decreased to about $\pm 0.1$~s for $\alpha \approx 1.0$. }
\label{fig:Shedding}
\end{center}
\end{figure}

Curves of nondimensional shedding period, $\overline{T}/T^*$, versus $\alpha$ are plotted in figure \ref{fig:Shedding} for $\epsilon = 0.24$, 0.36 and 0.43.  As can be seen from the figure, the curve for $\epsilon = 0.24$,  begins at $\alpha = 0.973$, the right side of the state II-III boundary in figure~\ref{fig:state_diagram}, with a shedding period of $\overline{T}/T^*\approx 6.8$.  The value of $\overline{T}/T^*$ then decreases monotonically to about 1.7 at $\alpha = 1.008$.  At the same time, the magnitude of the slope of the curve decreases monotonically, becoming nearly zero for $\alpha >1.0$.       The curves for $\epsilon = 0.36$ and 0.43 start at $\alpha \approx 0.96$ with $\overline{T}/T^*\approx 3.0$ and  are nearly identical.  As $\alpha$ increases, $\overline{T}/T^*$, decreases monotonically, reaching a value of about 1.4 at $\alpha = 1.008$.  These curves are also nearly level for $\alpha > 1.0$.  The fact that the two curves for the higher values of $\epsilon$ are nearly equal and very different from the curve for the lowest value of $\epsilon$ indicates  an insensitivity to the effects of nonlinearity beyond a certain level of forcing.  The fact that the curves for all three values of $\epsilon$ nearly converge to $\overline{T}/T^* \approx 1.5$ as $\alpha \rightarrow 1.0$, may indicate that the nonlinearity, which is of primary importance for creating the surface response at lower values of $\alpha$, causes  only a modification to linear theory for $\alpha \gtrsim 1$.  In fact,  for the cases with $\alpha > 1.0$ the oscillation may be primarily a starting transient, as is discussed further below.  

\begin{figure}
\begin{center}
\begin{tabular}{c}
(\textit{a})\\
  {\includegraphics[trim = 0 0.4in 0 0,clip=true,width=3.5in]{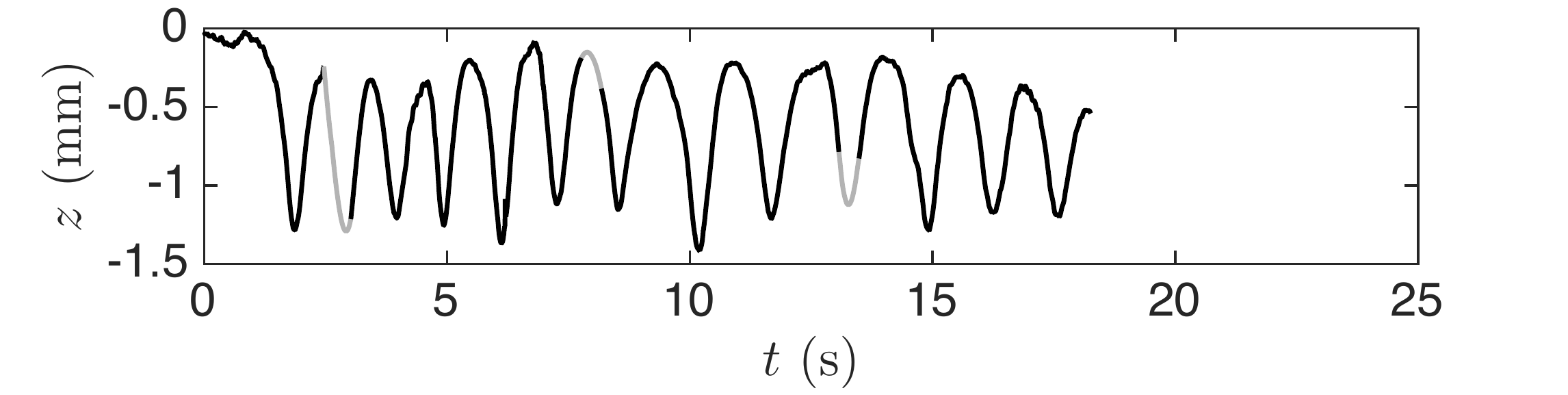}}\\
(\textit{b})\\
  {\includegraphics[trim = 0 0.40in 0 0,clip=true,width=3.5in]{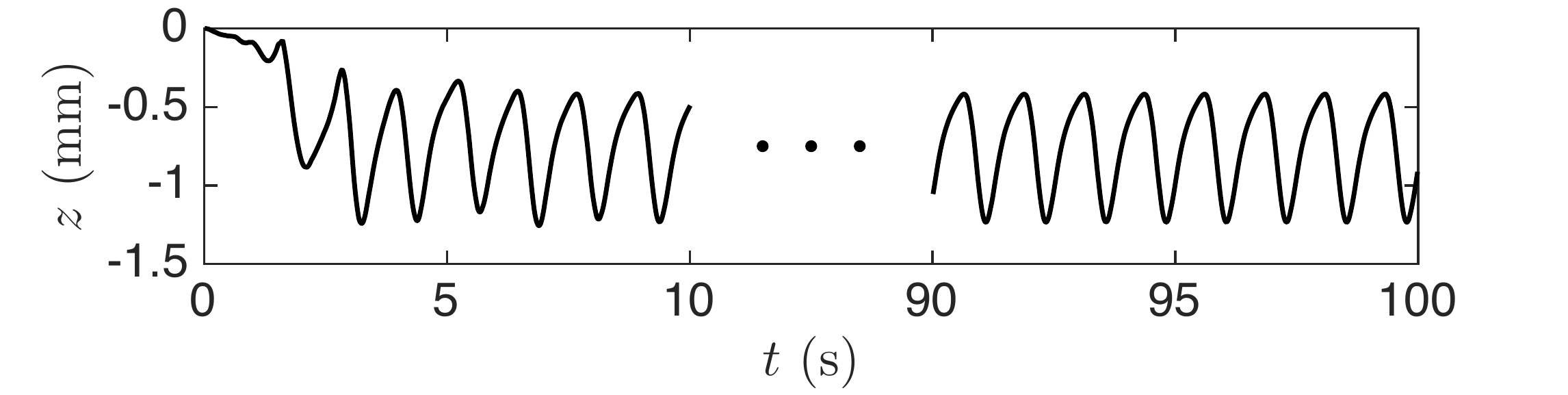}}\\
(\textit{c})\\
  {\includegraphics[trim = 0 0.40in 0 0,clip=true,width=3.5in]{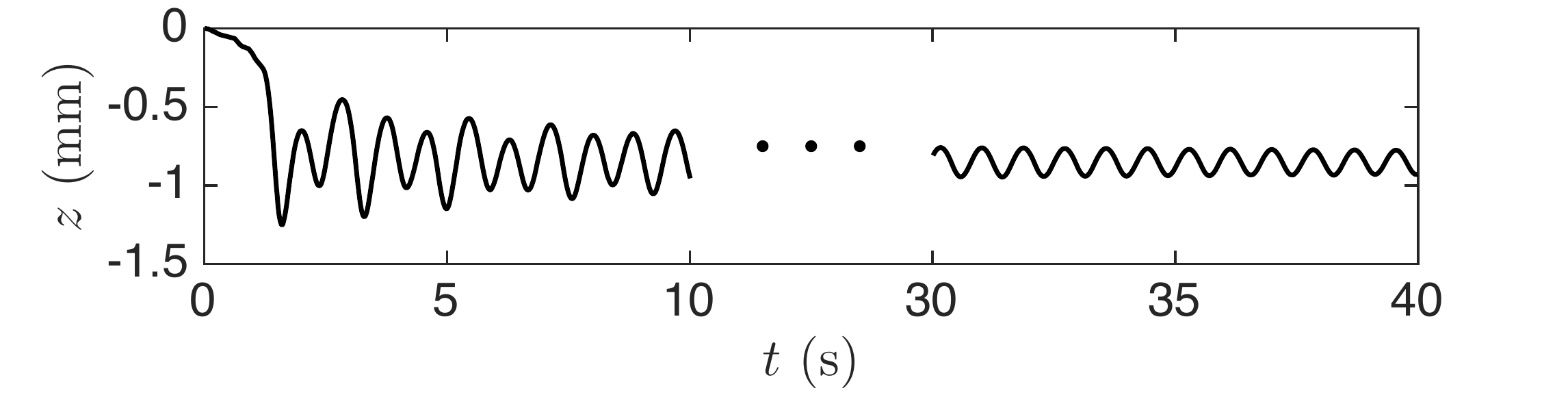}}\\
(\textit{d})\\
  {\includegraphics[width=3.5in]{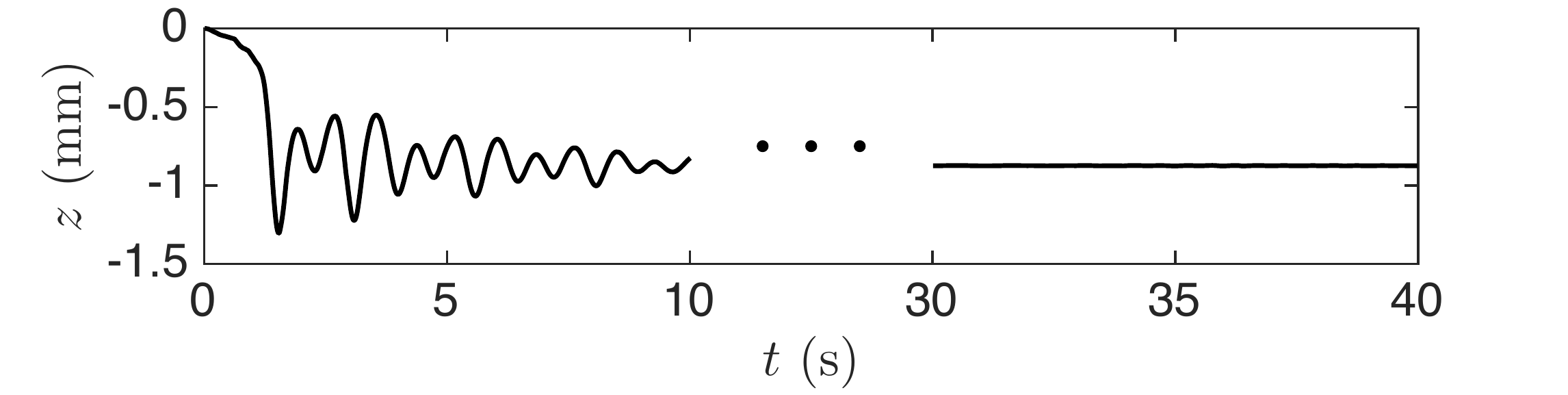}}\\
\end{tabular}
\end{center}
\caption{Minimum surface elevation at $y$=60~mm versus time.  (\textit{a}) Experiment with $\epsilon=0.24$ and $\alpha=0.994$. Numerical calculations of the model equation with $A=0.27$ and $\tilde{\nu}=2.4\tilde{\nu}_{0}$ for (\textit{b}) $\alpha=0.994$, (\textit{c}) $\alpha=1.012$ and (\textit{d}) $\alpha=1.016$. In the top plot, the grey lines are interpolated data for the locations where the camera view is blocked by elements of the tank structure.}
\label{fig:m_vs_t}
\end{figure}

The longevity of the lump shedding was addressed through the experimental measurements and the results from the model equation. Typical  results are shown in four plots of minimum surface height at the cross-stream position of $y=60$~mm versus time in figure \ref{fig:m_vs_t}.  The plot in figure \ref{fig:m_vs_t}(\textit{a}) is from an experimental run with $\epsilon=0.24$ and $\alpha=0.994$ and the oscillations in height correspond to the passage of lumps through the designated cross-stream location. These oscillations continue for  the entire length of the steady-speed part of the experimental run (18.2~s), which is limited by the present tank length. This result suggests that the shedding of lumps in state III may continue indefinitely.  To further investigate the shedding duration at longer times, the model equation described in \S \ref{sec:numerical_model} was used. The temporal evolution of the pattern was computed for 100 seconds with parameters $A=0.27$ and $\tilde{\nu}=2.4\tilde{\nu}_{0}$, where $\tilde{\nu}_{0}=0.003$ is the decay rate of linear sinusoidal waves.  For this set of parameters, the $\alpha$ values at the state II-III boundary in the model equation match the values in the  experiments when $\epsilon = 0.24$.  Calculation results are presented for three values of $\alpha=0.994$, 1.012 and  1.016, in figures~\ref{fig:m_vs_t}(\textit{b}), (\textit{c}) and (\textit{d}), respectively.
 As in the experiments, the minimum height of the pattern at the cross-stream position $y=60$~mm versus time is plotted in the figures.  For $\alpha = 0.994$, full field deformation maps (not shown here) indicate the response is clearly in State III and the plot in figure \ref{fig:m_vs_t}(\textit{b}) shows the short time ($0\leq t\leq10$ s) and the long time ($90\leq t\leq100$ s) part of the surface height record. It is observed that the oscillations become regular and well-defined after about 5 seconds and that the shedding of the depressions continues for the entire numerical run, which is about five times as long as the experimental run.  This lends strong support to the idea that the shedding in state III is not a transient phenomenon.

The model equation was also used to investigate the transition from periodic shedding in state III to a steady V-shape response for $\alpha>1$.  It was found that, for the parameters used in the simulations, the amplitude of the oscillations decreases as $\alpha$ is increased above 1, but the periodic behavior continues up to $\alpha=1.012$ (figure \ref{fig:m_vs_t}-\textit{c}), though, as can be seen in the plot, it takes a relatively long time to become established. For $\alpha=1.016$ some random oscillations are observed initially, but the response reaches a steady state after about 10 seconds (figure \ref{fig:m_vs_t}-\textit{d}). A thorough examination of the transition from the periodic state III to the steady linear pattern for $\alpha>1$ is left for future studies.

The periodic shedding of lumps in state III seems to be analogous to the generation of solitary waves upstream of a source moving near the critical speed of pure gravity waves in shallow water \cite[]{Wu1987}. Detailed numerical and experimental results for a surface pressure distribution and a bottom topography moving at trans-critical Froude numbers ($Fr=U/c_{0}$ where U is the source speed and $c_{0}=\sqrt{gH}$) and for a range of forcing parameters are reported in \cite{Lee1989}. In the following, some aspects of these two types of solitary waves (pure gravity solitary waves and gravity-capillary lumps) are compared and summarized.

In the shallow water case, according to linear theory, the surface response is singular as $Fr=1$ is approached and no wave response is predicted for $Fr>1$. Similarly, for the gravity-capillary case in deep water, no linear waves exist for $\alpha<1$ and the linear response becomes unbounded at $\alpha=1$. The mechanism of generation of solitary waves for a disturbance moving at a trans-critical speed is described qualitatively by \cite{Wu1987} as follows: ``In this trans-critical speed range, the dispersive effect is weak, so the velocity of propagating mechanical energy away (by means of radiating long waves) from the forcing disturbance is about equal to the velocity of the moving disturbance. The local wave will therefore grow as the energy acquired by local fluid at the rate of work by the moving disturbance keeps accumulating. When the local wave reaches a certain threshold magnitude, the increase in phase speed with increasing amplitude (due to nonlinear effects) will be sufficient to make the wave break away from the disturbance, thus `born free' as a new solitary wave propagating forward with a phase velocity appropriate to its own amplitude. The process is then repeated over a new cycle.'' A similar mechanism can be attributed to the periodic generation of lumps in state III. The main difference is that for gravity-capillary solitary waves, the phase speed decreases with increasing amplitude. Hence, as the local wave grows, the phase speed decreases and a solitary wave is generated downstream of the disturbance when the local wave reaches a certain threshold.

The similarities become more conspicuous using a differential speed parameter defined as $\alpha^\prime = \pm(U/c_{crit} -1)$ with the plus sign and $c_{crit} =c_0$ for pure gravity waves and the  minus sign and $c_{crit}=c_\mathrm{min}$ for gravity-capillary waves. Based on this definition, for both cases, the linear response becomes unbounded at the critical value $\alpha^\prime=0$ and no linear waves exist for $\alpha^\prime>0$.
According to our results for gravity-capillary waves and the findings reported in \cite{Lee1989} for shallow water, the generation of solitary waves continues indefinitely for a range of subcritical  and supercritical  values of $\alpha^\prime$ and the period of generation and the amplitude of the solitary waves increase with increasing $\alpha^\prime$. In the supercritical regime, there exists an upper bound for  $\alpha^\prime$ beyond which the phenomenon ceases to exist and only a local response is found. This upper bound becomes slightly larger when forcing is increased. In the subcritical regime, below a certain $\alpha^\prime$, the amplitude of the solitary waves diminishes with time and the periodic generation seems to be a transient feature of the response. It was also found that the period of generation of solitary waves decreases with increasing forcing.

\subsection{Asymmetric unsteady pattern}
\label{sec:stateII-III}

As discussed at the end of \S\ref{sec:State_diagram}, a new response state featuring an unsteady asymmetric  surface deformation pattern  was discovered experimentally in the boundary region between state II and state III. In \cite{ChoJFM}, the model equation described in \S \ref{sec:numerical_model} was successful in, at least qualitatively, capturing all the response states observed in the experiments described in \cite{DiorioJFM}. In the present work, the model is  used to further explore the  unsteady asymmetric behavior in the state II-III boundary region and the stability of state II.  To this end, calculations were performed with $A=0.30$, $\tilde{\nu}=2.4\tilde{\nu}_{0}$ and a fairly wide range of $\alpha$,  $0.800\leq\alpha\leq1.040$. For each $\alpha$, the evolution of the surface pattern  is calculated for a period of 20 seconds.

\begin{figure}
\begin{center}
\begin{tabular}{cc}
  (\textit{a})&(\textit{c})\\
  {\includegraphics[width=2.5in]{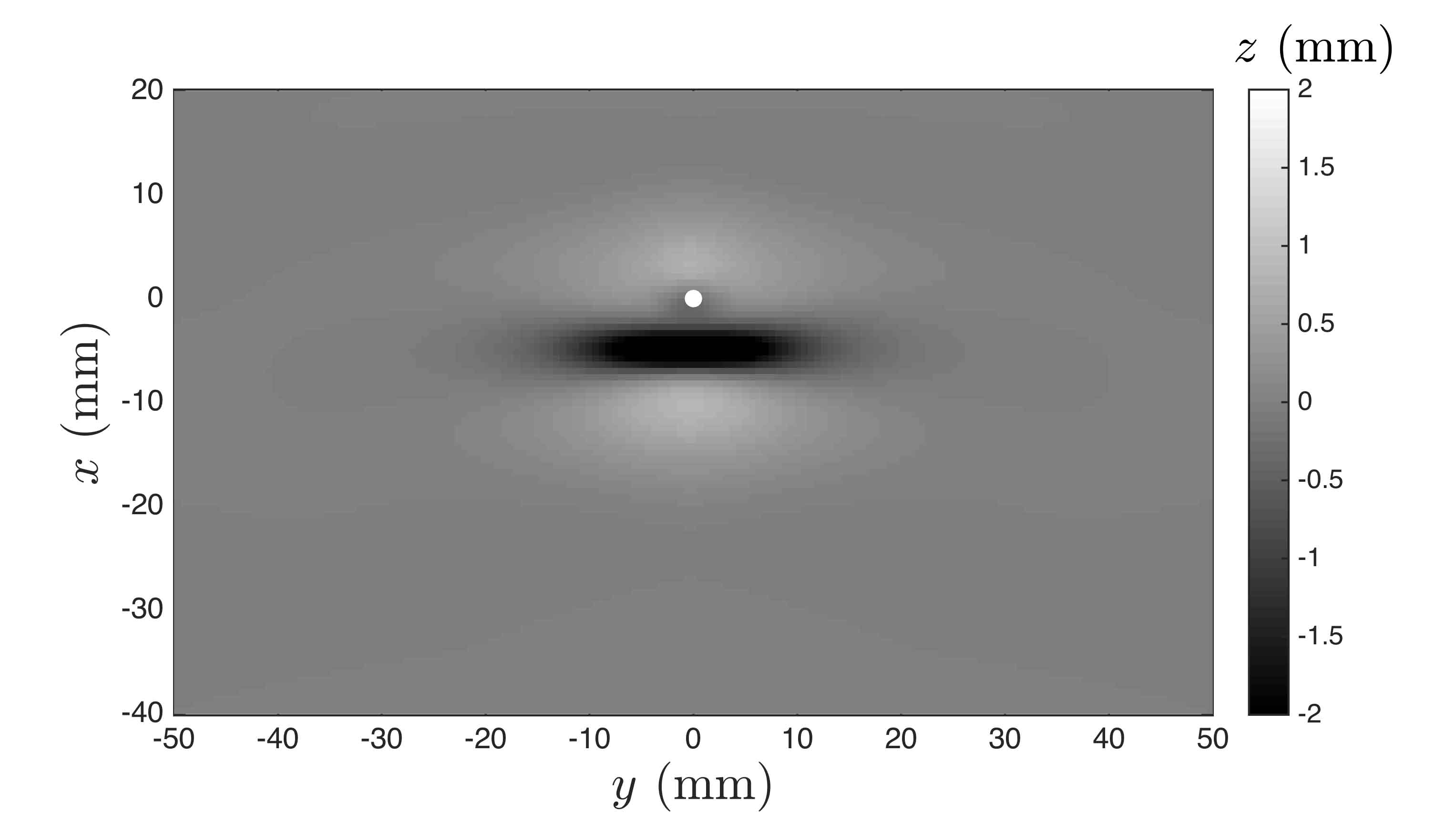}}&
  {\includegraphics[width=2.5in]{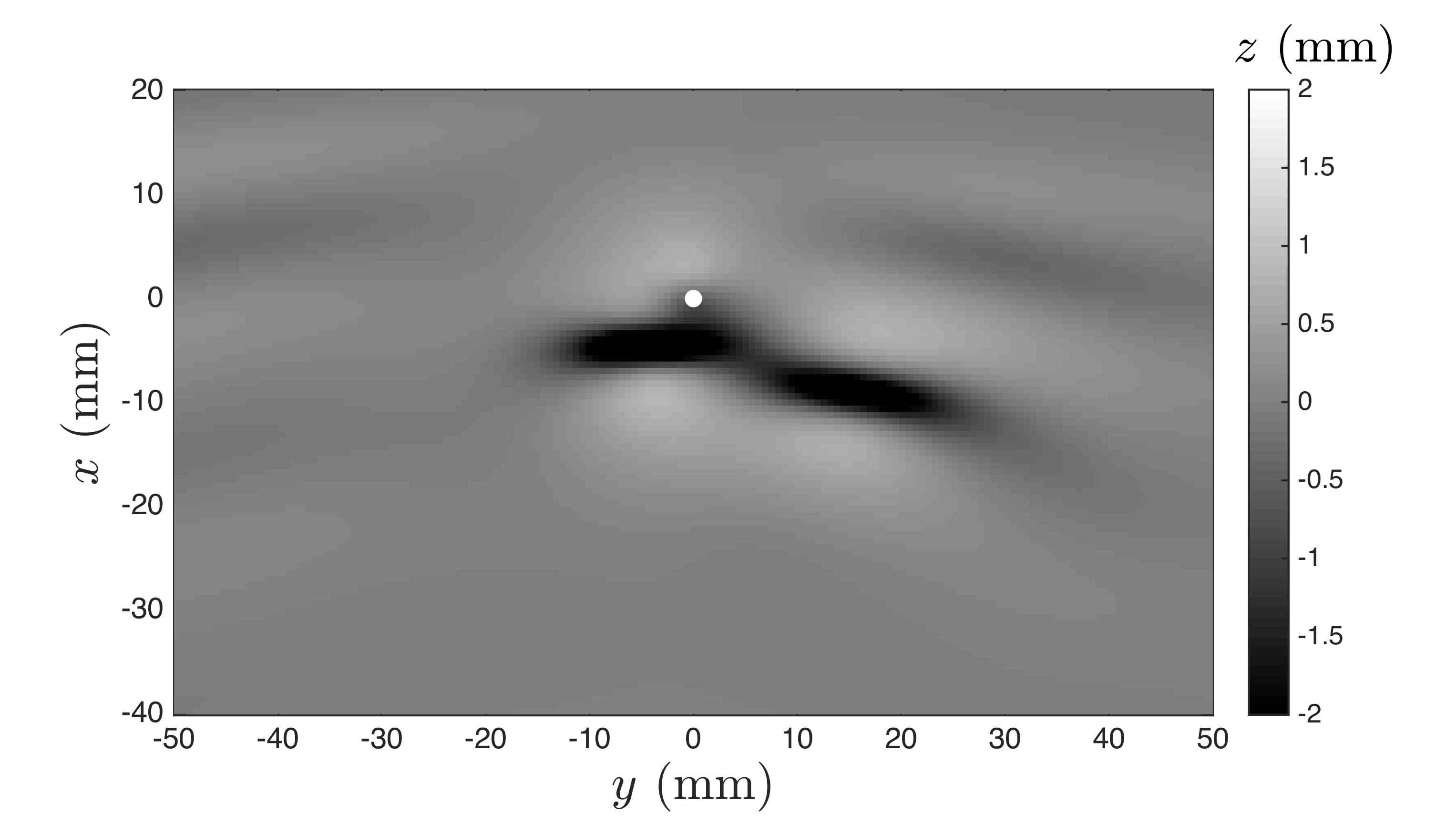}}\\
  (\textit{b})&(\textit{d})\\
  {\includegraphics[width=2.5in]{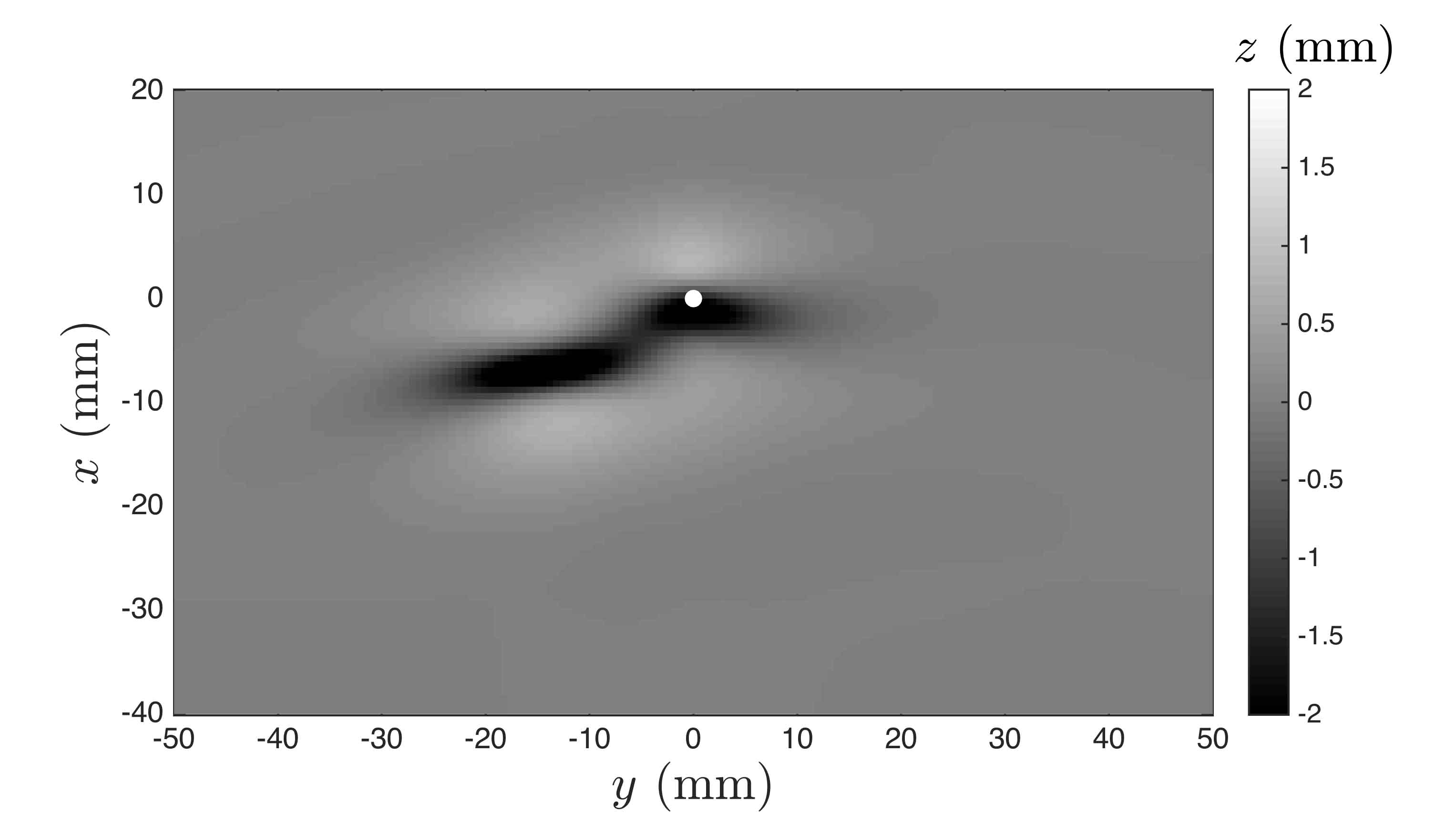}}&
  {\includegraphics[width=2.5in]{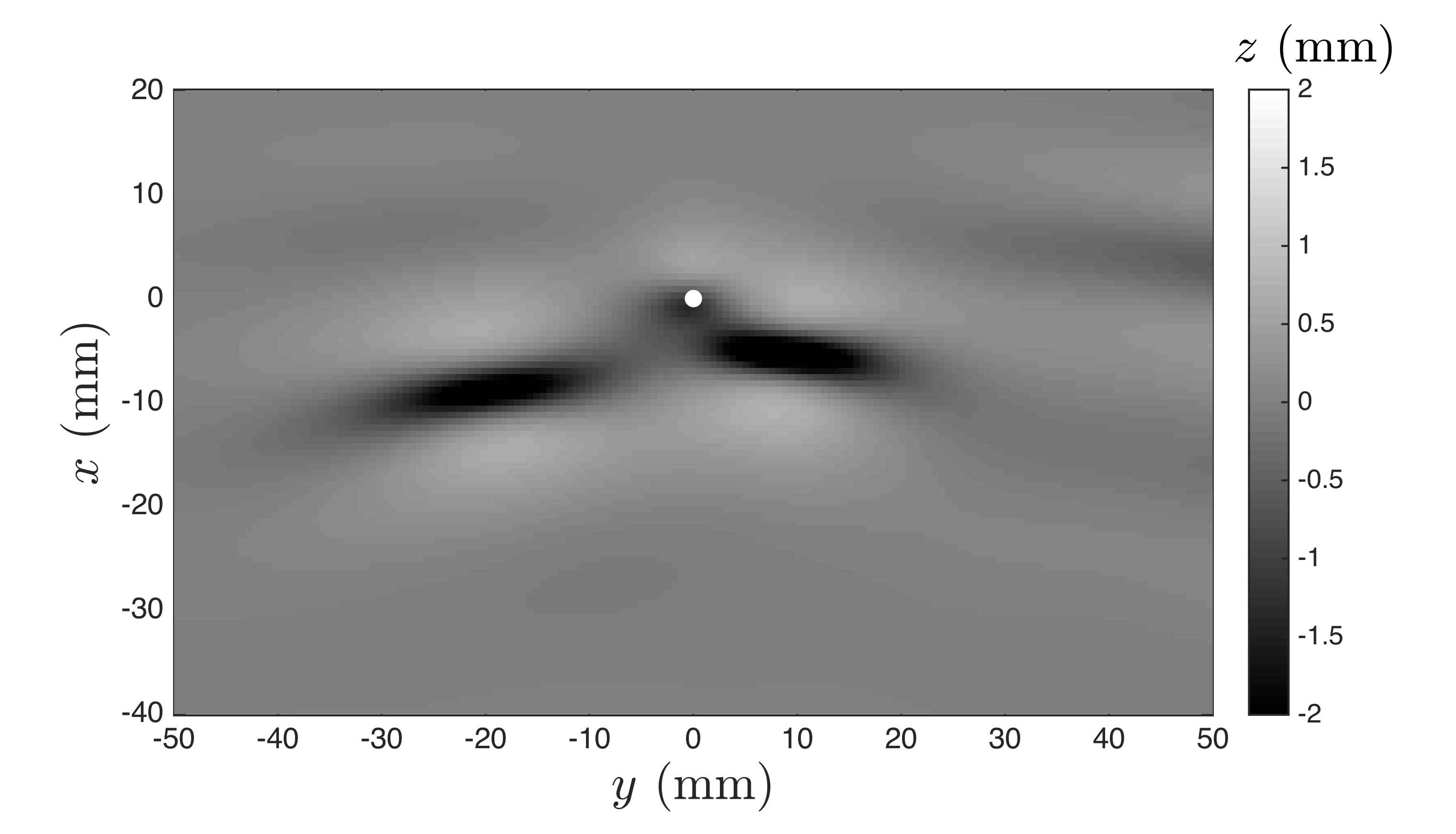}}\\
\end{tabular}
\end{center}
\caption{Surface elevation maps from simulations with $A=0.30$, $\alpha=0.908$ and $\tilde{\nu}_0=2.4\nu_{0}$. The pressure source is located at the origin and indicated by a white dot. (\textit{a}) $t$=6~s. (\textit{b}) $t$=7~s. (\textit{c}) $t$=8.5~s. (\textit{d}) $t$=9.5~s.}
\label{fig:SimBoundary}
\end{figure}

\begin{figure}
\begin{center}
\begin{tabular}{cc}
  (\textit{a})&(\textit{c})\\
  {\includegraphics[width=2.5in]{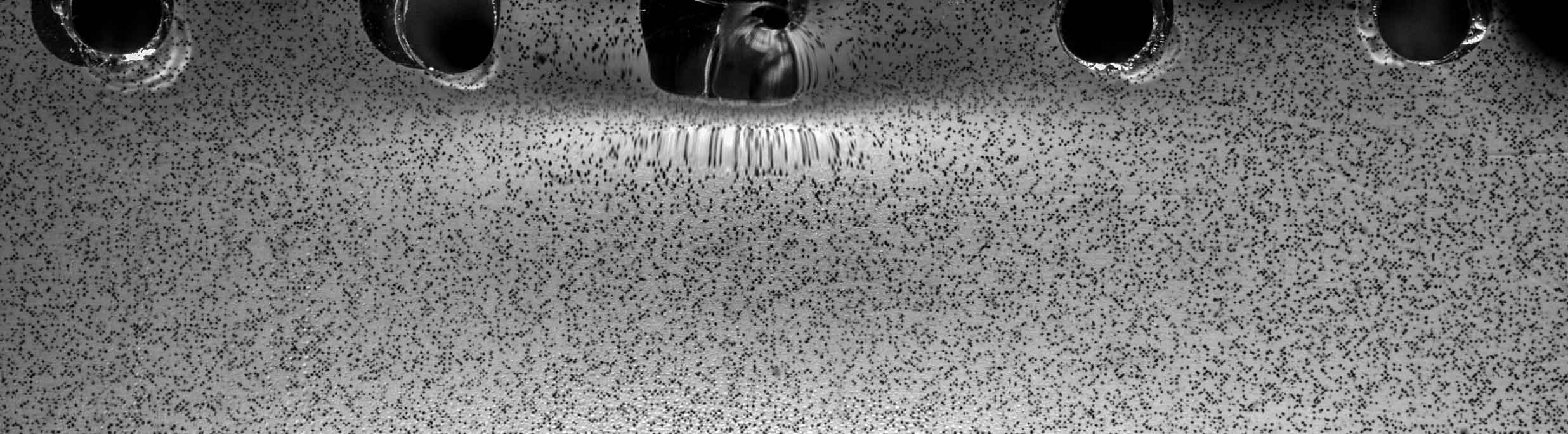}}&
  {\includegraphics[width=2.5in]{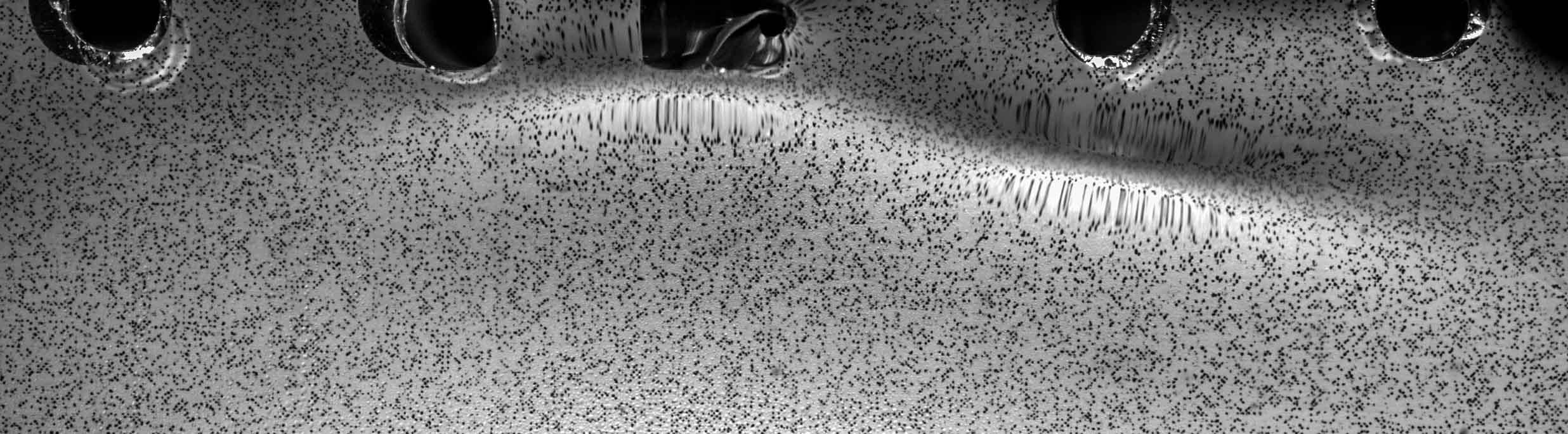}}\\
  (\textit{b})&(\textit{d})\\
  {\includegraphics[width=2.5in]{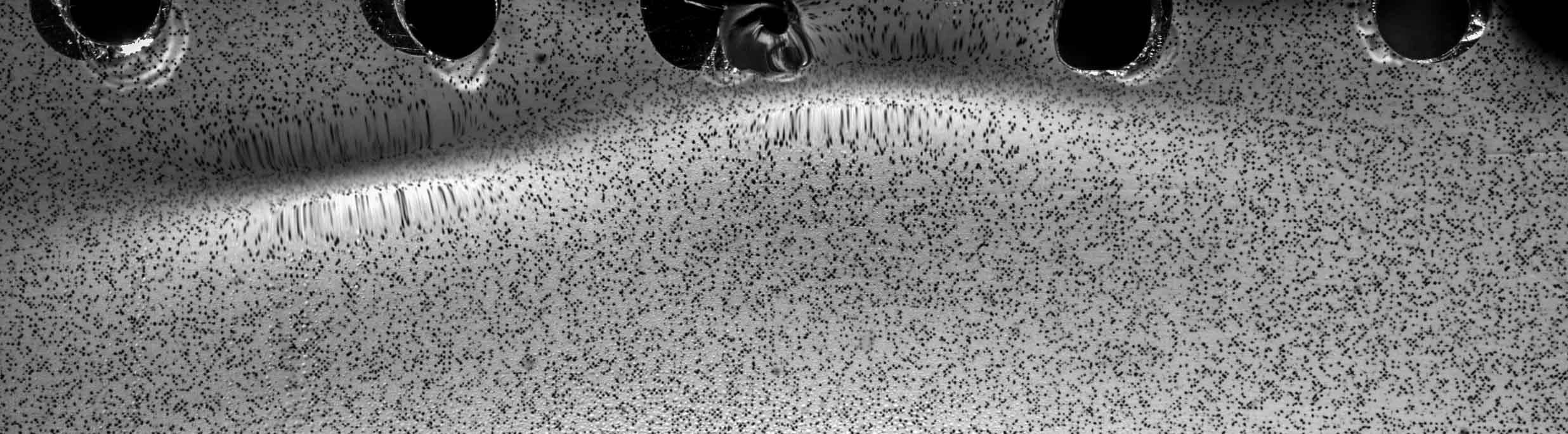}}&
  {\includegraphics[width=2.5in]{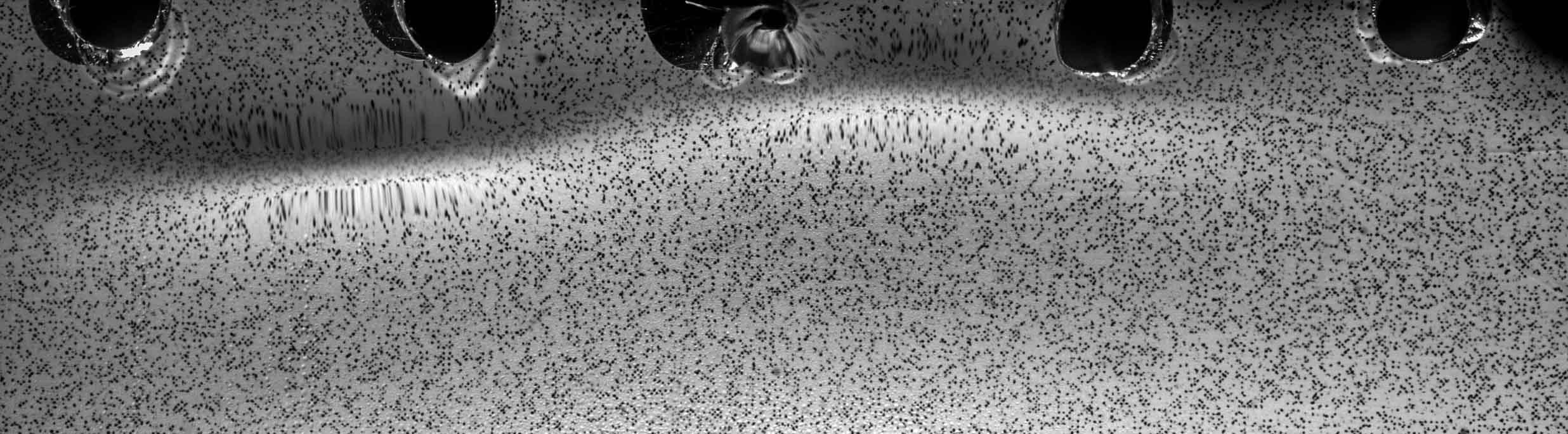}}\\
\end{tabular}
\end{center}
\caption{Refraction images  in an experiment with $\epsilon=0.24$ and $\alpha=0.964$. The air-jet tube is located in the top middle. The field of view is about 9~cm wide. (\textit{a})~$t=0.0$, (\textit{b})~$t=3.7$~s, (\textit{c})~$t$=7.1~s,  (\textit{d})~$t$=10.7~s.}
\label{fig:Boundary}
\end{figure}

As $\alpha$ is increased in successive numerical calculations, a sharp boundary is found between state I and II at $\alpha=0.873$ where a sudden jump in the maximum depth of the depression was observed. 
In the speed range $0.873\leq\alpha\leq0.962$, a steady state II   is formed after a few transient oscillations.  Both of these results were reported in \cite{ChoJFM}. However, in the present calculations it was found that after some time the steady response in state II breaks down into an unsteady and asymmetric pattern that is very similar to the asymmetric wave pattern observed in the experiments for conditions in the state II-III boundary.   This similarity can be seen by comparing the snapshots of the surface height field at late time in the computed wave pattern in state II,  $\alpha=0.908$,  given in figure~\ref{fig:SimBoundary},  to the series of refraction images showing the asymmetric deformation pattern in the state II-III region in the experiments,  given in figure~\ref{fig:Boundary}. In the numerical calculations, the surface pattern features alternate shedding of lumps from the right and left of the pressure distribution. This shedding behavior does not have a well defined period or left/right order especially at lower $\alpha$ values in the range. It was also found that the steady part of the response breaks down at larger times for higher $\alpha$ values (about 6 seconds for $\alpha$=0.890 to about 10 seconds for $\alpha$=0.962).  For an $\alpha$ value slightly above 0.962, periodic shedding of lumps in a symmetric pattern is observed. The shedding starts after about 2 seconds and continues for the length of the simulation.  This boundary between asymmetric shedding and the symmetric shedding in state III is quite sharp in both the simulations and the experiments.  Investigations into the possibility that the state II response in the experiments eventually breaks down into an asymmetric shedding pattern are left to future work using a longer tank. 

\section{Summary and conclusions}
\label{sec:Conclusions}

The unsteady wave pattern behind a surface disturbance moving with a constant speed ($U$)  close to the minimum phase speed of gravity-capillary water waves ($c_\mathrm{min}$) was investigated by using a combination of experiments and numerical calculations.  The experiments were performed in a long open-surface water tank in which the surface disturbance was created by downward air flow from the tip of a vertically oriented tube that was attached to a moving instrument carriage.  In the numerical calculations, the model equation described in \cite{ChoJFM} was used to extend the experimental results to times later than were possible in the experiments.

In previously published research, it was found that the water surface deformation pattern for $U\lesssim c_\mathrm{min}$ is determined by two external non-dimensional parameters: the strength of the forcing, $\epsilon=h_0/d$, where $h_0$ is the depth of the local depression under the air-jet tube when it is stationary and $d$ is the tube's internal diameter,  and the towing speed parameter $\alpha = U/c_\mathrm{min}$.  Depending on the values of $\epsilon$ and $\alpha$, the surface deformation pattern behind the source has three distinct response states: state I in which a steady axisymmetric dimple appears under the air jet tube, state II in which a steady gravity-capillary lump appears behind the air-jet tube and state III in which an unsteady V-shaped pattern appears and includes the shedding of isolated depressions from the tips of the V.  These depressions were found to decay rapidly as they propagated away from the source.  In the  previous experimental work, the steady response in states I and II as well as the unsteady response in the state I-II boundary region were studied in detail quantitatively using shadowgraph and laser induced fluorescence techniques, but the response in state III was observed only qualitatively.  

In the present experiments, state III and the boundary region between states II and III were studied in detail using a  combination of cinematic refraction-based and shadowgraph-based techniques.  It was found that the three-dimensional shape of the localized depressions in state III are similar to the shapes of steady, forced lumps of state II and the freely propagating lumps of inviscid potential theory. Measurements of the speed of these depressions during their decay revealed that they follow the speed-amplitude relation of steady lumps from inviscid potential theory. Based on these two findings, the isolated depressions appear to be freely propagating gravity-capillary lumps. The exponential decay rate of these lumps was found to be about 1.11 s$^{-1}$, indicating substantial decay over time scales on the order of one second, a time over which a typical lump travels only about 20~cm.
Numerical simulations of the model equation suggest that the periodic behavior in state III is not a transient phenomenon and continues indefinitely.  It was found in the experiments that the period of shedding ($\overline{T}$) in state III decreases with increasing towing speed  and approaches a value of about $1.5T^*$ as $\alpha$ approaches unity for all values of the forcing parameter, $\epsilon$.  The time scale $T^*$ is the period, as measured in the reference frame of the source, of a linear wave whose group velocity is equal to $c_\mathrm{min}$ and whose wavelength is longer than the wavelength of the wave with phase speed equal to $c_\mathrm{min}$. Numerical simulations of the model equation were used to explore the surface response pattern at times later than could be achieved in the experiments, due to the limited tank length.  It was found that in the simulations, once initiated,  the state III response continues unaltered over the longest numerical runs attempted, 100~s. 

In the experiments, it was found that there is a boundary region between state II and state III, in which the surface response is unsteady and features asymmetric irregular shedding of lumps from the two sides of the deformation pattern. A similar response state was found in the numerical simulations. In fact, in the simulations, the response for all state II conditions eventually breaks down into an asymmetric shedding pattern, similar to that found in the state II-III boundary region in the experiments.  The possibility that the state II response breaks down after a sufficiently long time in the experiments is left for future investigations.

It is noted that the above-described behavior bears a strong similarity to the upstream periodic shedding of two-dimensional solitary gravity waves from a source moving  at trans-critical speeds relative to the maximum speed of linear gravity waves in shallow water.  In this analogy,  the solitary wave shedding processes in the two systems are similar in many respects if one views the process in terms of the source strength and the differential speed parameter, $\alpha^\prime = \pm(U/c_{crit} -1)$, where  $U$ is the towing speed and $c_{crit}$ is the critical speed, with the minus sign and $c_{crit} = c_\mathrm{min}$ for gravity-capillary waves and the plus sign and $c_{crit} =\sqrt{gH}$ for shallow water gravity waves, where $H$ the water depth.

\bigskip

The authors gratefully acknowledge Yeunwoo Cho for helpful discussions about the numerical technique used to produce the unsteady solutions to the model equation.  This work was partially supported by the Office of Naval Research under grant N000141110029  and the National Science Foundation, Division of Ocean Sciences under grant OCE0751853.  JHD also acknowledges the support of the University of Maryland Elkins Professorship.

\bibliography{jfm-references}

\begin{thebibliography}{30}
\expandafter\ifx\csname natexlab\endcsname\relax\def\natexlab#1{#1}\fi
\def\au#1{#1} \def\ed#1{#1} \def\yr#1{#1}\def\at#1{#1}\def\jt#1{\textit{#1}}
  \def\bt#1{#1}\def\bvol#1{\textbf{#1}} \def\vol#1{#1} \def\pg#1{#1}
  \def\publ#1{#1}\def\arxiv#1{#1}\def\org#1{#1}\def\st#1{\textit{#1}}

\bibitem[Adrian \& Westerweel(2010)]{AdrianBook}
{\sc \au{Adrian, €œR.} \& \au{Westerweel, J.}} \yr{2010} {\em Particle Image
  Velocimetry\/}, 1st edn.  \publ{Cambridge University Press}.

\bibitem[Akers \& Milewski(2009)]{Akers2009}
{\sc \au{Akers, B} \& \au{Milewski, P.~A.}} \yr{2009}  \at{A model equation for
  wavepacket solitary waves arising from capillary-gravity flows}.  \jt{Studies
  in Applied Mathematics}  \bvol{122}~(3),  \pg{249--274}.

\bibitem[Akers \& Milewski(2010)]{Akers2010}
{\sc \au{Akers, B.} \& \au{Milewski, P.~A.}} \yr{2010}  \at{Dynamics of
  three-dimensional gravity-capillary solitary waves in deep water}.  \jt{SIAM
  Journal on Applied Mathematics}  \bvol{70}~(7),  \pg{2390--2408}.

\bibitem[Akylas(1984)]{Akylas1984}
{\sc \au{Akylas, T.~R.}} \yr{1984}  \at{On the excitation of long nonlinear
  water waves by a moving pressure distribution}.  \jt{Journal of Fluid
  Mechanics}  \bvol{141},  \pg{455--466}.

\bibitem[Akylas(1993)]{Akylas1993}
{\sc \au{Akylas, T.~R.}} \yr{1993}  \at{Envelope solitons with stationary
  crests}.  \jt{Physics of Fluids A: Fluid Dynamics (1989-1993)}  \bvol{5}~(4),
   \pg{789--791}.

\bibitem[Akylas \& Cho(2008)]{Akylas2008}
{\sc \au{Akylas, T.~R.} \& \au{Cho, Y.}} \yr{2008}  \at{On the stability of
  lumps and wave collapse in water waves}.  \jt{Philosophical Transactions of
  the Royal Society of London A: Mathematical, Physical and Engineering
  Sciences}  \bvol{366}~(1876),  \pg{2761--2774}.

\bibitem[Amini {\em et~al.\/}(1990)Amini, Weymouth \& Jain]{Amini1990}
{\sc \au{Amini, A.A.}, \au{Weymouth, T.E.} \& \au{Jain, R.C.}} \yr{1990}
  \at{Using dynamic programming for solving variational problems in vision}.
  \jt{Pattern Analysis and Machine Intelligence, IEEE Transactions on}
  \bvol{12}~(9),  \pg{855--867}.

\bibitem[Berger \& Milewski(2000)]{Berger2000}
{\sc \au{Berger, K.~M.} \& \au{Milewski, P.~A.}} \yr{2000}  \at{The generation
  and evolution of lump solitary waves in surface-tension-dominated flows}.
  \jt{SIAM Journal on Applied Mathematics}  \bvol{61}~(3),  \pg{731--750}.

\bibitem[Cho(2010)]{ChoThesis}
{\sc \au{Cho, Y.}} \yr{2010}  \at{Nonlinear dynamics of three-dimensional
  solitary waves}. PhD thesis, Massachusetts Institute of Technology.

\bibitem[Cho {\em et~al.\/}(2011)Cho, Diorio, Akylas \& Duncan]{ChoJFM}
{\sc \au{Cho, Y.}, \au{Diorio, J.~D.}, \au{Akylas, T.~R.} \& \au{Duncan,
  J.~H.}} \yr{2011}  \at{Resonantly forced gravity-capillary lumps on deep
  water. part 2. theoretical model}.  \jt{Journal of Fluid Mechanics}
  \bvol{672},  \pg{288--306}.

\bibitem[Dias \& Kharif(1999)]{DiasKharif1999}
{\sc \au{Dias, F.} \& \au{Kharif, C.}} \yr{1999}  \at{Nonlinear gravity and
  capillary-gravity waves}.  \jt{Annual Review of Fluid Mechanics}
  \bvol{31}~(1),  \pg{301--346}.

\bibitem[Diorio {\em et~al.\/}(2009)Diorio, Cho, Duncan \& Akylas]{DiorioPRL}
{\sc \au{Diorio, J.}, \au{Cho, Y.}, \au{Duncan, J.~H.} \& \au{Akylas, T.~R.}}
  \yr{2009}  \at{Gravity-capillary lumps generated by a moving pressure
  source}.  \jt{Phys. Rev. Lett.}  \bvol{103},  \pg{214502}.

\bibitem[Diorio {\em et~al.\/}(2011)Diorio, Cho, Duncan \& Akylas]{DiorioJFM}
{\sc \au{Diorio, J.~D.}, \au{Cho, Y.}, \au{Duncan, J.~H.} \& \au{Akylas,
  T.~R.}} \yr{2011}  \at{Resonantly forced gravity-capillary lumps on deep
  water. part 1. experiments}.  \jt{Journal of Fluid Mechanics}  \bvol{672},
  \pg{268--287}.

\bibitem[Duncan {\em et~al.\/}(1999)Duncan, Qiao, Philomin \& Wenz]{Duncan1999}
{\sc \au{Duncan, J.~H.}, \au{Qiao, H.}, \au{Philomin, V.} \& \au{Wenz, A.}}
  \yr{1999}  \at{Gentle spilling breakers: crest profile evolution}.
  \jt{Journal of Fluid Mechanics}  \bvol{379},  \pg{191--222}.

\bibitem[Falcon {\em et~al.\/}(2002)Falcon, Laroche \& Fauve]{Falcon2002}
{\sc \au{Falcon, \'E.}, \au{Laroche, C.} \& \au{Fauve, S.}} \yr{2002}
  \at{Observation of depression solitary surface waves on a thin fluid layer}.
  \jt{Phys. Rev. Lett.}  \bvol{89},  \pg{204501}.

\bibitem[Kass {\em et~al.\/}(1988)Kass, Witkin \& Terzopoulos]{Kaas1988}
{\sc \au{Kass, M.}, \au{Witkin, A.} \& \au{Terzopoulos, D.}} \yr{1988}
  \at{Snakes: Active contour models}.  \jt{International Journal of Computer
  Vision}  \bvol{1}~(4),  \pg{321--331}.

\bibitem[Kim \& Akylas(2005)]{Kim2005}
{\sc \au{Kim, B.} \& \au{Akylas, T.~R.}} \yr{2005}  \at{On gravity-capillary
  lumps}.  \jt{Journal of Fluid Mechanics}  \bvol{540},  \pg{337--351}.

\bibitem[Korteweg \& de~Vries(1895)]{KdV}
{\sc \au{Korteweg, D.~J.} \& \au{de~Vries, G.}} \yr{1895}  \at{On the change of
  form of long waves advancing in a rectangular canal, and on a new type of
  long stationary waves}.  \jt{Philosophical Magazine}  \bvol{39}~(240),
  \pg{422--443}.

\bibitem[Lee {\em et~al.\/}(1989)Lee, Yates \& Wu]{Lee1989}
{\sc \au{Lee, S-J.}, \au{Yates, G.~T.} \& \au{Wu, T.~Y.}} \yr{1989}
  \at{Experiments and analyses of upstream-advancing solitary waves generated
  by moving disturbances}.  \jt{Journal of Fluid Mechanics}  \bvol{199},
  \pg{569--593}.

\bibitem[Longuet-Higgins(1989)]{LH1989}
{\sc \au{Longuet-Higgins, M.~S.}} \yr{1989}  \at{Capillary-gravity waves of
  solitary type on deep water}.  \jt{Journal of Fluid Mechanics}  \bvol{200},
  \pg{451--470}.

\bibitem[Longuet-Higgins(1993)]{LH1993}
{\sc \au{Longuet-Higgins, M.~S.}} \yr{1993}  \at{Capillary-gravity waves of
  solitary type and envelope solitons on deep water}.  \jt{Journal of Fluid
  Mechanics}  \bvol{252},  \pg{703--711}.

\bibitem[Longuet-Higgins \& Zhang(1997)]{LHZhang1997}
{\sc \au{Longuet-Higgins, M.~S.} \& \au{Zhang, X.}} \yr{1997}  \at{Experiments
  on capillary-gravity waves of solitary type on deep water}.  \jt{Physics of
  Fluids (1994-present)}  \bvol{9}~(7),  \pg{1963--1968}.

\bibitem[Milewski(2005)]{Milewski2005}
{\sc \au{Milewski, P.~A.}} \yr{2005}  \at{Three-dimensional localized solitary
  gravity-capillary waves}.  \jt{Commun. Math. Sci.}  \bvol{3}~(1),
  \pg{89--99}.

\bibitem[Moisy {\em et~al.\/}(2009)Moisy, Rabaud \& Salsac]{Moisy2009}
{\sc \au{Moisy, F.}, \au{Rabaud, M.} \& \au{Salsac, K.}} \yr{2009}  \at{A
  synthetic schlieren method for the measurement of the topography of a liquid
  interface}.  \jt{Experiments in Fluids}  \bvol{46}~(6),  \pg{1021--1036}.

\bibitem[P\u{a}r\u{a}u {\em et~al.\/}(2005)P\u{a}r\u{a}u, Vanden-Broeck \&
  Cooker]{Parau2005}
{\sc \au{P\u{a}r\u{a}u, E.~I.}, \au{Vanden-Broeck, J.-M.} \& \au{Cooker,
  M.~J.}} \yr{2005}  \at{Nonlinear three-dimensional gravity-capillary solitary
  waves}.  \jt{Journal of Fluid Mechanics}  \bvol{536},  \pg{99--105}.

\bibitem[P\u{a}r\u{a}u {\em et~al.\/}(2007)P\u{a}r\u{a}u, Vanden-Broeck \&
  Cooker]{Parau2007}
{\sc \au{P\u{a}r\u{a}u, E.~I.}, \au{Vanden-Broeck, J.-M.} \& \au{Cooker,
  M.~J.}} \yr{2007}  \at{Three-dimensional capillary-gravity waves generated by
  a moving disturbance}.  \jt{Physics of Fluids (1994-present)}  \bvol{19}~(8).

\bibitem[Russell(1844)]{Russell}
{\sc \au{Russell, J.~S.}} \yr{1844}  \at{Reports on waves}.  \jt{Report of the
  fourteenth meeting of the British Association for the Advancement of Science}
   \pg{pp. 311--390}.

\bibitem[Wang \& Milewski(2012)]{Milewski2012}
{\sc \au{Wang, Zhan} \& \au{Milewski, P.~A.}} \yr{2012}  \at{Dynamics of
  gravity-capillary solitary waves in deep water}.  \jt{Journal of Fluid
  Mechanics}  \bvol{708},  \pg{480--501}.

\bibitem[Wu(1987)]{Wu1987}
{\sc \au{Wu, T.~Y.}} \yr{1987}  \at{Generation of upstream advancing solitons
  by moving disturbances}.  \jt{Journal of Fluid Mechanics}  \bvol{184},
  \pg{75--99}.

\bibitem[Zhang(1995)]{Zhang1995}
{\sc \au{Zhang, X.}} \yr{1995}  \at{Capillary-gravity and capillary waves
  generated in a wind wave tank: observations and theories}.  \jt{Journal of
  Fluid Mechanics}  \bvol{289},  \pg{51--82}.

\end{thebibliography}
\bibliographystyle{jfm}

\end{document}